\newcommand{\kreis}[3]{\put(#1
   ,0){\setlength{\unitlength}{7.8cm}\begin{picture}(0,0)(0,-0.05)
   \put(0,#2){\circle*{   0.02}}
   \put(0,#2){\line(0,1){#3}}
   \put(0,#2){\line(0,-1){#3}}
   \end{picture}}}
\newcommand{\kreisa}[3]{\put(#1
   ,0){\setlength{\unitlength}{7.8cm}\begin{picture}(0,0)(0,-0.05)
   \thicklines
   \put(0,#2){\circle{   0.02}}
   \thinlines
   \put(0,#2){\line(0,1){#3}}
   \put(0,#2){\line(0,-1){#3}}
   \end{picture}}}
\newcommand{\rfahne}[3]{\put(#1
   ,0){\setlength{\unitlength}{7.8cm}\begin{picture}(0,0)(0,-0.05)
   \put(0,#2){\makebox(0,0)[l]{\large{\hspace*{-0.12ex}$\triangleright$}}}
   \put(0,#2){\line(0,1){#3}}
   \put(0,#2){\line(0,-1){#3}}
   \end{picture}}}
\newcommand{\lfahne}[3]{\put(#1
   ,0){\setlength{\unitlength}{7.8cm}\begin{picture}(0,0)(0,-0.05)
   \put(0,#2){\makebox(0,0)[r]{\large{$\triangleleft$\hspace*{-0.2ex}}}}
   \put(0,#2){\line(0,1){#3}}
   \put(0,#2){\line(0,-1){#3}}
   \end{picture}}}
\newcommand{\quadrat}[3]{\put(#1
   ,0){\setlength{\unitlength}{7.8cm}\begin{picture}(0,0)(0,-0.05)
   \put(0,#2){\begin{picture}(0,0)(0,0)
               \thicklines
               \multiput(-0.008
               ,-0.008)(0.0008,0){20}{\line(0,1){0.016}}
               \end{picture}}
   \thinlines
   \put(0,#2){\line(0,1){#3}}
   \put(0,#2){\line(0,-1){#3}}
   \end{picture}}}
\newcommand{\quadrata}[3]{\put(#1
   ,0){\setlength{\unitlength}{7.8cm}\begin{picture}(0,0)(0,-0.05)
   \put(0,#2){\begin{picture}(0,0)(0,0)
               \thicklines
               \put(-0.008
               ,-0.008){\line(0,1){0.016}}
               \put(0.008
               ,-0.008){\line(0,1){0.016}}
               \put(-0.008
               ,-0.008){\line(1,0){0.016}}
               \put(-0.008
               ,0.008){\line(1,0){0.016}}
               \end{picture}}
   \thinlines
   \put(0,#2){\line(0,1){#3}}
   \put(0,#2){\line(0,-1){#3}}
   \end{picture}}}
\newcommand{\punkt}[9]{\put(#3
   ,0){\setlength{\unitlength}{5cm}\begin{picture}(0,0)(0,-0.25)
   \put(0,#4){\circle*{   0.015}}
   \put(0,#4){\line(0,1){#5}}
   \put(0,#4){\line(0,-1){#5}}
   \end{picture}}}
\newcommand{\punkta}[9]{\put(#3
   ,0){\setlength{\unitlength}{5cm}\begin{picture}(0,0)(0,-0.25)
   \put(0,#4){\circle{   0.03}}
   \end{picture}}}
\newcommand{\punktb}[9]{\put(#3
   ,0){\setlength{\unitlength}{20cm}\begin{picture}(0,0)(0,-0.125)
   \put(0,#4){\circle*{   0.005}}
   \put(0,#4){\line(0,1){#5}}
   \put(0,#4){\line(0,-1){#5}}
   \end{picture}}}
\newcommand{\punktc}[9]{\put(#3
   ,0){\setlength{\unitlength}{5cm}\begin{picture}(0,0)(0,-0.75)
   \put(0,#4){\line(0,1){#5}}
   \put(0,#4){\line(0,-1){#5}}
   \end{picture}}}
\newcommand{\punktd}[9]{\put(#3
   ,0){\setlength{\unitlength}{5cm}\begin{picture}(0,0)(0,-0.75)
   \put(0,#4){\circle*{   0.02}}
   \put(0,#4){\line(0,1){#5}}
   \put(0,#4){\line(0,-1){#5}}
   \end{picture}}}

\newcommand{\T}[1]{\rule[1.5ex]{0cm}{#1cm}}
\newcommand{\U}[1]{\rule[-#1cm]{0cm}{#1cm}}

\documentstyle{laa}
\begin{document}

\thesaurus{03(03.20.4; 11.17.3; 12.07.1)}
\title{The Hamburg Quasar Monitoring Program (HQM) at Calar Alto\thanks{Based
    on observations collected at the
    German-Spanish Astronomical Centre, Calar Alto,
    operated by the Max-Planck-Institut f\"ur Astronomie (MPIA), Heidelberg,
    jointly with the Spanish National Commission for Astronomy}:\\
    I. Low amplitude variability in quasars}
\author{U.~Borgeest
     \and
        K.-J.~Schramm}
\institute{Hamburger Sternwarte, Gojenbergsweg 112,
           D-W\,2050 Hamburg 80, Germany }
\date{Received date; accepted date}
\maketitle

\begin{abstract}
HQM is an optical broad-band photometric monitoring program carried out
since Sept.~1988. Our main intention is to search for indications of
microlensing in a sample of $\sim$\,100 selected quasars; however, we also
want to study the intrinsic variability. We use a CCD camera equipped to
the MPIA 1.2$\,$m telescope. Fully automatic photometric reduction relative
to stars in the frame is done within a few minutes after each exposure,
thus interesting brightness changes can be followed in detail. The typical
photometric error is 1--2\,\% for a 17.5\,mag quasar, making HQM the most
accurate long-term quasar monitoring program yet carried out. The main
results of HQM which we discuss here are: (1) Concerning variability,
quasars form two, clearly distinct classes, optically violent variables
(OVVs) and non-OVVs. (2) All OVVs are radio loud and probably belong to the
blazar class. (3) Non-OVVs have lightcurve gradients of at most several
0.1\,mag\,yr$^{-1}$ in the quasar restframe and can be well fitted by
polynomials of low order. (4) Although our data cover only a relatively
short timespan, we conclude that there is a large fraction of quasars
which would be undetectable in photographic surveys using optical
variability alone as the selection criterium. (5) A broad class of
flat-spectrum radio quasars are no blazars, they are even less variable in
the optical than radio quiet objects. (6) There is some statistical
evidence for microlensing in our sample; if present it does, however, lead
only to low lightcurve gradients.
\end{abstract}

\section{Introduction}

For the HQM program (e.g.\  Borgeest et al.\ 1991a,b;
Borgeest \& Schramm \cite{BS92}; Schramm \& Borgeest \cite{SB92};
v.~Linde et~al.\ \cite{LBSG93}; Schramm et al.\ \cite{SBCW93}), we mainly
concentrate on quasars which have a good chance of being influenced by
gravitational microlensing. Our intentions are to detect and analyse
characteristic microlensing features in the lightcurves and to determine
the time delays between the images of some multiply macrolensed quasars.
Microlensing is probably a very rare phenomenon so that most of the
results can be used ``only'' for a study of the intrinsic variability.
To distinguish microlensing flares from intrinsic events in a measured
lightcurve, high-quality data are required. Most published optical
variability data are based on photographic measurements and
therefore have a typical relative error of $ \sim \! 0.1 \,$mag.

In this paper, being the first one in a series, we exclude from the
discussion those quasars of our sample which were known to be optically
violent variables (OVVs) prior to our program. We {\em define\/} an object
as an OVV if it shows variations $\ga$\,0.5\,mag with gradients
$\ga$\,5\,mag\,yr$^{-1}$ in the quasar restframe. We here discuss
objects with measurements sufficiently spread over a time-span
$\ga$\,2\,yrs. Some interesting properties of these quasars are given in
Table~1; Table~4 lists values of some reasonable parameters through which
the lightcurve shapes can be quantified. In Fig.~1, we present a selection of
our lightcurves; the other curves will be presented in Schramm et~al.\
(\cite{SBKL93a}, hereafter Paper~II). A special discussion of the flaring
characteristics of some BL\,Lac objects and other OVVs will appear
in Scramm et al.\ (\cite{SBKL93b}, see also Borgeest \& Schramm
\cite{BS92}). One of the most interesting OVVs in our sample is 3C\,345;
our lightcurve together with a detailed discussion of possible variability
mechanisms can be found in a separate publication (Schramm et al.
\cite{SBCW93}; cf.\ also Borgeest \& Schramm \cite{BS92} and Schramm \&
Borgeest \cite{SB92}). For another two objects, we already published
photometric data separately: GC\,0248+430 (cf.\ Fig.~1), a quasar behind a
tidal tail of a merger galaxy system (Borgeest et~al.\ \cite{BDHK91},
hereafter BDHKS), and 0836+710, a high redshift $\gamma$-ray source
(v.~Linde et~al.\ \cite{LBSG93}). The lightcurves of the multiply
macrolensed quasars in our sample, together with models for the time
delays, will be discussed in subsequent papers.

\section{Observations and reduction}

\subsection{Photometric measurements}

We use a CCD camera at the
Cassegrain focus of the MPIA 1.2m telescope which has been equipped with
different chips, in 1988 with an RCA 15$ \mu $ chip ($640 \! \times \!
1024$, pixel size 0.315$''$) and later various, but similar, coated GEC 22$
\mu $ chips ($410 \! \times \! 580$, pixel size 0.462$''$). We measure the
quasar fluxes through standard Johnson broad-band filters ($R\/$, $V\/$ and
$B\/$) relative to stars included in the frames. The data reduction is
carried out automatically, immediately after the observation, on a
$\mu$VAX\,3200 workstation. Thus interesting features in the lightcurves
can be followed with an adequate time resolution, provided
that observing time is available at the telescope.
Usually, the observations are made in the $R\/$-band, where the quantum
efficiency of the CCD chips is best; other filters are only used if
significant variability is recognized in $R\/$.
A 0.01\,mag accuracy (in relative photometry) in the lightcurve of a
$\sim 17.5\,$mag quasar could be reached in this way also for
``non-photometric'' conditions with a typical exposure time of 500 sec (in
$R\/$). The error bars ($\pm1\sigma$) given in Fig.~1 are calculated from
$\sigma^2=\sigma_{\rm ref}^2+\sigma_{\rm fit}^2$ where
$\sigma_{\rm ref}$ is given by the weighted mean of the deviations
of the reference star magnitudes from their average values and
$\sigma_{\rm fit}$ is the fit error obtained by fitting a two-dimensional
Gaussian function to the quasar profil. $\sigma$ is therefore
an upper limit for the real error as can also be seen by comparing the
$\sigma_j$ in Table~4 with $\bar{\sigma}$ or $\hat{\sigma}$ (cf.~Eq.~[2]).

\subsection{Distribution of observing campaigns}

For our program, 15\%--30\% of the observation time at the MPIA 1.2$\,$m
telescope was available during the last years. The program started with
regular observations in May 1989. A test period in Sept./Oct. 1988 gave
additional data (see e.g. Borgeest et al. \cite{BKRS91} and Schramm \&
Borgeest \cite{SB92}). In order to analyse long-term features in the
lightcurves, observations are almost equally distributed over the whole
year with a typical spacing of two to four weeks between the single
campaigns. For a more detailed study of OVVs, nightly photometry is carried
out once or twice a year, during campaigns of a few weeks. Until now we
obtained more than 7,000 useful photometric data points. Although we tried
to observe with regular spacings, a lot of gaps in the lightcurves occurred
due to bad weather conditions, closure of the observatory or technical
problems.

\subsection{Automatic reduction}

Since we want to react immediately in the case of a microlensing
high-amplification event, a $\mu$Vax$\,$3200-workstation is connected
to the MPIA computer which produces the raw frames.
The software package ``HQM'' has been developed in Hamburg;
it is much faster than standard image
processing software: the photometric reduction of one frame is carried out
within 2 to 3\,min, during the following exposure. Generally,
each quasar is observed twice (e.g.\ 100 and 500\,sec) in the Johnson $R\/$
filter band for one point in the lightcurve. The shorter exposure is made
first and then used for telescope aquisition; hence, there is an angular
shift between the exposures. The shorter exposure is very useful for
recognizing photometric errors due to cosmic-ray events in one object or chip
defects. If a significant difference is found, a third measurement
is carried out. In the case of a real brightness change, additional
exposures in $B\/$ and $V\/$ are made.
The HQM-calculations include the following steps: bias subtraction,
flatfield correction, object identification and two-dimensional Gaussian
fits for the quasar and all stars that can be used as photometric
references. For a 500\,sec exposure (in $R\/$) of a 17.5\,mag quasar, this
procedure leads to an error in relative photometry typically between 0.005
and 0.02\,mag, depending on seeing, atmospheric
transmission, moon light and number of bright reference stars.
All these sources of error influence the photon
statistics, which has been proved to give the best estimate for the
total error (the chip read-out noise can be neglected in most cases). Other
sources of error, like cosmic ray events in one of the objects, chip
defects or variable reference stars, are rare and the software is designed
to recognize and correct for them in most cases.
Some data have been taken under very bad conditions, e.g.\
observation through clouds, seeing $ \ga \! 3''$, full moon near zenith
(which causes scattered light on the chip). Even for these cases, the errors
in relative photometry ($ \la \! 0.05 \, $mag) are of the order of (or even
better than) those for photographic measurements under good conditions.
Such ``very bad'' data have been included in Fig.~1 only if no better
measurements were made during the same night.

\subsection{POSS photometry}

We used the CCD frames obtained during our program to calibrate the
fields around the quasars on $R\/$-prints of the Palomar Observatory Sky
Survey (POSS). The corresponding Schmidt plates had been obtained during
the years 1949 to 1955 so that we were able to determine optical variations
for the $\sim$\,40\,yr period from $\sim$\,1950 to $\sim$\,1990. The
calibration was carried out by fitting a linear function to the pairs ($R_i
\,$,$\, x_i$), where $R_i$ are the $R\/$ magnitudes of the stars from the CCD
frames and $x_i$ are the diameters of the black stellar disks on the
prints. Each star and quasar diameter has been measured four times, twice
in S--N and W--E directions, with the left and right eye, respectively. In
Table~4, we list the results of the POSS photometry. The given errors
result from the fit errors of linear regression and the internal errors
of diameter determination. For some quasars which are marked in
Table~4, the errors are rather large since the photometric sequence
contains $\leq\,$4 stars or the star fluxes differ strongly from that
of the quasar (for more details see Linnert 1992).

\section{Candidates for microlensing}

Our sample includes more than 100 objects selected with respect to
different criteria and observed with different priority. Tables~1 and 4
list the HQM-quasars which are discussed in this article.

Multiply macrolensed quasars are the only objects for which one can
unambiguously separate microlensing from the intrinsic variability. Since
the macrolenses are very probably massive galaxies, one may
expect relatively high values for the microlensing optical depth
and the shear (cf.\ Kayser et al.\ \cite{KRS86}, hereafter KRS).
Monitoring of these objects is also very important to obtain the time
delays between the components from which limits on Hubble's constant
(Borgeest \& Refsdal \cite{BR84}, Falco et al. \cite{FGS91}) and a
relative accurate estimate of the lens mass (Borgeest \cite{Bor86}) can be
deduced. Unfortunately, most of these objects are not well resolved under
average conditions on Calar Alto and the angular scale of the CCD chips
which were used does not yield a sufficient sampling, so that the
photometric reduction is not straight forward. The work is still in process;
we shall discuss our data on multiply lensed quasars in subsequent papers.

\begin{table*}
\caption[ ]{HQM-quasars discussed in this article.
            Following the name, the selection
            technique is indicated: O (optical), R (radio), X (X-ray);
            $V$-magnitudes marked with ``\#'' are from the HST
            snapshot survey (Maoz et al.\ \cite{MBSB93}),
            otherwise from V\'eron-Cetty \&
            V\'eron (\cite{VV91}, hereafter VV);
            $S_6$ is the radio flux at 6cm, $\alpha$ the radio spectral
            index ($S\propto \nu^{-\alpha}$, from VV);
            $S_{\rm X}$ the Einstein
            X-ray flux in $10^{-13}$erg\,s$^{-1}$cm$^{-2}$
            between 0.3 and 3.5\,keV (for Refs
            see Hewitt \& Burbidge \cite{HB89}); in the
            following columns, redshifts and restframe equivalent
            widths (together for both lines) of MgII-absorption
            systems are given; the last columns list properties of
            foreground galaxies, $\theta_{\rm a}$ is the
            approximate radius of a galaxy on a POSS blue print. All
            references are given in Table~2. Abbreviations: str --
            strong, med -- medium, wk -- weak, nd -- not detected, x --
            detected}
\scriptsize
\begin{flushleft}
\begin{tabular}{lllllccrr|ccr|lccclr}
\hline
\hline
           &       &             & &&&&&
           &                    &                     &
           &                    &&
           &                    &                     &\\

Object     & Name & &$z_{\rm em}$ & \multicolumn{1}{c}{$V$} & $-M_V$
           & $S_6$ & \multicolumn{1}{c}{$\alpha$}
           & \multicolumn{1}{c|}{$S_{\rm X}$}
           & $z_{\rm abs}$ & \multicolumn{1}{c}{$W_{\rm r}$}
           & \multicolumn{1}{c|}{Ref}
           & $z_{\rm gal}$
           & \multicolumn{2}{c}{$\theta_{\rm gal}$}
           & $r_{\rm gal}$      & \multicolumn{1}{c}{$V_{\rm gal}$}&Ref\\
           &         &           &&&&[Jy]&
           &
           &                    & \multicolumn{1}{c}{[\AA]}         &
           &                    & \multicolumn{1}{c}{[\arcsec]}
           & $\!\!\!$[$\theta_{\rm a}$]$\!\!\!$
           & $\!\!\!\!$[kpc]$\!\!\!\!$\U{0.1}    & &\\
\hline
\hline
0003+158   & PHL$\,$658&ORX & 0.450 & 16.4 & 26.0 & 0.34  & 0.59 & 32.2\T{0.1}
           & --                 &                     &
           & --                 &&
           &                    &                     &\\
0007$-$000 & UM$\,$208&O  & 2.31  & 18.7$^{\#}$ & 27.8 & --    & -- & --
           & --                 &                     &
           &                    & 25&1.3
           &                    & 18                  & \ref{32}\\
0013$-$004 & UM$\,$224&O  & 2.086 & 18.2$^{\#}$ & 28.4 & --    & -- & --
           & 0.447              & 1.1                 & \ref{d}
           & --                 &&
           &                    &                     &\\
0014+813   & S5     &R     & 3.380 & 16.5$^{\#}$ & 30.8 & 0.55  & 0.17 & --
           & 1.111              & 1.5                 & \ref{3}
           & --                 &&
           &                    &                     &\\
&           &                    &  &&&&&
           & 1.113              & 5.0                 &
           & --                 &&
           &                    &                     &\\
0038$-$019 & PKS    &RX    & 1.690 & 18.5 & 27.4 & 0.28  & 1.10 & 3.6
           & --              &                 &
           & 0.017              & 60 & 1.4
\U{0.1}    & 30                 & 14.4                    & \ref{10}\\
\hline
0058+019   & PHL$\,$938&O & 1.955 & 17.1$^{\#}$ & 29.3 & --    & -- & --\T{0.1}
           & 0.613              & 3.2                 & \ref{d}
           & --                 &&
           &                    &                     &\\
0104+318   & 1E       &X  & 2.027 & 18.9 & 27.6 & --    & -- & 4.2
           & nd                 &                 & \ref{4}
           & 0.111              & 10 & 0.9
           & 27                 & 17.5                & \ref{10}\\
0151+045   & PHL$\,$1226&O& 0.404 & 17.9$^{\#}$ & 24.2 & --    & -- & --
           & 0.160              & 3.1                 & \ref{5b}
           & 0.160              & 6&$\sim$\,1
           & 21                 & 19.1                & \ref{6}\\
&           &                    & &&&&&
           &                    &                     &
           & 0.160              & 11&2.5
           & 40                 & 20.2                &\\
&           &                    & &&&&&
           &                    &                     &
           & 0.018              & 55&1.8
           & 29                 &                     &\\
0153+744   & S5        &R & 2.338 & 18.0$^{\#}$ & 28.5 & 1.51  & 0.32 & --
           & --                 &                     &
           & --                 &&
           &                    &                     &\\
0248+430   & GC        &R & 1.316 & 17.6$^{\#}$ & 27.6 & 1.21  & $-$0.38 & --
           & 0.394              &                     &  \ref{8a}
           & 0.051              & 15&0.9
\U{0.1}    & 23                 & 15.5                & \ref{9}\\
\hline
0446$-$208 & MC1       &R & 1.896 & 17.0 & 29.3 & --  & -- & --\T{0.1}
           & --                 &                     &
           & 0.067              & 13&$\sim$\,1
           & 23                 &                     & \ref{9b}\\
0454+039   & PKS        &R& 1.345 & 16.5 & 28.8 & 0.43  & $-$0.12 & --
           & 0.860              & 2.8                 &  \ref{9c}
           & --                 &&
           &                    &                     &\\
0731+653   & W1         &R& 3.035 & 18.2$^{\#}$ & 29.3 & 0.04  & -- & --
           & 0.932              & 1.4                 & \ref{9e}
           & --                 & &
           &                    &                     & \\
0745+557   & 1E         &X& 0.174 & 17.8 & 22.2 & --    & -- & 8.5
           & --                 &                 &
           & 0.004              & 78 & 2.4
           & 9                & 15.3                & \ref{10}\\
0805+046   & 4C$\,$05.34&RX& 2.877 & 18.2 & 29.3 & 0.31  & 0.62 & x
           & 0.703              & 1.1                 & \ref{10a}
           & --                 &&
           &                    &                     &\\
&           &                    & &&&&&
           & 0.960              & 2.0                 &
           &                    & &
           &                    &                     &\\
&           &                    & &&&&&
           & 1.014              & 1.6                    &
           &                    & &
\U{0.1}    &                    &                     &\\
\hline
0809+483   & 3C$\,$196  &RX& 0.871 & 17.8 & 26.2 & 4.36  & 0.91 & 3.0\T{0.1}
           & 0.437              & 4.7                 & \ref{11}
           &                    & 1.2&
           &                    & 21.9                & \ref{11}\\
&           &                    &  &&&&&
           & 0.871              &                     & \ref{11aa}
           &                    & 1.7&
           &                    & 20.7                & \\
0903+175   & H          &O& 2.756 & 18.0$^{\#}$ & 29.2 & --    & -- & --
           & --                 &                     &
           & 0.127              & 4&
           & 12                & 18                  & \ref{15a}\\
0955+326   & Ton$\,$469 &RX& 0.533 & 15.8 & 27.1 & 0.85  & 0.28 & x
           & 0.513              & 0.2                 &  \ref{15aa}
           & 0.005              & 114&2.0
           & 16                 &                     & \ref{9b}\\
1011+250   & Ton$\,$490 &ORX& 1.631 & 16.6 & 29.3 & 0.61  & $-$0.26 & x
           & 0.258              & wk                  & \ref{15e}
           & --                 &&
           &                    &                     &\\
1109+357   & 1E         &X& 0.909 & 18.1 & 26.0 & --    & -- & 3.8
           & --                 &                 &
           & 0.027              & 26 & 0.7
\U{0.1}    & 21                 & 14.5                & \ref{10}\\
\hline
1150+497   & LB$\,$2136 &OR& 0.334 & 17.1 & 24.6 & 1.12  & 0.57 & --\T{0.1}
           & --                 &                     &
           & 0.290              & 7&
           & 38                 & 20                  & \ref{17}\\
1209+107   & KP$\,$9    &O& 2.191 & 18.1$^{\#}$ & 28.6 & --    & -- & --
           & 0.630              & 5.0                 & \ref{18}
           & 0.63?              & 1.4&
           & 11                & 22.3                & \ref{19}\\
&           &                    & &&&&&
           & 0.393              & 1.6                 & \ref{19}
           & 0.393              & 7&
           & 45                 & 21.2                &\\
1219+755   & Mkn$\,$205 &OX& 0.072 & 15.2 & 22.9 & 0.00  & -- & 170.0
           & --                 &                     &
           & 0.006              & 42&0.5
           & 7                &                     & \ref{10}\\
1222+228   & Ton\,1530  &OX& 2.051 & 16.6 & 29.9 & 0.01  & -- & 2.1
           & 0.669              & 0.6                 & \ref{d}
           & --                 & &
           &                    &                     & \\
1332+552   & 4C$\,$55.27&R& 1.249 & 16.0 & 29.1 & 0.13  & 0.67 & --
           & 0.374              & str                 & \ref{21d}
           & 0.374
           & 5 &
\U{0.1}    & 32                 & 20.7                & \ref{21d}\\
\hline
1421+330   & Mkn$\,$679 &O& 1.904 & 16.7 & 29.6 & --    & -- & --\T{0.1}
           & 0.456              & 0.4                 & \ref{13a}
           & --                 &&
           &                    &                     &\\
1435+638   & S4         &R& 2.060 & 15.0 & 31.5 & 1.24  & 0.21 & --
           & nd                 &                     & \ref{d}
           & --                 &&
           &                    &                     &\\
1520+413   & SP$\,$43   &O& 3.1   & 18.7$^{\#}$ & 28.6 & --    & -- & --
           & --                 &                     &
           & --                 &&
           &                    &                     & \\
1522+101   & PG         &O& 1.321 & 16.2$^{\#}$ & 29.1 & 0.00  & -- & --
           & --                 &                     &
           &                    & 0.0 &
           & 0                   &                     & \ref{26aa}\\
1604+290   & KP$\,$63   &OR& 1.97  & 17.0 & 29.4 & 0.00  & 1.49 & --
           & --                 &                     &
           &
           & 7.5&
\U{0.1}    &                    & 19.5                & \ref{32}\\
\hline
1630+377   & PG         &O& 1.471 & 16.1 & 29.5 & 0.00  & -- & --\T{0.1}
           & --                 &                     &
           &                    & 9.5&
           &                    & 21                  & \ref{32}\\
&           &                    & &&&&&
           &                    &                     &
           &                    & 15&
           &                    & 20                  &\\
1633+267   & KP$\,$83   &O& 1.84  & 17.0 & 29.2 & --    & -- & --
           & --                 &                     &
           &                    & 39&
           &                    & 18                  & \ref{32}\\
1634+706   & PG         &OX& 1.334 & 14.9 & 30.5 & 0.00  & -- & 8.0
           & 0.993              & 1.0                 & \ref{1}
           & --                 &&
           &                    &                     &\\
&           &                    & &&&&&
           & 1.046              & 0.1                 &
           &                    &&
           &                    &                     &\\
1640+396   & 1E         &XR& 0.540 & 18.3 & 24.6 & 0.03  & --& 6.6
           & --              &                 &
           & 0.034              & 200 & 4.7
           & 200                & 15.2                & \ref{10}\\
1700+642   & HS         &O& 2.72  & 16.1 & 31.1 & --    & -- & --
           & --                 &                     &
           & 0.086              & 11&
           & 24                 & 18.8                & \ref{27}\\
&           &                    & &&&&&
           &                    &                     &
           & 0.19               & 18&
\U{0.1}    & 75                 &                     &\\
\hline
1701+610   &            &X& 0.164 & 17.0 & 22.9 & --    & -- & x\T{0.1}
           & --                 &                     &
           & 0.052              & 29&$\sim$\,4
           & 40                 &                     & \ref{27aa}\\
&           &                    & &&&&&
           &                    &                     &
           & 0.052              & 38&$\sim$\,5
           & 53                 &                     &\\
1704+608   & 3C$\,$351  &RX& 0.371 & 15.3 & 26.6 & 1.21  & 0.84 & 10.4
           & 0.163              &                     & \ref{27a}
           & --                 &&
           &                    &                     & \\
&           &                    & &&&&&
           & 0.222              &                     &
           &                    &&
           &                    &                     &\\
1715+535   & PG         &O& 1.929 & 16.5 & 29.8 & 0.00  & -- & --
           & 0.367              & 0.6                 & \ref{d}
           & --                 &&
           &                    &                     & \\
1718+481   & PG         &O& 1.084 & 14.7 & 29.9 & 0.14  & -- & --
           & 0.713              &                     & \ref{1}
           & --                 &&
           &                    &                     &\\
1821+643   & E          &X& 0.297 & 14.2 & 27.1 & --    & -- & x
           & --                 &                     &
           & --
           &&
\U{0.1}    &                    &                     & \\
\hline
1857+566   & 4C$\,$56.28&R& 1.595 & 17.3 & 28.5 & 0.23  & 1.44 & --\T{0.1}
           & 0.715              & 1.3                 & \ref{28aa}
           &                    &3&
           &                    &                     & \ref{15}\\
&           &                    &        &&&&&
           & 1.106              & 0.6                    &
           &                    &&
           &                    &                     & \\
&           &                    &        &&&&&
           & 1.235              & 1.5                 &
           &                    &&
           &                    &                     & \\
2126$-$158 & PKS        &RX& 3.266 & 17.0$^{\#}$ & 30.3 & 1.24  & $-$0.09 &
26.0
           & 2.022              & 1.0                 & \ref{28a}
           & --                 &&
           &                    &                     &\\
2134+004   & PHL$\,$61  &OXR& 1.936 & 17.1$^{\#}$ & 29.2 & 11.49 & $-$0.67 & x
           & --                 &                     &
           & --                 &&
           &                    &                     &\\
2215$-$037 &            &X& 0.241 & 17.2 & 23.7 & --    & -- & x
           & --                 &                     &
           & 0.061              & 14&$\sim$\,2
           & 23                 &                     & \ref{28d}\\
2251+244   & 4C$\,$24.61&R& 2.328 & 18.5$^{\#}$ & 28.0 & 0.79  & 0.93 & --
           & 1.090              & 0.9                 & \ref{28aa}
           &                    & 0.8 &
\U{0.1}    &                    &                     & \ref{15}\\
\hline
2308+098   & 4C$\,$09.72&R& 0.432 & 16.0 & 26.3 & 0.25  & 0.59 & --\T{0.1}
           & --                 &                     &
           & 0.173              & 9&
           & 35                 &                     & \ref{17}\\
2354+144   & PKS        &R& 1.813 & 18.2 & 27.9 & 0.37  & 0.81 & --
           & 1.576              & 1.6                 & \ref{28aa}
           &                    & 4.4&
\U{0.1}    &                    &                     & \ref{15}\\
\hline
\hline
\end{tabular}
\normalsize
\end{flushleft}
\end{table*}

\subsection{Quasar-galaxy associations}

We have selected from the literature many quasars with closely associated
foreground galaxies. In our opinion, only for
those foreground galaxies which lie not too far (i.e.\ $\la
50\,$kpc)\footnote{We use $H_0 = 50\,$km\,s$^{-1}\,$Mpc$^{-1}$, $q_0=0$
throughout this paper.} from the line of sight to a quasar, a relatively high
optical depth for microlensing can be expected. Also interesting
are those cases where a quasar shines through a foreground galaxy within
their optical isophotes (cf.\ column~15 of Table~1). We have denoted these
objects by {\em G1\/}; other quasars with associated foreground galaxies
are denoted by {\em G2\/}, see Table~3.

\begin{table}[t]
\begin{flushleft}
\caption{References for Table~1}
\begin{tabular}{l}
\hline
\parbox{8cm}{%
\begin{enumerate}
\item \label{32} This work
\item \label{1} Bechtold et al.\ \cite{BGWS84}
\item \label{d} Sargent et. al.\ \cite{SBS88}
\item \label{3} K\"uhr et al.\ \cite{KMRS84}
\item \label{10} Stocke et al.\ \cite{SSMG87}
\item \label{4} Blades \cite{Bla88}
\item \label{5b} Bergeron et al.\ \cite{BBKT88}
\item \label{6} Bergeron \cite{Ber88a}
\item \label{8a} Sargent \& Steidel\ \cite{SS90}
\item \label{9} Borgeest et al.\ \cite{BDHK91}
\item \label{9b} Blades et al.\ \cite{BHM81}
\item \label{9c} Burbidge et al.\ \cite{BSWW77}
\item \label{9e} Sargent et. al.\ \cite{SSB89}
\item \label{10a} Chen et. al.\ \cite{CMPW81}
\item \label{11} Boiss\'e \& Boulade \cite{BB90}
\item \label{11aa} Foltz et al.\ \cite{FCW88}
\item \label{13a} Foltz et al.\ \cite{FWPS86}
\item \label{15} Crampton et al.\ \cite{CMFH89}
\item \label{15a} Djorgovski \& McCarthy \cite{DM85}
\item \label{15aa} Boksenberg \& Sargent \cite{BS78}
\item \label{15e} Carswell et al.\ \cite{CCSW76}
\item \label{17} Stockton \cite{Sto78}
\item \label{18} Cristiani \cite{Cri87}
\item \label{19} Arnaud et al.\ \cite{AHJL88}
\item \label{21d} Miller et al.\ \cite{MGS87}
\item \label{26aa} Magain et al.\ \cite{MRSS90}
\item \label{27} Reimers et al.\ \cite{RCGE89}
\item \label{27aa} Hutchings et al.\ \cite{HHR86}
\item \label{27a} Jenkins et al.\ \cite{JCWB87}
\item \label{28aa} Barthel et al. \cite{BTT90}
\item \label{28a} Usher \cite{Ush78}
\item \label{28d} Heckman et al.\ \cite{HBBS84}
\end{enumerate}} \\
\hline
\end{tabular}
\end{flushleft}
\normalsize
\end{table}

Stocke et al.\ (\cite{SSMG87}, hereafter SSMG) searched for bright ($ m_V
\la 18 $) galaxies near the members of a sample of more than 200 X-ray
selected AGN on POSS plates. They found that the redshift
distribution of those 10 AGN lying within 3 optical radii of foreground
galaxies have a much higher percentage of high redshift objects than the
total sample. They interpret this result in terms of an amplification bias
due to microlensing. The total sample includes a subsample of 56 AGN
complete in X-ray flux (Maccacaro et al. \cite{MGS84}) which was observed
by Rix \& Hogan (\cite{RH88}) by CCD imaging (search limit for galaxies $
m_R \simeq 20.5 $); they found 5 additional AGN near galaxies, but could
not verify the result of SSMG. The largest problem is the small
number of objects near foreground galaxies in both studies, therefore
statistically significant results are difficult to obtain out of their
redshift distributions. In our opinion it is, however, very interesting
that the two objects with the highest X-ray luminosity ($ L_X \ga 10^{28}
{\rm ergs Hz^{-1} s^{-1}} $) in the complete sample have a foreground
galaxy, compared to only 6 of the 54 less luminous ones. Both objects,
1E\,0038$-$019 and 1E\,0104+318, are highly luminous in the optical, too;
and the first one is known to be radio loud. These might be hints on a
triple or double amplification bias, respectively (cf.\ Borgeest et al.\
\cite{BLR91}).

We have included 6 objects of the SSMG sample in the HQM program
(0038$-$019, 0104+318, 0745+557, 1109+357, 1219+755, 1640+396) to search
for indications of microlensing variability.

\begin{figure*}

\vspace*{0.4cm}

\begin{picture}(18 ,2.5 )(0,0)
\put(0,0){\setlength{\unitlength}{0.01059cm}%
\begin{picture}(1700, 236.111)(7350,0)
\put(7350,0){\framebox(1700, 236.111)[tl]{\begin{picture}(0,0)(0,0)
        \put(1700,0){\makebox(0,0)[tr]{\bf{0014+813}\T{0.4}
                                 \hspace*{0.5cm}}}
    \end{picture}}}

\thicklines
\put(7350,0){\setlength{\unitlength}{5cm}\begin{picture}(0,0)(0,-0.25)
   \put(0,0){\setlength{\unitlength}{1cm}\begin{picture}(0,0)(0,0)
        \put(0,0){\line(1,0){0.3}}
        \end{picture}}
   \end{picture}}

\put(9050,0){\setlength{\unitlength}{5cm}\begin{picture}(0,0)(0,-0.25)
   \put(0,0){\setlength{\unitlength}{1cm}\begin{picture}(0,0)(0,0)
        \put(0,0){\line(-1,0){0.3}}
        \end{picture}}
   \end{picture}}

\thinlines
\put(7350,0){\setlength{\unitlength}{5cm}\begin{picture}(0,0)(0,-0.25)
   \multiput(0,0)(0,0.1){3}{\setlength{\unitlength}{1cm}%
\begin{picture}(0,0)(0,0)
        \put(0,0){\line(1,0){0.12}}
        \end{picture}}
   \end{picture}}

\put(7350,0){\setlength{\unitlength}{5cm}\begin{picture}(0,0)(0,-0.25)
   \multiput(0,0)(0,-0.1){3}{\setlength{\unitlength}{1cm}%
\begin{picture}(0,0)(0,0)
        \put(0,0){\line(1,0){0.12}}
        \end{picture}}
   \end{picture}}

\put(9050,0){\setlength{\unitlength}{5cm}\begin{picture}(0,0)(0,-0.25)
   \multiput(0,0)(0,0.1){3}{\setlength{\unitlength}{1cm}%
\begin{picture}(0,0)(0,0)
        \put(0,0){\line(-1,0){0.12}}
        \end{picture}}
   \end{picture}}

\put(9050,0){\setlength{\unitlength}{5cm}\begin{picture}(0,0)(0,-0.25)
   \multiput(0,0)(0,-0.1){3}{\setlength{\unitlength}{1cm}%
\begin{picture}(0,0)(0,0)
        \put(0,0){\line(-1,0){0.12}}
        \end{picture}}
   \end{picture}}

   \put(7527.5, 236.111){\setlength{\unitlength}{1cm}\begin{picture}(0,0)(0,0)
        \put(0,0){\line(0,-1){0.2}}
        \put(0,0.2){\makebox(0,0)[b]{\bf 1989}}
   \end{picture}}
   \put(7892.5, 236.111){\setlength{\unitlength}{1cm}\begin{picture}(0,0)(0,0)
        \put(0,0){\line(0,-1){0.2}}
        \put(0,0.2){\makebox(0,0)[b]{\bf 1990}}
   \end{picture}}
   \put(8257.5, 236.111){\setlength{\unitlength}{1cm}\begin{picture}(0,0)(0,0)
        \put(0,0){\line(0,-1){0.2}}
        \put(0,0.2){\makebox(0,0)[b]{\bf 1991}}
   \end{picture}}
   \put(8622.5, 236.111){\setlength{\unitlength}{1cm}\begin{picture}(0,0)(0,0)
        \put(0,0){\line(0,-1){0.2}}
        \put(0,0.2){\makebox(0,0)[b]{\bf 1992}}
   \end{picture}}
   \put(8987.5, 236.111){\setlength{\unitlength}{1cm}\begin{picture}(0,0)(0,0)
        \put(0,0){\line(0,-1){0.2}}
        \put(0,0.2){\makebox(0,0)[b]{\bf 1993}}
   \end{picture}}
    \multiput(7350,0)(50,0){33}%
        {\setlength{\unitlength}{1cm}\begin{picture}(0,0)(0,0)
        \put(0,0){\line(0,1){0.12}}
    \end{picture}}
    \put(7500,0){\setlength{\unitlength}{1cm}\begin{picture}(0,0)(0,0)
        \put(0,0){\line(0,1){0.2}}
    \end{picture}}
    \put(7750,0){\setlength{\unitlength}{1cm}\begin{picture}(0,0)(0,0)
        \put(0,0){\line(0,1){0.2}}
    \end{picture}}
    \put(8000,0){\setlength{\unitlength}{1cm}\begin{picture}(0,0)(0,0)
        \put(0,0){\line(0,1){0.2}}
    \end{picture}}
    \put(8250,0){\setlength{\unitlength}{1cm}\begin{picture}(0,0)(0,0)
        \put(0,0){\line(0,1){0.2}}
    \end{picture}}
    \put(8500,0){\setlength{\unitlength}{1cm}\begin{picture}(0,0)(0,0)
        \put(0,0){\line(0,1){0.2}}
    \end{picture}}
    \put(8750,0){\setlength{\unitlength}{1cm}\begin{picture}(0,0)(0,0)
        \put(0,0){\line(0,1){0.2}}
    \end{picture}}
    \put(9000,0){\setlength{\unitlength}{1cm}\begin{picture}(0,0)(0,0)
        \put(0,0){\line(0,1){0.2}}
    \end{picture}}

\punkt{01/09/88}{01:29}{7405.562}{ -0.010}{0.031}{ 600}{2.68}{
6698}{1.2CA}
\punkt{04/09/88}{00:54}{7408.538}{ -0.023}{0.028}{ 600}{1.46}{
730}{1.2CA}
\punkt{05/10/88}{22:08}{7440.422}{ -0.038}{0.019}{ 500}{1.38}{
707}{1.2CA}
\punkt{24/06/89}{03:09}{7701.632}{ -0.040}{0.024}{ 500}{1.42}{
1258}{1.2CA}
\punkt{25/06/89}{02:48}{7702.617}{ -0.015}{0.015}{ 500}{1.74}{
934}{1.2CA}
\punkt{26/06/89}{02:52}{7703.620}{ -0.023}{0.014}{ 500}{1.31}{
1026}{1.2CA}
\punkt{27/06/89}{03:20}{7704.640}{ -0.018}{0.015}{ 500}{2.24}{
928}{1.2CA}
\punkt{21/07/89}{02:25}{7728.601}{ -0.043}{0.018}{1000}{1.86}{
4125}{1.2CA}
\punkt{23/07/89}{03:01}{7730.626}{ -0.022}{0.039}{ 500}{1.99}{
1328}{1.2CA}
\punkt{11/08/89}{14:13}{7750.093}{ -0.014}{0.011}{1000}{2.29}{
1105}{1.2CA}
\punkt{11/08/89}{15:00}{7750.125}{ -0.029}{0.016}{1000}{2.33}{
990}{1.2CA}
\punkt{11/08/89}{15:41}{7750.154}{ -0.030}{0.018}{1000}{2.19}{
1342}{1.2CA}
\punkt{31/08/89}{23:39}{7770.486}{ -0.010}{0.014}{ 500}{2.31}{
1097}{1.2CA}
\punkt{02/09/89}{01:13}{7771.551}{ -0.031}{0.013}{ 500}{2.26}{
965}{1.2CA}
\punkt{03/09/89}{02:19}{7772.597}{ -0.011}{0.009}{ 500}{2.05}{
806}{1.2CA}
\punkt{01/10/89}{12:33}{7801.024}{ -0.034}{0.026}{ 500}{1.97}{
553}{1.2CA}
\punkt{16/10/89}{23:13}{7816.468}{ -0.037}{0.015}{1000}{2.64}{
1872}{1.2CA}
\punkt{31/10/89}{21:27}{7831.394}{ -0.008}{0.012}{ 500}{1.45}{
543}{1.2CA}
\punkt{01/11/89}{20:46}{7832.366}{ -0.014}{0.012}{ 500}{1.53}{
609}{1.2CA}
\punkt{05/12/89}{19:16}{7866.303}{ -0.029}{0.016}{ 500}{2.03}{
832}{1.2CA}
\punkt{14/12/89}{23:36}{7875.483}{  0.020}{0.030}{ 500}{3.14}{
1117}{1.2CA}
\punkt{19/12/89}{20:36}{7880.359}{  0.000}{0.006}{ 500}{1.40}{
616}{1.2CA}
\punkt{22/12/89}{19:25}{7883.309}{ -0.008}{0.015}{ 500}{2.69}{
628}{1.2CA}
\punkt{29/01/90}{18:34}{7921.274}{  0.018}{0.025}{ 500}{3.10}{
858}{1.2CA}
\punkt{08/02/90}{18:52}{7931.286}{ -0.003}{0.030}{ 500}{2.21}{
1217}{1.2CA}
\punkt{22/05/90}{03:29}{8033.646}{ -0.006}{0.022}{ 400}{1.87}{
2243}{1.2CA}
\punkt{01/08/90}{03:32}{8104.648}{ -0.011}{0.010}{ 500}{1.42}{
746}{1.2CA}
\punkt{25/09/90}{01:34}{8159.566}{  0.018}{0.030}{ 500}{1.57}{
868}{1.2CA}
\punkt{26/09/90}{01:34}{8160.566}{  0.017}{0.030}{ 500}{1.56}{
869}{1.2CA}
\punkt{27/09/90}{23:06}{8162.463}{ -0.026}{0.011}{ 500}{1.62}{
1161}{1.2CA}
\punkt{30/09/90}{00:02}{8164.502}{ -0.016}{0.030}{ 216}{2.32}{
992}{1.2CA}
\punkt{03/10/90}{22:22}{8168.432}{ -0.008}{0.035}{ 405}{2.79}{
3108}{1.2CA}
\punkt{19/10/90}{22:05}{8184.420}{ -0.002}{0.023}{ 500}{2.12}{
826}{1.2CA}
\punkt{21/10/90}{23:16}{8186.470}{ -0.010}{0.019}{ 500}{3.24}{
852}{1.2CA}
\punkt{22/10/90}{23:16}{8187.470}{ -0.007}{0.014}{ 500}{3.23}{
852}{1.2CA}
\punkt{22/12/90}{09:50}{8247.910}{ -0.019}{0.034}{ 500}{2.60}{
677}{1.2CA}
\punkt{22/02/91}{20:18}{8310.346}{  0.004}{0.013}{ 500}{2.23}{
1122}{1.2CA}
\punkt{25/05/91}{01:47}{8401.575}{ -0.008}{0.035}{ 500}{2.73}{
999}{1.2CA}
\punkt{26/07/91}{03:12}{8463.634}{  0.012}{0.019}{ 500}{2.29}{
1616}{1.2CA}
\punkt{02/08/91}{04:01}{8470.668}{  0.006}{0.010}{ 300}{1.09}{
782}{1.2CA}
\punkt{19/09/91}{02:04}{8518.586}{  0.010}{0.015}{ 500}{1.72}{
482}{1.2CA}
\punkt{19/09/91}{22:10}{8519.424}{  0.000}{0.008}{ 500}{1.36}{
1289}{1.2CA}
\punkt{22/09/91}{01:54}{8521.580}{ -0.001}{0.029}{ 500}{1.93}{
1561}{1.2CA}
\punkt{23/09/91}{00:14}{8522.510}{  0.014}{0.038}{ 500}{1.84}{
2257}{1.2CA}
\punkt{17/10/91}{22:28}{8547.436}{  0.001}{0.010}{ 500}{1.89}{
988}{1.2CA}
\punkt{28/10/91}{02:46}{8557.615}{  0.006}{0.020}{ 500}{2.26}{
1323}{1.2CA}
\punkt{22/09/92}{01:34}{8887.565}{  0.016}{0.027}{ 150}{2.78}{
552}{1.2CA}
\punkt{22/09/92}{01:34}{8887.565}{  0.016}{0.027}{ 150}{2.78}{
552}{1.2CA}

\end{picture}}

\end{picture}

\vspace*{-0.02cm}

\begin{picture}(18 ,2.5 )(0,0)
\put(0,0){\setlength{\unitlength}{0.01059cm}%
\begin{picture}(1700, 236.111)(7350,0)
\put(7350,0){\framebox(1700, 236.111)[tl]{\begin{picture}(0,0)(0,0)
        \put(1700,-132){\makebox(0,0)[tr]{\bf{0038-$\!$-019}\T{0.4}
                                 \hspace*{0.5cm}}}
    \end{picture}}}

\thicklines
\put(7350,0){\setlength{\unitlength}{5cm}\begin{picture}(0,0)(0,-0.25)
   \put(0,0){\setlength{\unitlength}{1cm}\begin{picture}(0,0)(0,0)
        \put(0,0){\line(1,0){0.3}}
        \end{picture}}
   \end{picture}}

\put(9050,0){\setlength{\unitlength}{5cm}\begin{picture}(0,0)(0,-0.25)
   \put(0,0){\setlength{\unitlength}{1cm}\begin{picture}(0,0)(0,0)
        \put(0,0){\line(-1,0){0.3}}
        \end{picture}}
   \end{picture}}

\thinlines
\put(7350,0){\setlength{\unitlength}{5cm}\begin{picture}(0,0)(0,-0.25)
   \multiput(0,0)(0,0.1){3}{\setlength{\unitlength}{1cm}%
\begin{picture}(0,0)(0,0)
        \put(0,0){\line(1,0){0.12}}
        \end{picture}}
   \end{picture}}

\put(7350,0){\setlength{\unitlength}{5cm}\begin{picture}(0,0)(0,-0.25)
   \multiput(0,0)(0,-0.1){3}{\setlength{\unitlength}{1cm}%
\begin{picture}(0,0)(0,0)
        \put(0,0){\line(1,0){0.12}}
        \end{picture}}
   \end{picture}}

\put(9050,0){\setlength{\unitlength}{5cm}\begin{picture}(0,0)(0,-0.25)
   \multiput(0,0)(0,0.1){3}{\setlength{\unitlength}{1cm}%
\begin{picture}(0,0)(0,0)
        \put(0,0){\line(-1,0){0.12}}
        \end{picture}}
   \end{picture}}

\put(9050,0){\setlength{\unitlength}{5cm}\begin{picture}(0,0)(0,-0.25)
   \multiput(0,0)(0,-0.1){3}{\setlength{\unitlength}{1cm}%
\begin{picture}(0,0)(0,0)
        \put(0,0){\line(-1,0){0.12}}
        \end{picture}}
   \end{picture}}

   \put(7527.5, 236.111){\setlength{\unitlength}{1cm}\begin{picture}(0,0)(0,0)
        \put(0,0){\line(0,-1){0.2}}
   \end{picture}}
   \put(7892.5, 236.111){\setlength{\unitlength}{1cm}\begin{picture}(0,0)(0,0)
        \put(0,0){\line(0,-1){0.2}}
   \end{picture}}
   \put(8257.5, 236.111){\setlength{\unitlength}{1cm}\begin{picture}(0,0)(0,0)
        \put(0,0){\line(0,-1){0.2}}
   \end{picture}}
   \put(8622.5, 236.111){\setlength{\unitlength}{1cm}\begin{picture}(0,0)(0,0)
        \put(0,0){\line(0,-1){0.2}}
   \end{picture}}
   \put(8987.5, 236.111){\setlength{\unitlength}{1cm}\begin{picture}(0,0)(0,0)
        \put(0,0){\line(0,-1){0.2}}
   \end{picture}}
    \multiput(7350,0)(50,0){33}%
        {\setlength{\unitlength}{1cm}\begin{picture}(0,0)(0,0)
        \put(0,0){\line(0,1){0.12}}
    \end{picture}}
    \put(7500,0){\setlength{\unitlength}{1cm}\begin{picture}(0,0)(0,0)
        \put(0,0){\line(0,1){0.2}}
    \end{picture}}
    \put(7750,0){\setlength{\unitlength}{1cm}\begin{picture}(0,0)(0,0)
        \put(0,0){\line(0,1){0.2}}
    \end{picture}}
    \put(8000,0){\setlength{\unitlength}{1cm}\begin{picture}(0,0)(0,0)
        \put(0,0){\line(0,1){0.2}}
    \end{picture}}
    \put(8250,0){\setlength{\unitlength}{1cm}\begin{picture}(0,0)(0,0)
        \put(0,0){\line(0,1){0.2}}
    \end{picture}}
    \put(8500,0){\setlength{\unitlength}{1cm}\begin{picture}(0,0)(0,0)
        \put(0,0){\line(0,1){0.2}}
    \end{picture}}
    \put(8750,0){\setlength{\unitlength}{1cm}\begin{picture}(0,0)(0,0)
        \put(0,0){\line(0,1){0.2}}
    \end{picture}}
    \put(9000,0){\setlength{\unitlength}{1cm}\begin{picture}(0,0)(0,0)
        \put(0,0){\line(0,1){0.2}}
    \end{picture}}

\punkt{03/09/88}{02:23}{7407.600}{  0.078}{0.037}{2000}{2.14}{
1913}{1.2CA}
\punkt{06/09/88}{00:43}{7410.530}{  0.006}{0.032}{ 500}{1.91}{
688}{1.2CA}
\punkta{11/10/88}{21:28}{7446.394}{  0.089}{0.066}{ 500}{2.77}{
1132}{1.2CA}
\punkt{13/08/89}{01:16}{7751.553}{ -0.032}{0.010}{1000}{2.02}{
989}{1.2CA}
\punkt{13/08/89}{01:38}{7751.568}{ -0.019}{0.009}{1000}{2.02}{
984}{1.2CA}
\punkt{13/08/89}{01:57}{7751.582}{ -0.013}{0.009}{1000}{2.20}{
985}{1.2CA}
\punkt{14/08/89}{03:09}{7752.631}{ -0.029}{0.009}{1000}{2.16}{
899}{1.2CA}
\punkta{01/09/89}{02:16}{7770.595}{ -0.001}{0.019}{ 499}{3.45}{
877}{1.2CA}
\punkt{02/10/89}{11:48}{7801.992}{ -0.050}{0.035}{1000}{2.03}{
679}{1.2CA}
\punkt{31/10/89}{21:45}{7831.406}{ -0.040}{0.015}{ 500}{1.80}{
536}{1.2CA}
\punkt{01/11/89}{21:21}{7832.390}{ -0.059}{0.014}{ 500}{1.70}{
598}{1.2CA}
\punkta{05/12/89}{19:37}{7866.318}{  0.039}{0.058}{ 500}{3.04}{
1345}{1.2CA}
\punkt{19/12/89}{19:48}{7880.325}{ -0.030}{0.020}{ 500}{1.86}{
648}{1.2CA}
\punkt{29/01/90}{18:53}{7921.287}{ -0.045}{0.038}{ 500}{3.07}{
762}{1.2CA}
\punkt{31/07/90}{03:31}{8103.647}{ -0.050}{0.008}{ 500}{1.52}{
612}{1.2CA}
\punkt{25/09/90}{00:51}{8159.536}{ -0.027}{0.013}{ 500}{2.07}{
854}{1.2CA}
\punkt{26/09/90}{00:43}{8160.531}{ -0.009}{0.019}{ 100}{1.77}{
566}{1.2CA}
\punkt{26/09/90}{00:51}{8160.536}{ -0.024}{0.013}{ 500}{2.05}{
851}{1.2CA}
\punkt{22/10/90}{23:41}{8187.487}{ -0.039}{0.029}{ 500}{3.24}{
923}{1.2CA}
\punkta{23/10/90}{23:41}{8188.487}{ -0.043}{0.056}{ 500}{3.22}{
922}{1.2CA}
\punkta{}{}{8464.635}{0.048}{0.05}{}{}{}{}
\punkt{13/08/91}{02:54}{8481.621}{  0.020}{0.020}{ 500}{1.62}{
546}{1.2CA}
\punkt{22/08/91}{02:52}{8490.620}{ -0.013}{0.003}{ 500}{2.07}{
720}{1.2CA}
\punkt{20/09/91}{00:33}{8519.523}{  0.008}{0.018}{1000}{1.06}{
3108}{1.2CA}
\punkt{21/09/92}{00:34}{8886.524}{  0.107}{0.023}{ 300}{0.91}{
763}{1.2CA}

\end{picture}}

\end{picture}

\vspace*{-0.02cm}

\begin{picture}(18 ,2.5 )(0,0)
\put(0,0){\setlength{\unitlength}{0.01059cm}%
\begin{picture}(1700, 236.111)(7350,0)
\put(7350,0){\framebox(1700, 236.111)[tl]{\begin{picture}(0,0)(0,0)
        \put(1700,0){\makebox(0,0)[tr]{\bf{0153+744}\T{0.4}
                                 \hspace*{0.5cm}}}
    \end{picture}}}

\thicklines
\put(7350,0){\setlength{\unitlength}{5cm}\begin{picture}(0,0)(0,-0.25)
   \put(0,0){\setlength{\unitlength}{1cm}\begin{picture}(0,0)(0,0)
        \put(0,0){\line(1,0){0.3}}
        \end{picture}}
   \end{picture}}

\put(9050,0){\setlength{\unitlength}{5cm}\begin{picture}(0,0)(0,-0.25)
   \put(0,0){\setlength{\unitlength}{1cm}\begin{picture}(0,0)(0,0)
        \put(0,0){\line(-1,0){0.3}}
        \end{picture}}
   \end{picture}}

\thinlines
\put(7350,0){\setlength{\unitlength}{5cm}\begin{picture}(0,0)(0,-0.25)
   \multiput(0,0)(0,0.1){3}{\setlength{\unitlength}{1cm}%
\begin{picture}(0,0)(0,0)
        \put(0,0){\line(1,0){0.12}}
        \end{picture}}
   \end{picture}}

\put(7350,0){\setlength{\unitlength}{5cm}\begin{picture}(0,0)(0,-0.25)
   \multiput(0,0)(0,-0.1){3}{\setlength{\unitlength}{1cm}%
\begin{picture}(0,0)(0,0)
        \put(0,0){\line(1,0){0.12}}
        \end{picture}}
   \end{picture}}

\put(9050,0){\setlength{\unitlength}{5cm}\begin{picture}(0,0)(0,-0.25)
   \multiput(0,0)(0,0.1){3}{\setlength{\unitlength}{1cm}%
\begin{picture}(0,0)(0,0)
        \put(0,0){\line(-1,0){0.12}}
        \end{picture}}
   \end{picture}}

\put(9050,0){\setlength{\unitlength}{5cm}\begin{picture}(0,0)(0,-0.25)
   \multiput(0,0)(0,-0.1){3}{\setlength{\unitlength}{1cm}%
\begin{picture}(0,0)(0,0)
        \put(0,0){\line(-1,0){0.12}}
        \end{picture}}
   \end{picture}}

   \put(7527.5, 236.111){\setlength{\unitlength}{1cm}\begin{picture}(0,0)(0,0)
        \put(0,0){\line(0,-1){0.2}}
   \end{picture}}
   \put(7892.5, 236.111){\setlength{\unitlength}{1cm}\begin{picture}(0,0)(0,0)
        \put(0,0){\line(0,-1){0.2}}
   \end{picture}}
   \put(8257.5, 236.111){\setlength{\unitlength}{1cm}\begin{picture}(0,0)(0,0)
        \put(0,0){\line(0,-1){0.2}}
   \end{picture}}
   \put(8622.5, 236.111){\setlength{\unitlength}{1cm}\begin{picture}(0,0)(0,0)
        \put(0,0){\line(0,-1){0.2}}
   \end{picture}}
   \put(8987.5, 236.111){\setlength{\unitlength}{1cm}\begin{picture}(0,0)(0,0)
        \put(0,0){\line(0,-1){0.2}}
   \end{picture}}
    \multiput(7350,0)(50,0){33}%
        {\setlength{\unitlength}{1cm}\begin{picture}(0,0)(0,0)
        \put(0,0){\line(0,1){0.12}}
    \end{picture}}
    \put(7500,0){\setlength{\unitlength}{1cm}\begin{picture}(0,0)(0,0)
        \put(0,0){\line(0,1){0.2}}
    \end{picture}}
    \put(7750,0){\setlength{\unitlength}{1cm}\begin{picture}(0,0)(0,0)
        \put(0,0){\line(0,1){0.2}}
    \end{picture}}
    \put(8000,0){\setlength{\unitlength}{1cm}\begin{picture}(0,0)(0,0)
        \put(0,0){\line(0,1){0.2}}
    \end{picture}}
    \put(8250,0){\setlength{\unitlength}{1cm}\begin{picture}(0,0)(0,0)
        \put(0,0){\line(0,1){0.2}}
    \end{picture}}
    \put(8500,0){\setlength{\unitlength}{1cm}\begin{picture}(0,0)(0,0)
        \put(0,0){\line(0,1){0.2}}
    \end{picture}}
    \put(8750,0){\setlength{\unitlength}{1cm}\begin{picture}(0,0)(0,0)
        \put(0,0){\line(0,1){0.2}}
    \end{picture}}
    \put(9000,0){\setlength{\unitlength}{1cm}\begin{picture}(0,0)(0,0)
        \put(0,0){\line(0,1){0.2}}
    \end{picture}}

\punkt{05/09/88}{01:15}{7409.552}{ -0.036}{0.063}{1000}{2.20}{
977}{1.2CA}
\punkt{06/09/88}{01:19}{7410.556}{  0.026}{0.042}{2000}{1.31}{
1635}{1.2CA}
\punkt{05/10/88}{03:24}{7439.642}{ -0.017}{0.068}{ 500}{1.10}{
746}{1.2CA}
\punkt{08/10/88}{02:20}{7442.598}{  0.020}{0.071}{ 500}{1.28}{
691}{1.2CA}
\punkt{17/10/88}{01:09}{7451.548}{  0.121}{0.083}{ 500}{2.84}{
672}{1.2CA}
\punkt{24/06/89}{03:25}{7701.643}{  0.040}{0.061}{ 500}{1.41}{
1208}{1.2CA}
\punkt{24/06/89}{15:06}{7702.129}{  0.043}{0.064}{ 500}{2.01}{
978}{1.2CA}
\punkt{26/06/89}{03:08}{7703.631}{  0.009}{0.016}{ 500}{1.51}{
919}{1.2CA}
\punkt{24/07/89}{03:14}{7731.635}{ -0.013}{0.064}{ 500}{1.78}{
982}{1.2CA}
\punkt{13/08/89}{02:26}{7751.601}{ -0.003}{0.018}{1000}{1.95}{
1043}{1.2CA}
\punkt{13/08/89}{03:13}{7751.635}{ -0.018}{0.028}{1000}{1.71}{
1108}{1.2CA}
\punkt{13/08/89}{03:53}{7751.662}{ -0.025}{0.017}{1000}{1.71}{
1128}{1.2CA}
\punkt{01/09/89}{03:09}{7770.632}{ -0.012}{0.070}{ 500}{1.98}{
767}{1.2CA}
\punkt{01/11/89}{21:01}{7832.376}{ -0.037}{0.032}{ 500}{1.69}{
581}{1.2CA}
\punkt{22/12/89}{19:08}{7883.297}{ -0.032}{0.079}{ 500}{2.65}{
611}{1.2CA}
\punkt{01/08/90}{03:53}{8104.662}{ -0.022}{0.064}{ 500}{1.53}{
867}{1.2CA}
\punkt{25/09/90}{21:53}{8160.412}{  0.037}{0.039}{ 500}{2.09}{
1059}{1.2CA}
\punkt{26/09/90}{21:53}{8161.412}{  0.017}{0.042}{ 500}{2.02}{
1062}{1.2CA}
\punkta{04/10/90}{01:08}{8168.547}{ -0.017}{0.053}{ 500}{2.48}{
2083}{1.2CA}
\punkt{22/02/91}{20:37}{8310.360}{  0.019}{0.047}{ 500}{1.56}{
1143}{1.2CA}
\punkt{28/10/91}{02:58}{8557.624}{  0.011}{0.088}{ 252}{1.83}{
960}{1.2CA}
\punkt{22/09/92}{01:23}{8887.558}{  0.035}{0.042}{ 150}{2.33}{
486}{1.2CA}
\punkt{22/09/92}{01:29}{8887.562}{  0.006}{0.052}{ 150}{2.56}{
486}{1.2CA}
\punkt{22/09/92}{01:29}{8887.562}{  0.006}{0.052}{ 150}{2.56}{
486}{1.2CA}

\end{picture}}

\end{picture}

\vspace*{-0.02cm}

\begin{picture}(18 ,2.5 )(0,0)
\put(0,0){\setlength{\unitlength}{0.01059cm}%
\begin{picture}(1700, 236.111)(7350,0)
\put(7350,0){\framebox(1700, 236.111)[tl]{\begin{picture}(0,0)(0,0)
        \put(1700,0){\makebox(0,0)[tr]{\bf{0248+430}\T{0.4}
                                 \hspace*{0.5cm}}}
    \end{picture}}}

\thicklines
\put(7350,0){\setlength{\unitlength}{5cm}\begin{picture}(0,0)(0,-0.25)
   \put(0,0){\setlength{\unitlength}{1cm}\begin{picture}(0,0)(0,0)
        \put(0,0){\line(1,0){0.3}}
        \end{picture}}
   \end{picture}}

\put(9050,0){\setlength{\unitlength}{5cm}\begin{picture}(0,0)(0,-0.25)
   \put(0,0){\setlength{\unitlength}{1cm}\begin{picture}(0,0)(0,0)
        \put(0,0){\line(-1,0){0.3}}
        \end{picture}}
   \end{picture}}

\thinlines
\put(7350,0){\setlength{\unitlength}{5cm}\begin{picture}(0,0)(0,-0.25)
   \multiput(0,0)(0,0.1){3}{\setlength{\unitlength}{1cm}%
\begin{picture}(0,0)(0,0)
        \put(0,0){\line(1,0){0.12}}
        \end{picture}}
   \end{picture}}

\put(7350,0){\setlength{\unitlength}{5cm}\begin{picture}(0,0)(0,-0.25)
   \multiput(0,0)(0,-0.1){3}{\setlength{\unitlength}{1cm}%
\begin{picture}(0,0)(0,0)
        \put(0,0){\line(1,0){0.12}}
        \end{picture}}
   \end{picture}}

\put(9050,0){\setlength{\unitlength}{5cm}\begin{picture}(0,0)(0,-0.25)
   \multiput(0,0)(0,0.1){3}{\setlength{\unitlength}{1cm}%
\begin{picture}(0,0)(0,0)
        \put(0,0){\line(-1,0){0.12}}
        \end{picture}}
   \end{picture}}

\put(9050,0){\setlength{\unitlength}{5cm}\begin{picture}(0,0)(0,-0.25)
   \multiput(0,0)(0,-0.1){3}{\setlength{\unitlength}{1cm}%
\begin{picture}(0,0)(0,0)
        \put(0,0){\line(-1,0){0.12}}
        \end{picture}}
   \end{picture}}

   \put(7527.5, 236.111){\setlength{\unitlength}{1cm}\begin{picture}(0,0)(0,0)
        \put(0,0){\line(0,-1){0.2}}
   \end{picture}}
   \put(7892.5, 236.111){\setlength{\unitlength}{1cm}\begin{picture}(0,0)(0,0)
        \put(0,0){\line(0,-1){0.2}}
   \end{picture}}
   \put(8257.5, 236.111){\setlength{\unitlength}{1cm}\begin{picture}(0,0)(0,0)
        \put(0,0){\line(0,-1){0.2}}
   \end{picture}}
   \put(8622.5, 236.111){\setlength{\unitlength}{1cm}\begin{picture}(0,0)(0,0)
        \put(0,0){\line(0,-1){0.2}}
   \end{picture}}
   \put(8987.5, 236.111){\setlength{\unitlength}{1cm}\begin{picture}(0,0)(0,0)
        \put(0,0){\line(0,-1){0.2}}
   \end{picture}}
    \multiput(7350,0)(50,0){33}%
        {\setlength{\unitlength}{1cm}\begin{picture}(0,0)(0,0)
        \put(0,0){\line(0,1){0.12}}
    \end{picture}}
    \put(7500,0){\setlength{\unitlength}{1cm}\begin{picture}(0,0)(0,0)
        \put(0,0){\line(0,1){0.2}}
    \end{picture}}
    \put(7750,0){\setlength{\unitlength}{1cm}\begin{picture}(0,0)(0,0)
        \put(0,0){\line(0,1){0.2}}
    \end{picture}}
    \put(8000,0){\setlength{\unitlength}{1cm}\begin{picture}(0,0)(0,0)
        \put(0,0){\line(0,1){0.2}}
    \end{picture}}
    \put(8250,0){\setlength{\unitlength}{1cm}\begin{picture}(0,0)(0,0)
        \put(0,0){\line(0,1){0.2}}
    \end{picture}}
    \put(8500,0){\setlength{\unitlength}{1cm}\begin{picture}(0,0)(0,0)
        \put(0,0){\line(0,1){0.2}}
    \end{picture}}
    \put(8750,0){\setlength{\unitlength}{1cm}\begin{picture}(0,0)(0,0)
        \put(0,0){\line(0,1){0.2}}
    \end{picture}}
    \put(9000,0){\setlength{\unitlength}{1cm}\begin{picture}(0,0)(0,0)
        \put(0,0){\line(0,1){0.2}}
    \end{picture}}

\punkt{06/10/88}{21:25}{7441.393}{ -0.104}{0.039}{ 500}{1.89}{
556}{1.2CA}
\punkt{06/10/88}{22:01}{7441.418}{ -0.124}{0.047}{1000}{1.91}{
684}{1.2CA}
\punkt{06/10/88}{23:06}{7441.463}{ -0.102}{0.046}{1000}{1.55}{
1013}{1.2CA}
\punkt{07/10/88}{01:32}{7441.564}{ -0.132}{0.039}{1000}{1.38}{
992}{1.2CA}
\punkt{07/10/88}{03:46}{7441.658}{ -0.125}{0.022}{1000}{1.35}{
1025}{1.2CA}
\punkt{08/10/88}{00:22}{7442.516}{ -0.135}{0.033}{1000}{1.01}{
777}{1.2CA}
\punkt{17/10/88}{02:45}{7451.615}{ -0.132}{0.064}{ 500}{1.90}{
633}{1.2CA}
\punkta{24/06/89}{03:39}{7701.653}{  0.027}{0.046}{ 415}{1.68}{
1963}{1.2CA}
\punkt{25/07/89}{02:47}{7732.616}{ -0.026}{0.013}{ 500}{1.49}{
1120}{1.2CA}
\punkt{11/08/89}{02:58}{7749.624}{ -0.062}{0.021}{ 500}{1.47}{
678}{1.2CA}
\punkt{14/08/89}{14:22}{7753.099}{ -0.025}{0.052}{ 500}{1.74}{
875}{1.2CA}
\punkt{03/09/89}{03:35}{7772.650}{ -0.034}{0.057}{ 500}{1.45}{
778}{1.2CA}
\punkt{01/10/89}{14:03}{7801.086}{ -0.032}{0.024}{ 500}{1.39}{
501}{1.2CA}
\punkta{17/10/89}{00:46}{7816.532}{ -0.021}{0.021}{1000}{1.91}{
5076}{1.2CA}
\punkt{31/10/89}{23:23}{7831.474}{ -0.027}{0.049}{1000}{1.27}{
637}{1.2CA}
\punkt{01/11/89}{23:28}{7832.478}{ -0.030}{0.047}{1000}{1.45}{
629}{1.2CA}
\punkt{07/12/89}{21:47}{7868.408}{ -0.007}{0.048}{ 100}{1.81}{
545}{1.2CA}
\punkta{13/12/89}{23:20}{7874.472}{ -0.125}{0.081}{ 500}{1.73}{
3700}{1.2CA}
\punkt{19/12/89}{21:11}{7880.383}{ -0.021}{0.094}{ 250}{1.44}{
519}{1.2CA}
\punkt{19/12/89}{21:17}{7880.387}{ -0.021}{0.016}{ 250}{1.61}{
511}{1.2CA}
\punkt{20/12/89}{00:28}{7880.520}{ -0.028}{0.031}{ 200}{1.88}{
505}{1.2CA}
\punkt{20/12/89}{19:07}{7881.297}{ -0.016}{0.019}{ 500}{1.61}{
619}{1.2CA}
\punkt{20/12/89}{20:57}{7881.373}{ -0.023}{0.013}{ 500}{1.57}{
597}{1.2CA}
\punkt{08/02/90}{19:52}{7931.328}{  0.007}{0.045}{ 507}{1.76}{
1139}{1.2CA}
\punkt{09/02/90}{20:38}{7932.360}{  0.015}{0.025}{ 500}{1.38}{
879}{1.2CA}
\punkt{26/02/90}{20:59}{7949.375}{ -0.001}{0.030}{ 500}{1.91}{
773}{1.2CA}
\punkt{27/02/90}{19:29}{7950.312}{  0.005}{0.028}{ 500}{1.28}{
671}{1.2CA}
\punkt{28/02/90}{19:19}{7951.305}{  0.007}{0.063}{ 500}{1.29}{
670}{1.2CA}
\punkt{31/07/90}{04:05}{8103.670}{  0.046}{0.051}{ 500}{1.13}{
795}{1.2CA}
\punkta{13/08/90}{03:40}{8116.653}{  0.035}{0.011}{ 200}{1.09}{
5996}{1.2CA}
\punkt{14/08/90}{03:45}{8117.657}{  0.047}{0.011}{ 100}{1.12}{
917}{1.2CA}
\punkt{15/08/90}{04:05}{8118.671}{  0.030}{0.008}{ 100}{0.98}{
577}{1.2CA}
\punkt{25/09/90}{02:58}{8159.624}{  0.043}{0.038}{ 500}{1.66}{
839}{1.2CA}
\punkt{25/09/90}{22:32}{8160.439}{  0.062}{0.012}{ 500}{1.87}{
860}{1.2CA}
\punkt{26/09/90}{02:58}{8160.624}{  0.031}{0.010}{ 500}{1.64}{
838}{1.2CA}
\punkt{26/09/90}{22:32}{8161.439}{  0.064}{0.040}{ 500}{1.85}{
861}{1.2CA}
\punkt{28/09/90}{00:09}{8162.507}{  0.051}{0.012}{ 500}{1.45}{
892}{1.2CA}
\punkt{18/10/90}{01:01}{8182.542}{  0.050}{0.086}{ 100}{1.99}{
561}{1.2CA}
\punkt{19/10/90}{02:38}{8183.610}{  0.024}{0.038}{ 500}{1.21}{
805}{1.2CA}
\punkt{20/10/90}{02:38}{8184.610}{  0.026}{0.038}{ 500}{1.20}{
803}{1.2CA}
\punkt{15/01/91}{18:15}{8272.260}{  0.023}{0.023}{  50}{1.25}{
1565}{1.2CA}
\punkta{}{}{8297.275}{0.033}{0.05}{}{}{}{}
\punkt{23/02/91}{20:34}{8311.357}{  0.032}{0.026}{ 300}{1.54}{
2494}{1.2CA}
\punkt{19/03/91}{19:38}{8335.319}{  0.022}{0.051}{ 100}{1.76}{
358}{1.2CA}
\punkt{25/07/91}{03:24}{8462.642}{ -0.009}{0.019}{ 100}{1.40}{
437}{1.2CA}
\punkt{26/07/91}{03:46}{8463.657}{ -0.019}{0.043}{ 500}{1.65}{
1294}{1.2CA}
\punkta{29/07/91}{03:53}{8466.662}{ -0.038}{0.027}{ 300}{1.94}{
4290}{1.2CA}
\punkt{01/08/91}{04:04}{8469.670}{ -0.009}{0.043}{ 500}{1.09}{
1401}{1.2CA}
\punkt{06/08/91}{03:22}{8474.640}{  0.021}{0.058}{ 500}{1.13}{
962}{1.2CA}
\punkt{08/08/91}{03:34}{8476.649}{  0.006}{0.041}{ 500}{1.46}{
528}{1.2CA}
\punkt{12/08/91}{03:13}{8480.634}{  0.003}{0.043}{ 500}{1.74}{
559}{1.2CA}
\punkt{24/08/91}{04:00}{8492.667}{  0.006}{0.019}{ 500}{1.41}{
909}{1.2CA}
\punkt{25/08/91}{00:57}{8493.540}{  0.001}{0.048}{ 500}{1.76}{
3017}{1.2CA}
\punkt{19/09/91}{04:35}{8518.691}{  0.002}{0.042}{ 500}{1.52}{
473}{1.2CA}
\punkt{20/09/91}{03:47}{8519.658}{ -0.008}{0.044}{ 500}{1.17}{
431}{1.2CA}
\punkt{06/10/91}{04:04}{8535.669}{ -0.005}{0.024}{ 100}{1.17}{
1351}{1.2CA}
\punkt{19/10/91}{03:36}{8548.650}{  0.004}{0.015}{ 500}{1.66}{
550}{1.2CA}
\punkt{}{}{8550.475}{-0.012}{0.010}{}{}{}{}
\punkt{27/10/91}{23:51}{8557.494}{ -0.017}{0.011}{ 500}{1.39}{
1129}{1.2CA}
\punkt{}{}{8557.525}{0.015}{0.03}{}{}{}{}
\punkt{13/11/91}{02:02}{8573.585}{  0.009}{0.044}{ 500}{1.42}{
816}{1.2CA}
\punkt{07/02/92}{19:52}{8660.328}{  0.007}{0.010}{ 500}{1.39}{
540}{1.2CA}
\punkt{08/02/92}{19:30}{8661.313}{  0.003}{0.009}{ 500}{1.43}{
678}{1.2CA}
\punkt{14/02/92}{19:17}{8667.304}{ -0.017}{0.020}{ 500}{1.20}{
4553}{1.2CA}
\punkt{21/09/92}{02:48}{8886.617}{ -0.016}{0.013}{ 100}{0.74}{
490}{1.2CA}
\punkt{21/09/92}{02:51}{8886.619}{ -0.016}{0.012}{ 100}{0.68}{
489}{1.2CA}
\punkt{22/09/92}{02:14}{8887.593}{ -0.032}{0.021}{ 300}{1.54}{
717}{1.2CA}
\punkt{24/09/92}{02:19}{8889.597}{ -0.015}{0.029}{ 150}{1.02}{
457}{1.2CA}
\punkt{25/09/92}{05:07}{8890.714}{ -0.006}{0.029}{ 150}{0.87}{
1230}{1.2CA}
\punkt{25/09/92}{05:07}{8890.714}{ -0.006}{0.029}{ 150}{0.87}{
1230}{1.2CA}

\end{picture}}

\end{picture}

\vspace*{-0.02cm}

\begin{picture}(18 ,2.5 )(0,0)
\put(0,0){\setlength{\unitlength}{0.01059cm}%
\begin{picture}(1700, 236.072)(7350,0)
\put(7350,0){\framebox(1700, 236.072)[tl]{\begin{picture}(0,0)(0,0)
        \put(1700,-132){\makebox(0,0)[tr]{\bf{0446-$\!$-208}\T{0.4}
                                 \hspace*{0.5cm}}}
    \end{picture}}}

\thicklines
\put(7350,0){\setlength{\unitlength}{5cm}\begin{picture}(0,0)(0,-0.25)
   \put(0,0){\setlength{\unitlength}{1cm}\begin{picture}(0,0)(0,0)
        \put(0,0){\line(1,0){0.3}}
        \end{picture}}
   \end{picture}}

\put(9050,0){\setlength{\unitlength}{5cm}\begin{picture}(0,0)(0,-0.25)
   \put(0,0){\setlength{\unitlength}{1cm}\begin{picture}(0,0)(0,0)
        \put(0,0){\line(-1,0){0.3}}
        \end{picture}}
   \end{picture}}

\thinlines
\put(7350,0){\setlength{\unitlength}{5cm}\begin{picture}(0,0)(0,-0.25)
   \multiput(0,0)(0,0.1){3}{\setlength{\unitlength}{1cm}%
\begin{picture}(0,0)(0,0)
        \put(0,0){\line(1,0){0.12}}
        \end{picture}}
   \end{picture}}

\put(7350,0){\setlength{\unitlength}{5cm}\begin{picture}(0,0)(0,-0.25)
   \multiput(0,0)(0,-0.1){3}{\setlength{\unitlength}{1cm}%
\begin{picture}(0,0)(0,0)
        \put(0,0){\line(1,0){0.12}}
        \end{picture}}
   \end{picture}}

\put(9050,0){\setlength{\unitlength}{5cm}\begin{picture}(0,0)(0,-0.25)
   \multiput(0,0)(0,0.1){3}{\setlength{\unitlength}{1cm}%
\begin{picture}(0,0)(0,0)
        \put(0,0){\line(-1,0){0.12}}
        \end{picture}}
   \end{picture}}

\put(9050,0){\setlength{\unitlength}{5cm}\begin{picture}(0,0)(0,-0.25)
   \multiput(0,0)(0,-0.1){3}{\setlength{\unitlength}{1cm}%
\begin{picture}(0,0)(0,0)
        \put(0,0){\line(-1,0){0.12}}
        \end{picture}}
   \end{picture}}

   \put(7527.5, 236.072){\setlength{\unitlength}{1cm}\begin{picture}(0,0)(0,0)
        \put(0,0){\line(0,-1){0.2}}
   \end{picture}}
   \put(7892.5, 236.072){\setlength{\unitlength}{1cm}\begin{picture}(0,0)(0,0)
        \put(0,0){\line(0,-1){0.2}}
   \end{picture}}
   \put(8257.5, 236.072){\setlength{\unitlength}{1cm}\begin{picture}(0,0)(0,0)
        \put(0,0){\line(0,-1){0.2}}
   \end{picture}}
   \put(8622.5, 236.072){\setlength{\unitlength}{1cm}\begin{picture}(0,0)(0,0)
        \put(0,0){\line(0,-1){0.2}}
   \end{picture}}
   \put(8987.5, 236.072){\setlength{\unitlength}{1cm}\begin{picture}(0,0)(0,0)
        \put(0,0){\line(0,-1){0.2}}
   \end{picture}}
    \multiput(7350,0)(50,0){33}%
        {\setlength{\unitlength}{1cm}\begin{picture}(0,0)(0,0)
        \put(0,0){\line(0,1){0.12}}
    \end{picture}}
    \put(7500,0){\setlength{\unitlength}{1cm}\begin{picture}(0,0)(0,0)
        \put(0,0){\line(0,1){0.2}}
    \end{picture}}
    \put(7750,0){\setlength{\unitlength}{1cm}\begin{picture}(0,0)(0,0)
        \put(0,0){\line(0,1){0.2}}
    \end{picture}}
    \put(8000,0){\setlength{\unitlength}{1cm}\begin{picture}(0,0)(0,0)
        \put(0,0){\line(0,1){0.2}}
    \end{picture}}
    \put(8250,0){\setlength{\unitlength}{1cm}\begin{picture}(0,0)(0,0)
        \put(0,0){\line(0,1){0.2}}
    \end{picture}}
    \put(8500,0){\setlength{\unitlength}{1cm}\begin{picture}(0,0)(0,0)
        \put(0,0){\line(0,1){0.2}}
    \end{picture}}
    \put(8750,0){\setlength{\unitlength}{1cm}\begin{picture}(0,0)(0,0)
        \put(0,0){\line(0,1){0.2}}
    \end{picture}}
    \put(9000,0){\setlength{\unitlength}{1cm}\begin{picture}(0,0)(0,0)
        \put(0,0){\line(0,1){0.2}}
    \end{picture}}

\punkt{08/10/88}{02:53}{7442.620}{  0.101}{0.016}{ 500}{1.67}{
601}{1.2CA}
\punkt{17/10/88}{03:39}{7451.652}{  0.098}{0.062}{ 500}{2.49}{
628}{1.2CA}
\punkt{02/09/89}{03:48}{7771.659}{  0.077}{0.014}{ 500}{2.06}{
794}{1.2CA}
\punkt{01/10/89}{15:54}{7801.163}{  0.083}{0.037}{ 500}{2.07}{
573}{1.2CA}
\punkt{01/11/89}{01:08}{7831.548}{  0.043}{0.013}{ 500}{2.03}{
546}{1.2CA}
\punkt{02/11/89}{00:39}{7832.527}{  0.051}{0.060}{ 500}{2.15}{
550}{1.2CA}
\punkt{19/12/89}{23:16}{7880.470}{  0.054}{0.014}{ 500}{2.46}{
602}{1.2CA}
\punkta{26/01/90}{20:31}{7918.355}{ -0.148}{0.042}{ 500}{3.51}{
2666}{1.2CA}
\punkta{08/02/90}{20:57}{7931.373}{ -0.025}{0.050}{ 500}{2.86}{
1690}{1.2CA}
\punkt{09/02/90}{20:20}{7932.347}{  0.039}{0.060}{ 500}{1.63}{
847}{1.2CA}
\punkt{28/02/90}{19:37}{7951.317}{  0.023}{0.054}{ 500}{2.07}{
688}{1.2CA}
\punkt{25/09/90}{03:59}{8159.666}{  0.055}{0.024}{ 100}{2.40}{
567}{1.2CA}
\punkt{26/09/90}{03:59}{8160.666}{  0.035}{0.079}{ 100}{2.38}{
567}{1.2CA}
\punkt{30/09/90}{03:33}{8164.648}{ -0.005}{0.064}{ 500}{3.03}{
908}{1.2CA}
\punkt{17/10/90}{04:12}{8181.675}{  0.009}{0.045}{ 500}{2.84}{
873}{1.2CA}
\punkt{18/10/90}{04:12}{8182.675}{  0.003}{0.045}{ 500}{2.84}{
872}{1.2CA}
\punkt{19/10/90}{04:59}{8183.708}{ -0.004}{0.035}{ 500}{2.29}{
910}{1.2CA}
\punkt{20/10/90}{04:59}{8184.708}{ -0.003}{0.017}{ 500}{2.28}{
909}{1.2CA}
\punkt{27/11/90}{01:01}{8222.542}{  0.007}{0.008}{ 500}{2.92}{
1220}{1.2CA}
\punkt{22/12/90}{10:29}{8247.937}{ -0.032}{0.050}{ 500}{3.12}{
670}{1.2CA}
\punkt{07/02/91}{20:27}{8295.352}{ -0.031}{0.097}{ 164}{3.40}{
397}{1.2CA}
\punkt{14/02/91}{20:54}{8302.371}{ -0.032}{0.024}{ 500}{3.08}{
639}{1.2CA}
\punkt{22/02/91}{20:00}{8310.334}{ -0.022}{0.015}{ 500}{1.83}{
1231}{1.2CA}
\punkt{23/02/91}{19:50}{8311.327}{ -0.052}{0.015}{ 300}{1.43}{
1590}{1.2CA}
\punkt{17/10/91}{03:33}{8546.648}{ -0.076}{0.027}{ 500}{2.88}{
714}{1.2CA}
\punkt{18/10/91}{03:16}{8547.636}{ -0.071}{0.032}{ 500}{1.90}{
697}{1.2CA}
\punkt{21/10/91}{03:11}{8550.633}{ -0.067}{0.016}{ 500}{1.48}{
1442}{1.2CA}
\punkt{31/01/92}{21:27}{8653.394}{ -0.037}{0.017}{ 500}{2.16}{
612}{1.2CA}
\punkt{02/02/92}{20:01}{8655.334}{ -0.056}{0.012}{ 500}{1.87}{
634}{1.2CA}
\punkt{07/02/92}{20:34}{8660.357}{ -0.053}{0.046}{ 500}{1.58}{
600}{1.2CA}
\punkt{06/03/92}{00:49}{8687.534}{  0.014}{0.011}{ 600}{1.26}{
454}{1.2CA}
\punkt{09/03/92}{01:10}{8690.549}{  0.006}{0.017}{ 300}{1.75}{
314}{1.2CA}
\punkt{10/03/92}{01:14}{8691.551}{  0.001}{0.016}{ 300}{1.65}{
363}{1.2CA}
\punkt{11/03/92}{00:48}{8692.534}{ -0.019}{0.012}{ 300}{1.49}{
390}{1.2CA}
\punkt{13/03/92}{00:59}{8694.542}{  0.021}{0.015}{ 500}{1.41}{
1207}{1.2CA}
\punkt{14/03/92}{00:27}{8695.519}{  0.001}{0.035}{ 150}{2.19}{
356}{1.2CA}
\punkt{15/03/92}{00:21}{8696.515}{  0.005}{0.026}{ 150}{1.57}{
445}{1.2CA}
\punkt{16/03/92}{00:28}{8697.520}{  0.004}{0.014}{ 300}{1.54}{
569}{1.2CA}
\punkt{17/03/92}{00:46}{8698.533}{  0.019}{0.024}{ 300}{1.50}{
614}{1.2CA}
\punkt{18/03/92}{00:51}{8699.536}{  0.032}{0.020}{ 300}{1.96}{
694}{1.2CA}
\punkt{19/03/92}{00:15}{8700.511}{ -0.017}{0.028}{ 500}{1.87}{
1817}{1.2CA}
\punkt{20/03/92}{00:12}{8701.509}{  0.000}{0.021}{ 300}{1.66}{
543}{1.2CA}
\punkt{21/03/92}{01:55}{8702.580}{  0.055}{0.060}{ 200}{2.45}{
524}{1.2CA}
\punkt{21/03/92}{02:00}{8702.583}{  0.067}{0.049}{ 200}{2.50}{
539}{1.2CA}
\punkt{24/03/92}{00:35}{8705.524}{  0.000}{0.021}{ 300}{1.82}{
305}{1.2CA}
\punkt{25/03/92}{00:34}{8706.524}{  0.011}{0.027}{ 300}{1.68}{
307}{1.2CA}
\punkt{21/09/92}{03:12}{8886.634}{  0.066}{0.026}{ 100}{1.37}{
700}{1.2CA}
\punkt{21/09/92}{03:15}{8886.636}{  0.060}{0.024}{ 100}{1.23}{
732}{1.2CA}
\punkt{23/09/92}{02:16}{8888.595}{  0.020}{0.034}{ 150}{2.43}{
576}{1.2CA}
\punkt{23/09/92}{02:19}{8888.597}{  0.081}{0.039}{ 100}{2.24}{
489}{1.2CA}
\punkt{23/09/92}{02:19}{8888.597}{  0.081}{0.039}{ 100}{2.24}{
489}{1.2CA}

\end{picture}}

\end{picture}

\vspace*{-0.02cm}

\begin{picture}(18 ,2.5 )(0,0)
\put(0,0){\setlength{\unitlength}{0.01059cm}%
\begin{picture}(1700, 236.072)(7350,0)
\put(7350,0){\framebox(1700, 236.072)[tl]{\begin{picture}(0,0)(0,0)
        \put(1700,-132){\makebox(0,0)[tr]{\bf{0454+039}\T{0.4}
                                 \hspace*{0.5cm}}}
    \end{picture}}}

\thicklines
\put(7350,0){\setlength{\unitlength}{5cm}\begin{picture}(0,0)(0,-0.25)
   \put(0,0){\setlength{\unitlength}{1cm}\begin{picture}(0,0)(0,0)
        \put(0,0){\line(1,0){0.3}}
        \end{picture}}
   \end{picture}}

\put(9050,0){\setlength{\unitlength}{5cm}\begin{picture}(0,0)(0,-0.25)
   \put(0,0){\setlength{\unitlength}{1cm}\begin{picture}(0,0)(0,0)
        \put(0,0){\line(-1,0){0.3}}
        \end{picture}}
   \end{picture}}

\thinlines
\put(7350,0){\setlength{\unitlength}{5cm}\begin{picture}(0,0)(0,-0.25)
   \multiput(0,0)(0,0.1){3}{\setlength{\unitlength}{1cm}%
\begin{picture}(0,0)(0,0)
        \put(0,0){\line(1,0){0.12}}
        \end{picture}}
   \end{picture}}

\put(7350,0){\setlength{\unitlength}{5cm}\begin{picture}(0,0)(0,-0.25)
   \multiput(0,0)(0,-0.1){3}{\setlength{\unitlength}{1cm}%
\begin{picture}(0,0)(0,0)
        \put(0,0){\line(1,0){0.12}}
        \end{picture}}
   \end{picture}}

\put(9050,0){\setlength{\unitlength}{5cm}\begin{picture}(0,0)(0,-0.25)
   \multiput(0,0)(0,0.1){3}{\setlength{\unitlength}{1cm}%
\begin{picture}(0,0)(0,0)
        \put(0,0){\line(-1,0){0.12}}
        \end{picture}}
   \end{picture}}

\put(9050,0){\setlength{\unitlength}{5cm}\begin{picture}(0,0)(0,-0.25)
   \multiput(0,0)(0,-0.1){3}{\setlength{\unitlength}{1cm}%
\begin{picture}(0,0)(0,0)
        \put(0,0){\line(-1,0){0.12}}
        \end{picture}}
   \end{picture}}

   \put(7527.5, 236.072){\setlength{\unitlength}{1cm}\begin{picture}(0,0)(0,0)
        \put(0,0){\line(0,-1){0.2}}
   \end{picture}}
   \put(7892.5, 236.072){\setlength{\unitlength}{1cm}\begin{picture}(0,0)(0,0)
        \put(0,0){\line(0,-1){0.2}}
   \end{picture}}
   \put(8257.5, 236.072){\setlength{\unitlength}{1cm}\begin{picture}(0,0)(0,0)
        \put(0,0){\line(0,-1){0.2}}
   \end{picture}}
   \put(8622.5, 236.072){\setlength{\unitlength}{1cm}\begin{picture}(0,0)(0,0)
        \put(0,0){\line(0,-1){0.2}}
   \end{picture}}
   \put(8987.5, 236.072){\setlength{\unitlength}{1cm}\begin{picture}(0,0)(0,0)
        \put(0,0){\line(0,-1){0.2}}
   \end{picture}}
    \multiput(7350,0)(50,0){33}%
        {\setlength{\unitlength}{1cm}\begin{picture}(0,0)(0,0)
        \put(0,0){\line(0,1){0.12}}
    \end{picture}}
    \put(7500,0){\setlength{\unitlength}{1cm}\begin{picture}(0,0)(0,0)
        \put(0,0){\line(0,1){0.2}}
    \end{picture}}
   \put(7750,0){\setlength{\unitlength}{1cm}\begin{picture}(0,0)(0,0)
        \put(0,0){\line(0,1){0.2}}
    \end{picture}}
    \put(8000,0){\setlength{\unitlength}{1cm}\begin{picture}(0,0)(0,0)
        \put(0,0){\line(0,1){0.2}}
    \end{picture}}
    \put(8250,0){\setlength{\unitlength}{1cm}\begin{picture}(0,0)(0,0)
        \put(0,0){\line(0,1){0.2}}
    \end{picture}}
    \put(8500,0){\setlength{\unitlength}{1cm}\begin{picture}(0,0)(0,0)
        \put(0,0){\line(0,1){0.2}}
    \end{picture}}
    \put(8750,0){\setlength{\unitlength}{1cm}\begin{picture}(0,0)(0,0)
        \put(0,0){\line(0,1){0.2}}
    \end{picture}}
    \put(9000,0){\setlength{\unitlength}{1cm}\begin{picture}(0,0)(0,0)
        \put(0,0){\line(0,1){0.2}}
    \end{picture}}

\punkt{01/10/89}{16:26}{7801.185}{ -0.119}{0.029}{ 500}{1.95}{
514}{1.2CA}
\punkt{01/11/89}{01:24}{7831.559}{ -0.096}{0.012}{ 500}{2.22}{
510}{1.2CA}
\punkt{02/11/89}{00:54}{7832.538}{ -0.073}{0.018}{ 500}{1.58}{
526}{1.2CA}
\punkt{29/01/90}{20:50}{7921.369}{ -0.106}{0.026}{ 500}{4.20}{
568}{1.2CA}
\punkt{28/02/90}{19:54}{7951.329}{ -0.110}{0.018}{ 500}{1.65}{
634}{1.2CA}
\punkt{18/10/90}{04:30}{8182.688}{ -0.043}{0.012}{ 500}{2.64}{
808}{1.2CA}
\punkt{19/10/90}{04:24}{8183.683}{ -0.033}{0.013}{ 500}{2.12}{
800}{1.2CA}
\punkt{20/10/90}{03:17}{8184.637}{ -0.031}{0.013}{ 500}{2.60}{
759}{1.2CA}
\punkt{20/10/90}{04:24}{8184.683}{ -0.030}{0.013}{ 500}{2.11}{
799}{1.2CA}
\punkt{21/10/90}{03:17}{8185.637}{ -0.033}{0.012}{ 500}{2.63}{
758}{1.2CA}
\punkt{23/10/90}{03:36}{8187.650}{ -0.038}{0.043}{ 500}{3.57}{
826}{1.2CA}
\punkt{25/02/91}{00:03}{8312.503}{ -0.044}{0.061}{ 100}{1.27}{
2179}{1.2CA}
\punkt{17/10/91}{03:50}{8546.660}{ -0.004}{0.020}{ 500}{3.00}{
583}{1.2CA}
\punkt{31/01/92}{21:49}{8653.409}{  0.030}{0.040}{ 500}{2.07}{
504}{1.2CA}
\punkt{07/02/92}{20:57}{8660.373}{  0.017}{0.008}{ 500}{1.52}{
496}{1.2CA}
\punkt{09/02/92}{20:19}{8662.347}{  0.023}{0.007}{ 500}{1.43}{
691}{1.2CA}
\punkt{07/03/92}{00:56}{8688.539}{  0.062}{0.016}{ 500}{1.47}{
401}{1.2CA}
\punkt{10/03/92}{00:18}{8691.513}{  0.060}{0.033}{ 500}{1.96}{
1350}{1.2CA}
\punkt{18/03/92}{00:58}{8699.540}{  0.078}{0.044}{ 200}{2.50}{
611}{1.2CA}
\punkt{18/03/92}{01:02}{8699.544}{  0.024}{0.037}{ 200}{2.08}{
620}{1.2CA}
\punkt{23/03/92}{00:17}{8704.512}{  0.043}{0.047}{ 300}{1.52}{
364}{1.2CA}
\punkt{22/09/92}{03:54}{8887.663}{  0.024}{0.046}{ 200}{2.95}{
625}{1.2CA}
\punkt{23/09/92}{03:44}{8888.656}{  0.045}{0.026}{ 150}{1.36}{
522}{1.2CA}
\punkt{23/09/92}{03:47}{8888.658}{  0.050}{0.021}{ 100}{1.29}{
429}{1.2CA}
\punkt{23/09/92}{03:47}{8888.658}{  0.050}{0.021}{ 100}{1.29}{
429}{1.2CA}

\end{picture}}

\end{picture}

\vspace*{-0.02cm}

\begin{picture}(18 ,2.5 )(0,0)
\put(0,0){\setlength{\unitlength}{0.01059cm}%
\begin{picture}(1700, 236.111)(7350,0)
\put(7350,0){\framebox(1700, 236.111)[tl]{\begin{picture}(0,0)(0,0)
        \put(1700,0){\makebox(0,0)[tr]{\bf{0955+326}\T{0.4}
                                 \hspace*{0.5cm}}}
    \end{picture}}}

\thicklines
\put(7350,0){\setlength{\unitlength}{5cm}\begin{picture}(0,0)(0,-0.25)
   \put(0,0){\setlength{\unitlength}{1cm}\begin{picture}(0,0)(0,0)
        \put(0,0){\line(1,0){0.3}}
        \end{picture}}
   \end{picture}}

\put(9050,0){\setlength{\unitlength}{5cm}\begin{picture}(0,0)(0,-0.25)
   \put(0,0){\setlength{\unitlength}{1cm}\begin{picture}(0,0)(0,0)
        \put(0,0){\line(-1,0){0.3}}
        \end{picture}}
   \end{picture}}

\thinlines
\put(7350,0){\setlength{\unitlength}{5cm}\begin{picture}(0,0)(0,-0.25)
   \multiput(0,0)(0,0.1){3}{\setlength{\unitlength}{1cm}%
\begin{picture}(0,0)(0,0)
        \put(0,0){\line(1,0){0.12}}
        \end{picture}}
   \end{picture}}

\put(7350,0){\setlength{\unitlength}{5cm}\begin{picture}(0,0)(0,-0.25)
   \multiput(0,0)(0,-0.1){3}{\setlength{\unitlength}{1cm}%
\begin{picture}(0,0)(0,0)
        \put(0,0){\line(1,0){0.12}}
        \end{picture}}
   \end{picture}}

\put(9050,0){\setlength{\unitlength}{5cm}\begin{picture}(0,0)(0,-0.25)
   \multiput(0,0)(0,0.1){3}{\setlength{\unitlength}{1cm}%
\begin{picture}(0,0)(0,0)
        \put(0,0){\line(-1,0){0.12}}
        \end{picture}}
   \end{picture}}

\put(9050,0){\setlength{\unitlength}{5cm}\begin{picture}(0,0)(0,-0.25)
   \multiput(0,0)(0,-0.1){3}{\setlength{\unitlength}{1cm}%
\begin{picture}(0,0)(0,0)
        \put(0,0){\line(-1,0){0.12}}
        \end{picture}}
   \end{picture}}

   \put(7527.5, 236.111){\setlength{\unitlength}{1cm}\begin{picture}(0,0)(0,0)
        \put(0,0){\line(0,-1){0.2}}
   \end{picture}}
   \put(7892.5, 236.111){\setlength{\unitlength}{1cm}\begin{picture}(0,0)(0,0)
        \put(0,0){\line(0,-1){0.2}}
   \end{picture}}
   \put(8257.5, 236.111){\setlength{\unitlength}{1cm}\begin{picture}(0,0)(0,0)
        \put(0,0){\line(0,-1){0.2}}
   \end{picture}}
   \put(8622.5, 236.111){\setlength{\unitlength}{1cm}\begin{picture}(0,0)(0,0)
        \put(0,0){\line(0,-1){0.2}}
   \end{picture}}
   \put(8987.5, 236.111){\setlength{\unitlength}{1cm}\begin{picture}(0,0)(0,0)
        \put(0,0){\line(0,-1){0.2}}
  \end{picture}}
    \multiput(7350,0)(50,0){33}%
        {\setlength{\unitlength}{1cm}\begin{picture}(0,0)(0,0)
        \put(0,0){\line(0,1){0.12}}
    \end{picture}}
    \put(7500,0){\setlength{\unitlength}{1cm}\begin{picture}(0,0)(0,0)
        \put(0,0){\line(0,1){0.2}}
    \end{picture}}
    \put(7750,0){\setlength{\unitlength}{1cm}\begin{picture}(0,0)(0,0)
        \put(0,0){\line(0,1){0.2}}
    \end{picture}}
    \put(8000,0){\setlength{\unitlength}{1cm}\begin{picture}(0,0)(0,0)
        \put(0,0){\line(0,1){0.2}}
    \end{picture}}
    \put(8250,0){\setlength{\unitlength}{1cm}\begin{picture}(0,0)(0,0)
        \put(0,0){\line(0,1){0.2}}
    \end{picture}}
    \put(8500,0){\setlength{\unitlength}{1cm}\begin{picture}(0,0)(0,0)
        \put(0,0){\line(0,1){0.2}}
    \end{picture}}
    \put(8750,0){\setlength{\unitlength}{1cm}\begin{picture}(0,0)(0,0)
        \put(0,0){\line(0,1){0.2}}
    \end{picture}}
    \put(9000,0){\setlength{\unitlength}{1cm}\begin{picture}(0,0)(0,0)
        \put(0,0){\line(0,1){0.2}}
    \end{picture}}

\punkt{01/11/89}{05:10}{7831.715}{ -0.075}{0.006}{ 500}{1.38}{
562}{1.2CA}
\punkt{06/01/90}{02:59}{7897.625}{ -0.075}{0.038}{ 500}{2.20}{
609}{1.2CA}
\punkt{27/02/90}{22:56}{7950.456}{ -0.048}{0.010}{ 500}{1.58}{
648}{1.2CA}
\punkt{27/11/90}{05:55}{8222.747}{ -0.018}{0.012}{ 300}{2.86}{
615}{1.2CA}
\punkt{22/12/90}{15:02}{8248.127}{ -0.033}{0.025}{ 500}{3.00}{
594}{1.2CA}
\punkt{02/02/91}{23:54}{8290.496}{  0.010}{0.040}{ 194}{3.82}{
687}{1.2CA}
\punkt{03/02/91}{23:43}{8291.488}{ -0.002}{0.019}{ 300}{3.37}{
988}{1.2CA}
\punkt{14/02/91}{22:53}{8302.454}{ -0.002}{0.025}{ 100}{3.79}{
328}{1.2CA}
\punkt{23/02/91}{00:45}{8310.531}{ -0.017}{0.006}{ 500}{1.16}{
1002}{1.2CA}
\punkt{23/02/91}{23:50}{8311.494}{ -0.015}{0.008}{ 300}{1.25}{
736}{1.2CA}
\punkt{25/02/91}{04:05}{8312.670}{ -0.006}{0.006}{ 500}{2.25}{
1338}{1.2CA}
\punkta{16/03/91}{01:26}{8331.560}{  0.009}{0.045}{ 500}{3.02}{
532}{1.2CA}
\punkt{16/03/91}{22:09}{8332.423}{ -0.017}{0.013}{ 500}{4.12}{
529}{1.2CA}
\punkt{18/03/91}{21:58}{8334.415}{ -0.021}{0.024}{ 500}{2.37}{
508}{1.2CA}
\punkt{18/03/91}{22:06}{8334.421}{ -0.016}{0.025}{ 500}{2.20}{
513}{1.2CA}
\punkt{19/03/91}{22:29}{8335.437}{ -0.014}{0.028}{ 500}{1.30}{
515}{1.2CA}
\punkt{23/05/91}{20:42}{8400.363}{ -0.019}{0.016}{ 500}{1.16}{
1420}{1.2CA}
\punkt{19/10/91}{05:00}{8548.709}{  0.008}{0.029}{ 500}{2.19}{
698}{1.2CA}
\punkt{21/10/91}{05:27}{8550.728}{  0.015}{0.018}{ 149}{2.28}{
758}{1.2CA}
\punkt{29/10/91}{04:38}{8558.693}{  0.008}{0.019}{ 300}{1.26}{
1002}{1.2CA}
\punkt{02/02/92}{03:15}{8654.636}{  0.020}{0.030}{ 500}{1.72}{
523}{1.2CA}
\punkta{03/02/92}{02:15}{8655.594}{  0.037}{0.051}{ 500}{1.57}{
557}{1.2CA}
\punkt{07/02/92}{01:23}{8659.558}{  0.015}{0.028}{ 500}{1.15}{
513}{1.2CA}
\punkt{10/02/92}{02:26}{8662.602}{  0.023}{0.029}{ 500}{1.11}{
503}{1.2CA}
\punkt{11/02/92}{02:03}{8663.586}{  0.017}{0.005}{ 500}{0.98}{
499}{1.2CA}
\punkt{12/02/92}{02:20}{8664.597}{  0.022}{0.027}{ 500}{1.75}{
494}{1.2CA}
\punkt{15/02/92}{00:10}{8667.507}{  0.046}{0.042}{ 500}{1.63}{
2010}{1.2CA}
\punkt{15/02/92}{00:10}{8667.507}{  0.046}{0.042}{ 500}{1.63}{
2010}{1.2CA}

\end{picture}}

\end{picture}

\vspace*{-0.02cm}

\begin{picture}(18 ,2.5 )(0,0)
\put(0,0){\setlength{\unitlength}{0.01059cm}%
\begin{picture}(1700, 236.111)(7350,0)
\put(7350,0){\framebox(1700, 236.111)[tl]{\begin{picture}(0,0)(0,0)
        \put(1700,0){\makebox(0,0)[tr]{\bf{1219+755}\T{0.4}
                                 \hspace*{0.5cm}}}
        \put(1700,-
236.111){\setlength{\unitlength}{1cm}\begin{picture}(0,0)(0,0)
            \put(0,-1){\makebox(0,0)[br]{\bf J.D.\,2,440,000\,+}}
        \end{picture}}
    \end{picture}}}

\thicklines
\put(7350,0){\setlength{\unitlength}{5cm}\begin{picture}(0,0)(0,-0.25)
   \put(0,0){\setlength{\unitlength}{1cm}\begin{picture}(0,0)(0,0)
        \put(0,0){\line(1,0){0.3}}
        \end{picture}}
   \end{picture}}

\put(9050,0){\setlength{\unitlength}{5cm}\begin{picture}(0,0)(0,-0.25)
   \put(0,0){\setlength{\unitlength}{1cm}\begin{picture}(0,0)(0,0)
        \put(0,0){\line(-1,0){0.3}}
        \end{picture}}
   \end{picture}}

\thinlines
\put(7350,0){\setlength{\unitlength}{5cm}\begin{picture}(0,0)(0,-0.25)
   \multiput(0,0)(0,0.1){3}{\setlength{\unitlength}{1cm}%
\begin{picture}(0,0)(0,0)
        \put(0,0){\line(1,0){0.12}}
        \end{picture}}
   \end{picture}}

\put(7350,0){\setlength{\unitlength}{5cm}\begin{picture}(0,0)(0,-0.25)
   \multiput(0,0)(0,-0.1){3}{\setlength{\unitlength}{1cm}%
\begin{picture}(0,0)(0,0)
        \put(0,0){\line(1,0){0.12}}
        \end{picture}}
   \end{picture}}

\put(9050,0){\setlength{\unitlength}{5cm}\begin{picture}(0,0)(0,-0.25)
   \multiput(0,0)(0,0.1){3}{\setlength{\unitlength}{1cm}%
\begin{picture}(0,0)(0,0)
        \put(0,0){\line(-1,0){0.12}}
        \end{picture}}
   \end{picture}}

\put(9050,0){\setlength{\unitlength}{5cm}\begin{picture}(0,0)(0,-0.25)
   \multiput(0,0)(0,-0.1){3}{\setlength{\unitlength}{1cm}%
\begin{picture}(0,0)(0,0)
        \put(0,0){\line(-1,0){0.12}}
        \end{picture}}
   \end{picture}}

   \put(7527.5, 236.111){\setlength{\unitlength}{1cm}\begin{picture}(0,0)(0,0)
        \put(0,0){\line(0,-1){0.2}}
   \end{picture}}
   \put(7892.5, 236.111){\setlength{\unitlength}{1cm}\begin{picture}(0,0)(0,0)
        \put(0,0){\line(0,-1){0.2}}
   \end{picture}}
   \put(8257.5, 236.111){\setlength{\unitlength}{1cm}\begin{picture}(0,0)(0,0)
        \put(0,0){\line(0,-1){0.2}}
   \end{picture}}
   \put(8622.5, 236.111){\setlength{\unitlength}{1cm}\begin{picture}(0,0)(0,0)
        \put(0,0){\line(0,-1){0.2}}
   \end{picture}}
   \put(8987.5, 236.111){\setlength{\unitlength}{1cm}\begin{picture}(0,0)(0,0)
        \put(0,0){\line(0,-1){0.2}}
   \end{picture}}
    \multiput(7350,0)(50,0){33}%
        {\setlength{\unitlength}{1cm}\begin{picture}(0,0)(0,0)
        \put(0,0){\line(0,1){0.12}}
    \end{picture}}
    \put(7500,0){\setlength{\unitlength}{1cm}\begin{picture}(0,0)(0,0)
        \put(0,0){\line(0,1){0.2}}
        \put(0,-0.2){\makebox(0,0)[t]{\bf 7500}}
    \end{picture}}
    \put(7750,0){\setlength{\unitlength}{1cm}\begin{picture}(0,0)(0,0)
        \put(0,0){\line(0,1){0.2}}
        \put(0,-0.2){\makebox(0,0)[t]{\bf 7750}}
    \end{picture}}
    \put(8000,0){\setlength{\unitlength}{1cm}\begin{picture}(0,0)(0,0)
        \put(0,0){\line(0,1){0.2}}
        \put(0,-0.2){\makebox(0,0)[t]{\bf 8000}}
    \end{picture}}
    \put(8250,0){\setlength{\unitlength}{1cm}\begin{picture}(0,0)(0,0)
        \put(0,0){\line(0,1){0.2}}
        \put(0,-0.2){\makebox(0,0)[t]{\bf 8250}}
    \end{picture}}
    \put(8500,0){\setlength{\unitlength}{1cm}\begin{picture}(0,0)(0,0)
        \put(0,0){\line(0,1){0.2}}
        \put(0,-0.2){\makebox(0,0)[t]{\bf 8500}}
    \end{picture}}
    \put(8750,0){\setlength{\unitlength}{1cm}\begin{picture}(0,0)(0,0)
        \put(0,0){\line(0,1){0.2}}
        \put(0,-0.2){\makebox(0,0)[t]{\bf 8750}}
    \end{picture}}
    \put(9000,0){\setlength{\unitlength}{1cm}\begin{picture}(0,0)(0,0)
        \put(0,0){\line(0,1){0.2}}
        \put(0,-0.2){\makebox(0,0)[t]{\bf 9000}}
    \end{picture}}

\punkt{04/09/88}{20:05}{7409.337}{ -0.026}{0.019}{ 600}{1.71}{
808}{1.2CA}
\punkt{11/06/89}{21:24}{7689.392}{  0.127}{0.009}{ 500}{1.33}{
1194}{1.2CA}
\punkta{12/06/89}{21:50}{7690.410}{  0.108}{0.026}{1000}{1.18}{
1942}{1.2CA}
\punkt{14/08/89}{20:12}{7753.342}{ -0.004}{0.018}{ 500}{1.77}{
2287}{1.2CA}
\punkt{01/11/89}{05:24}{7831.726}{ -0.105}{0.027}{ 500}{1.37}{
670}{1.2CA}
\punkt{09/02/90}{01:45}{7931.573}{ -0.122}{0.009}{ 500}{1.18}{
1398}{1.2CA}
\punkt{28/02/90}{02:20}{7950.597}{ -0.107}{0.026}{ 500}{1.30}{
686}{1.2CA}
\punkt{01/08/90}{20:37}{8105.359}{ -0.176}{0.026}{ 500}{1.77}{
1152}{1.2CA}
\punkt{15/02/91}{02:28}{8302.603}{  0.039}{0.034}{ 300}{1.70}{
453}{1.2CA}
\punkt{23/02/91}{01:31}{8310.563}{  0.060}{0.035}{ 300}{1.84}{
707}{1.2CA}
\punkt{24/02/91}{01:38}{8311.569}{  0.018}{0.030}{ 300}{1.33}{
709}{1.2CA}
\punkt{25/02/91}{04:39}{8312.694}{  0.042}{0.031}{ 300}{1.81}{
1093}{1.2CA}
\punkt{20/03/91}{01:02}{8335.543}{  0.045}{0.035}{ 500}{1.49}{
557}{1.2CA}
\punkt{23/05/91}{22:07}{8400.422}{  0.058}{0.015}{ 100}{1.12}{
424}{1.2CA}
\punkt{28/05/91}{22:22}{8405.433}{  0.106}{0.037}{ 100}{1.62}{
647}{1.2CA}
\punkt{29/10/91}{05:24}{8558.725}{  0.004}{0.047}{ 500}{1.28}{
1222}{1.2CA}
\punkt{02/02/92}{05:22}{8654.724}{ -0.026}{0.043}{ 100}{1.81}{
355}{1.2CA}
\punkt{08/02/92}{05:16}{8660.720}{ -0.025}{0.024}{ 500}{1.44}{
531}{1.2CA}
\punkta{09/02/92}{04:38}{8661.694}{ -0.054}{0.038}{ 500}{1.05}{
500}{1.2CA}
\punkt{12/02/92}{05:33}{8664.731}{ -0.008}{0.025}{ 500}{1.66}{
552}{1.2CA}
\punkt{15/02/92}{04:11}{8667.675}{  0.000}{0.018}{ 500}{1.64}{
720}{1.2CA}
\punkt{16/02/92}{03:46}{8668.657}{  0.004}{0.025}{ 500}{1.59}{
2443}{1.2CA}
\punkt{17/02/92}{02:11}{8669.592}{ -0.003}{0.022}{ 500}{1.48}{
2726}{1.2CA}
\punkt{17/02/92}{02:11}{8669.592}{ -0.003}{0.022}{ 500}{1.48}{
2726}{1.2CA}

\end{picture}}

\end{picture}

\vspace*{1cm}

\caption{HQM lightcurves in the $R\/$-band. Dashes on the vertical axes
represent 0.1\,mag steps. Plotted are variations $\Delta R$=$R_0$$-$$R$;
the reference magnitude $R_0$ is indicated by thick dashes. Measurements
obtained under very bad atmospheric conditions or those with only one
reference star are shown by open circles; reliable error bars can in these
cases not be given}
\end{figure*}
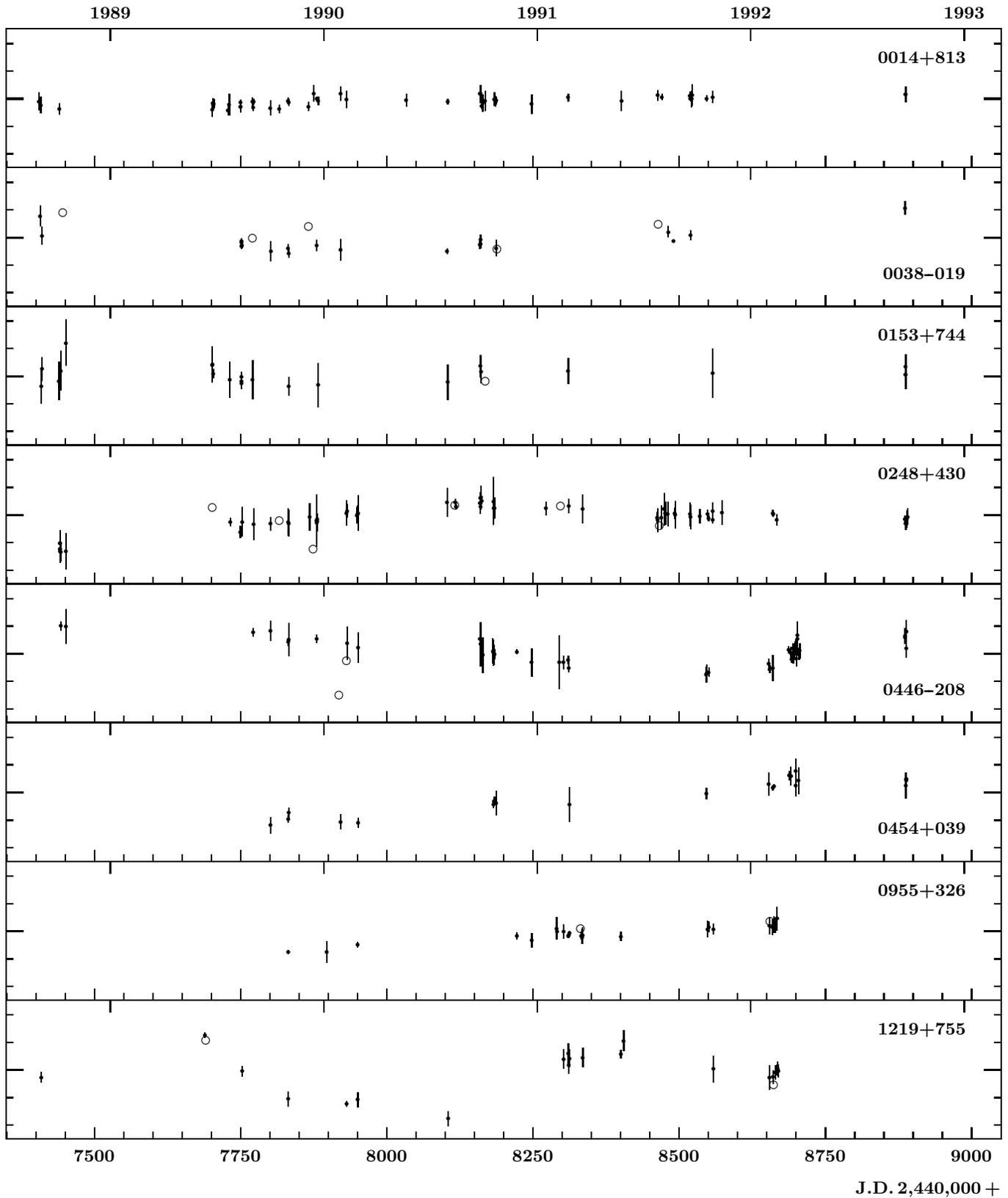

\begin{figure*}

\vspace*{0.4cm}

\begin{picture}(18 ,7.5 )(0,0)
\put(0,0){\setlength{\unitlength}{0.01059cm}%
\begin{picture}(1700, 708.215)(7350,0)
\put(7350,0){\framebox(1700, 708.215)[tl]{\begin{picture}(0,0)(0,0)
        \put(1700,0){\makebox(0,0)[tr]{\bf{1150+497}\T{0.4}
                                 \hspace*{0.5cm}}}
    \end{picture}}}

\thicklines
\put(7350,0){\setlength{\unitlength}{5cm}\begin{picture}(0,0)(0,-0.75)
   \put(0,0){\setlength{\unitlength}{1cm}\begin{picture}(0,0)(0,0)
        \put(0,0){\line(1,0){0.3}}
        \end{picture}}
   \end{picture}}

\put(9050,0){\setlength{\unitlength}{5cm}\begin{picture}(0,0)(0,-0.75)
   \put(0,0){\setlength{\unitlength}{1cm}\begin{picture}(0,0)(0,0)
        \put(0,0){\line(-1,0){0.3}}
        \end{picture}}
   \end{picture}}

\thinlines
\put(7350,0){\setlength{\unitlength}{5cm}\begin{picture}(0,0)(0,-0.75)
   \multiput(0,0)(0,0.1){8}{\setlength{\unitlength}{1cm}%
\begin{picture}(0,0)(0,0)
        \put(0,0){\line(1,0){0.12}}
        \end{picture}}
   \end{picture}}

\put(7350,0){\setlength{\unitlength}{5cm}\begin{picture}(0,0)(0,-0.75)
   \multiput(0,0)(0,-0.1){8}{\setlength{\unitlength}{1cm}%
\begin{picture}(0,0)(0,0)
        \put(0,0){\line(1,0){0.12}}
        \end{picture}}
   \end{picture}}

\put(9050,0){\setlength{\unitlength}{5cm}\begin{picture}(0,0)(0,-0.75)
   \multiput(0,0)(0,0.1){8}{\setlength{\unitlength}{1cm}%
\begin{picture}(0,0)(0,0)
        \put(0,0){\line(-1,0){0.12}}
        \end{picture}}
   \end{picture}}

\put(9050,0){\setlength{\unitlength}{5cm}\begin{picture}(0,0)(0,-0.75)
   \multiput(0,0)(0,-0.1){8}{\setlength{\unitlength}{1cm}%
\begin{picture}(0,0)(0,0)
        \put(0,0){\line(-1,0){0.12}}
        \end{picture}}
   \end{picture}}

   \put(7527.5, 708.215){\setlength{\unitlength}{1cm}\begin{picture}(0,0)(0,0)
        \put(0,0){\line(0,-1){0.2}}
        \put(0,0.2){\makebox(0,0)[b]{\bf 1989}}
   \end{picture}}
   \put(7892.5, 708.215){\setlength{\unitlength}{1cm}\begin{picture}(0,0)(0,0)
        \put(0,0){\line(0,-1){0.2}}
        \put(0,0.2){\makebox(0,0)[b]{\bf 1990}}
   \end{picture}}
   \put(8257.5, 708.215){\setlength{\unitlength}{1cm}\begin{picture}(0,0)(0,0)
        \put(0,0){\line(0,-1){0.2}}
        \put(0,0.2){\makebox(0,0)[b]{\bf 1991}}
   \end{picture}}
   \put(8622.5, 708.215){\setlength{\unitlength}{1cm}\begin{picture}(0,0)(0,0)
        \put(0,0){\line(0,-1){0.2}}
        \put(0,0.2){\makebox(0,0)[b]{\bf 1992}}
   \end{picture}}
   \put(8987.5, 708.215){\setlength{\unitlength}{1cm}\begin{picture}(0,0)(0,0)
        \put(0,0){\line(0,-1){0.2}}
        \put(0,0.2){\makebox(0,0)[b]{\bf 1993}}
   \end{picture}}
    \multiput(7350,0)(50,0){33}%
        {\setlength{\unitlength}{1cm}\begin{picture}(0,0)(0,0)
        \put(0,0){\line(0,1){0.12}}
    \end{picture}}
    \put(7500,0){\setlength{\unitlength}{1cm}\begin{picture}(0,0)(0,0)
        \put(0,0){\line(0,1){0.2}}
    \end{picture}}
    \put(7750,0){\setlength{\unitlength}{1cm}\begin{picture}(0,0)(0,0)
        \put(0,0){\line(0,1){0.2}}
    \end{picture}}
    \put(8000,0){\setlength{\unitlength}{1cm}\begin{picture}(0,0)(0,0)
        \put(0,0){\line(0,1){0.2}}
    \end{picture}}
    \put(8250,0){\setlength{\unitlength}{1cm}\begin{picture}(0,0)(0,0)
        \put(0,0){\line(0,1){0.2}}
    \end{picture}}
    \put(8500,0){\setlength{\unitlength}{1cm}\begin{picture}(0,0)(0,0)
        \put(0,0){\line(0,1){0.2}}
    \end{picture}}
    \put(8750,0){\setlength{\unitlength}{1cm}\begin{picture}(0,0)(0,0)
        \put(0,0){\line(0,1){0.2}}
    \end{picture}}
    \put(9000,0){\setlength{\unitlength}{1cm}\begin{picture}(0,0)(0,0)
        \put(0,0){\line(0,1){0.2}}
    \end{picture}}

\put(7000,0){\begin{picture}(0,0)(7000,-236.07177)
\punkt{12/06/89}{21:32}{7690.397}{ -0.248}{0.020}{ 500}{1.06}{
1416}{1.2CA}
\punkt{20/12/89}{04:50}{7880.702}{  0.102}{0.010}{ 500}{1.72}{
659}{1.2CA}
\punkt{21/12/89}{03:40}{7881.653}{  0.128}{0.013}{ 500}{1.92}{
616}{1.2CA}
\punkt{08/02/90}{23:41}{7931.487}{ -0.252}{0.020}{ 500}{1.54}{
1478}{1.2CA}
\punkt{28/02/90}{01:31}{7950.564}{ -0.271}{0.018}{ 500}{1.37}{
613}{1.2CA}
\punkt{03/02/91}{02:40}{8290.612}{ -0.176}{0.019}{ 500}{2.08}{
1096}{1.2CA}
\punkt{23/02/91}{02:36}{8310.609}{ -0.147}{0.011}{ 500}{1.06}{
772}{1.2CA}
\punkta{23/05/91}{21:47}{8400.408}{ -0.188}{0.036}{ 100}{1.35}{
407}{1.2CA}
\punkta{28/10/91}{05:34}{8557.732}{ -0.542}{0.086}{ 300}{3.31}{
1322}{1.2CA}
\punkt{18/01/92}{21:35}{8640.400}{ -0.577}{0.015}{ 500}{1.67}{
6324}{1.2CA}
\punkt{12/02/92}{04:46}{8664.699}{ -0.443}{0.009}{ 500}{1.14}{
474}{1.2CA}
\punkt{15/02/92}{02:55}{8667.622}{  0.244}{0.008}{ 500}{1.26}{
1180}{1.2CA}
\punkt{15/02/92}{03:34}{8667.649}{  0.287}{0.013}{ 500}{1.34}{
881}{1.2CA}
\punkt{15/02/92}{04:56}{8667.706}{  0.293}{0.013}{ 500}{1.79}{
507}{1.2CA}
\punkt{15/02/92}{23:50}{8668.493}{  0.534}{0.035}{ 500}{1.82}{
1849}{1.2CA}
\punkt{16/02/92}{00:38}{8668.527}{  0.469}{0.014}{ 500}{1.03}{
1808}{1.2CA}
\punkt{28/02/92}{02:54}{8680.621}{ -0.334}{0.026}{ 500}{3.02}{
524}{1.2CA}
\punkt{28/02/92}{02:54}{8680.621}{ -0.334}{0.026}{ 500}{3.02}{
524}{1.2CA}

\end{picture}}

\end{picture}}

\end{picture}

\vspace*{-0.02cm}

\begin{picture}(18 ,2.5 )(0,0)
\put(0,0){\setlength{\unitlength}{0.01059cm}%
\begin{picture}(1700, 236.111)(7350,0)
\put(7350,0){\framebox(1700, 236.111)[tl]{\begin{picture}(0,0)(0,0)
        \put(1700,0){\makebox(0,0)[tr]{\bf{1522+101}\T{0.4}
                                 \hspace*{0.5cm}}}
    \end{picture}}}

\thicklines
\put(7350,0){\setlength{\unitlength}{5cm}\begin{picture}(0,0)(0,-0.25)
   \put(0,0){\setlength{\unitlength}{1cm}\begin{picture}(0,0)(0,0)
        \put(0,0){\line(1,0){0.3}}
        \end{picture}}
   \end{picture}}

\put(9050,0){\setlength{\unitlength}{5cm}\begin{picture}(0,0)(0,-0.25)
   \put(0,0){\setlength{\unitlength}{1cm}\begin{picture}(0,0)(0,0)
        \put(0,0){\line(-1,0){0.3}}
        \end{picture}}
   \end{picture}}

\thinlines
\put(7350,0){\setlength{\unitlength}{5cm}\begin{picture}(0,0)(0,-0.25)
   \multiput(0,0)(0,0.1){3}{\setlength{\unitlength}{1cm}%
\begin{picture}(0,0)(0,0)
        \put(0,0){\line(1,0){0.12}}
        \end{picture}}
   \end{picture}}

\put(7350,0){\setlength{\unitlength}{5cm}\begin{picture}(0,0)(0,-0.25)
   \multiput(0,0)(0,-0.1){3}{\setlength{\unitlength}{1cm}%
\begin{picture}(0,0)(0,0)
        \put(0,0){\line(1,0){0.12}}
        \end{picture}}
   \end{picture}}

\put(9050,0){\setlength{\unitlength}{5cm}\begin{picture}(0,0)(0,-0.25)
   \multiput(0,0)(0,0.1){3}{\setlength{\unitlength}{1cm}%
\begin{picture}(0,0)(0,0)
        \put(0,0){\line(-1,0){0.12}}
        \end{picture}}
   \end{picture}}

\put(9050,0){\setlength{\unitlength}{5cm}\begin{picture}(0,0)(0,-0.25)
   \multiput(0,0)(0,-0.1){3}{\setlength{\unitlength}{1cm}%
\begin{picture}(0,0)(0,0)
        \put(0,0){\line(-1,0){0.12}}
        \end{picture}}
   \end{picture}}

   \put(7527.5, 236.111){\setlength{\unitlength}{1cm}\begin{picture}(0,0)(0,0)
        \put(0,0){\line(0,-1){0.2}}
   \end{picture}}
   \put(7892.5, 236.111){\setlength{\unitlength}{1cm}\begin{picture}(0,0)(0,0)
        \put(0,0){\line(0,-1){0.2}}
   \end{picture}}
   \put(8257.5, 236.111){\setlength{\unitlength}{1cm}\begin{picture}(0,0)(0,0)
        \put(0,0){\line(0,-1){0.2}}
   \end{picture}}
   \put(8622.5, 236.111){\setlength{\unitlength}{1cm}\begin{picture}(0,0)(0,0)
        \put(0,0){\line(0,-1){0.2}}
   \end{picture}}
   \put(8987.5, 236.111){\setlength{\unitlength}{1cm}\begin{picture}(0,0)(0,0)
        \put(0,0){\line(0,-1){0.2}}
   \end{picture}}
    \multiput(7350,0)(50,0){33}%
        {\setlength{\unitlength}{1cm}\begin{picture}(0,0)(0,0)
        \put(0,0){\line(0,1){0.12}}
    \end{picture}}
    \put(7500,0){\setlength{\unitlength}{1cm}\begin{picture}(0,0)(0,0)
        \put(0,0){\line(0,1){0.2}}
    \end{picture}}
    \put(7750,0){\setlength{\unitlength}{1cm}\begin{picture}(0,0)(0,0)
        \put(0,0){\line(0,1){0.2}}
    \end{picture}}
    \put(8000,0){\setlength{\unitlength}{1cm}\begin{picture}(0,0)(0,0)
        \put(0,0){\line(0,1){0.2}}
    \end{picture}}
    \put(8250,0){\setlength{\unitlength}{1cm}\begin{picture}(0,0)(0,0)
        \put(0,0){\line(0,1){0.2}}
    \end{picture}}
    \put(8500,0){\setlength{\unitlength}{1cm}\begin{picture}(0,0)(0,0)
        \put(0,0){\line(0,1){0.2}}
    \end{picture}}
    \put(8750,0){\setlength{\unitlength}{1cm}\begin{picture}(0,0)(0,0)
        \put(0,0){\line(0,1){0.2}}
    \end{picture}}
    \put(9000,0){\setlength{\unitlength}{1cm}\begin{picture}(0,0)(0,0)
        \put(0,0){\line(0,1){0.2}}
    \end{picture}}

\punkt{24/06/89}{23:50}{7702.493}{  0.066}{0.007}{ 500}{1.24}{
928}{1.2CA}
\punkt{29/06/89}{20:30}{7707.355}{  0.079}{0.012}{  30}{1.29}{
2151}{1.2CA}
\punkt{20/07/89}{22:09}{7728.423}{  0.063}{0.007}{ 500}{1.46}{
1526}{1.2CA}
\punkt{23/07/89}{22:37}{7731.443}{  0.040}{0.015}{ 500}{2.08}{
1131}{1.2CA}
\punkt{11/08/89}{20:55}{7750.372}{  0.028}{0.018}{ 500}{2.85}{
1697}{1.2CA}
\punkt{24/05/90}{00:44}{8035.531}{ -0.017}{0.034}{ 400}{3.14}{
1497}{1.2CA}
\punkt{30/07/90}{21:50}{8103.410}{ -0.037}{0.036}{ 500}{2.42}{
1198}{1.2CA}
\punkt{03/02/91}{05:30}{8290.729}{ -0.062}{0.021}{ 249}{3.03}{
669}{1.2CA}
\punkt{15/02/91}{05:28}{8302.728}{ -0.042}{0.016}{ 402}{2.02}{
473}{1.2CA}
\punkt{20/03/91}{03:58}{8335.665}{ -0.028}{0.016}{ 500}{1.17}{
505}{1.2CA}
\punkt{22/05/91}{00:00}{8398.501}{ -0.043}{0.028}{ 100}{1.79}{
376}{1.2CA}
\punkt{23/05/91}{00:01}{8399.501}{ -0.033}{0.029}{ 150}{1.85}{
435}{1.2CA}
\punkt{23/05/91}{23:43}{8400.489}{ -0.042}{0.013}{ 100}{0.99}{
418}{1.2CA}
\punkt{23/05/91}{23:46}{8400.490}{ -0.039}{0.017}{ 100}{1.07}{
416}{1.2CA}
\punkt{04/08/91}{21:46}{8473.408}{ -0.043}{0.021}{ 200}{1.98}{
353}{1.2CA}
\punkt{24/08/91}{20:56}{8493.372}{ -0.041}{0.033}{ 200}{2.05}{
966}{1.2CA}
\punkt{16/02/92}{05:26}{8668.727}{  0.040}{0.016}{ 500}{1.93}{
539}{1.2CA}
\punkt{28/02/92}{05:21}{8680.723}{  0.016}{0.014}{ 300}{2.69}{
470}{1.2CA}
\punkt{10/03/92}{09:17}{8691.887}{ -0.002}{0.018}{ 500}{1.34}{
388}{1.2CA}
\punkt{}{}{8697.827}{0.056}{0.025}{}{}{}{}
\punkt{}{}{8700.827}{0.039}{0.015}{}{}{}{}
%

\end{picture}}

\end{picture}

\vspace*{-0.02cm}

\begin{picture}(18 ,2.5 )(0,0)
\put(0,0){\setlength{\unitlength}{0.01059cm}%
\begin{picture}(1700, 236.111)(7350,0)
\put(7350,0){\framebox(1700, 236.111)[tl]{\begin{picture}(0,0)(0,0)
        \put(1700,0){\makebox(0,0)[tr]{\bf{1700+642}\T{0.4}
                                 \hspace*{0.5cm}}}
    \end{picture}}}

\thicklines
\put(7350,0){\setlength{\unitlength}{5cm}\begin{picture}(0,0)(0,-0.25)
   \put(0,0){\setlength{\unitlength}{1cm}\begin{picture}(0,0)(0,0)
        \put(0,0){\line(1,0){0.3}}
        \end{picture}}
   \end{picture}}

\put(9050,0){\setlength{\unitlength}{5cm}\begin{picture}(0,0)(0,-0.25)
   \put(0,0){\setlength{\unitlength}{1cm}\begin{picture}(0,0)(0,0)
        \put(0,0){\line(-1,0){0.3}}
        \end{picture}}
   \end{picture}}

\thinlines
\put(7350,0){\setlength{\unitlength}{5cm}\begin{picture}(0,0)(0,-0.25)
   \multiput(0,0)(0,0.1){3}{\setlength{\unitlength}{1cm}%
\begin{picture}(0,0)(0,0)
        \put(0,0){\line(1,0){0.12}}
        \end{picture}}
   \end{picture}}

\put(7350,0){\setlength{\unitlength}{5cm}\begin{picture}(0,0)(0,-0.25)
   \multiput(0,0)(0,-0.1){3}{\setlength{\unitlength}{1cm}%
\begin{picture}(0,0)(0,0)
        \put(0,0){\line(1,0){0.12}}
        \end{picture}}
   \end{picture}}

\put(9050,0){\setlength{\unitlength}{5cm}\begin{picture}(0,0)(0,-0.25)
   \multiput(0,0)(0,0.1){3}{\setlength{\unitlength}{1cm}%
\begin{picture}(0,0)(0,0)
        \put(0,0){\line(-1,0){0.12}}
        \end{picture}}
   \end{picture}}

\put(9050,0){\setlength{\unitlength}{5cm}\begin{picture}(0,0)(0,-0.25)
   \multiput(0,0)(0,-0.1){3}{\setlength{\unitlength}{1cm}%
\begin{picture}(0,0)(0,0)
        \put(0,0){\line(-1,0){0.12}}
        \end{picture}}
   \end{picture}}

   \put(7527.5, 236.111){\setlength{\unitlength}{1cm}\begin{picture}(0,0)(0,0)
        \put(0,0){\line(0,-1){0.2}}
   \end{picture}}
   \put(7892.5, 236.111){\setlength{\unitlength}{1cm}\begin{picture}(0,0)(0,0)
        \put(0,0){\line(0,-1){0.2}}
   \end{picture}}
   \put(8257.5, 236.111){\setlength{\unitlength}{1cm}\begin{picture}(0,0)(0,0)
        \put(0,0){\line(0,-1){0.2}}
   \end{picture}}
   \put(8622.5, 236.111){\setlength{\unitlength}{1cm}\begin{picture}(0,0)(0,0)
        \put(0,0){\line(0,-1){0.2}}
   \end{picture}}
   \put(8987.5, 236.111){\setlength{\unitlength}{1cm}\begin{picture}(0,0)(0,0)
        \put(0,0){\line(0,-1){0.2}}
   \end{picture}}
    \multiput(7350,0)(50,0){33}%
        {\setlength{\unitlength}{1cm}\begin{picture}(0,0)(0,0)
        \put(0,0){\line(0,1){0.12}}
    \end{picture}}
    \put(7500,0){\setlength{\unitlength}{1cm}\begin{picture}(0,0)(0,0)
        \put(0,0){\line(0,1){0.2}}
    \end{picture}}
    \put(7750,0){\setlength{\unitlength}{1cm}\begin{picture}(0,0)(0,0)
        \put(0,0){\line(0,1){0.2}}
    \end{picture}}
    \put(8000,0){\setlength{\unitlength}{1cm}\begin{picture}(0,0)(0,0)
        \put(0,0){\line(0,1){0.2}}
    \end{picture}}
    \put(8250,0){\setlength{\unitlength}{1cm}\begin{picture}(0,0)(0,0)
        \put(0,0){\line(0,1){0.2}}
    \end{picture}}
    \put(8500,0){\setlength{\unitlength}{1cm}\begin{picture}(0,0)(0,0)
        \put(0,0){\line(0,1){0.2}}
    \end{picture}}
    \put(8750,0){\setlength{\unitlength}{1cm}\begin{picture}(0,0)(0,0)
        \put(0,0){\line(0,1){0.2}}
    \end{picture}}
    \put(9000,0){\setlength{\unitlength}{1cm}\begin{picture}(0,0)(0,0)
        \put(0,0){\line(0,1){0.2}}
    \end{picture}}

\punkt{03/09/88}{20:20}{7408.348}{ -0.021}{0.026}{1000}{1.18}{
1090}{1.2CA}
\punkt{05/09/88}{21:53}{7410.412}{ -0.026}{0.019}{ 500}{1.45}{
679}{1.2CA}
\punkt{07/09/88}{20:11}{7412.341}{ -0.022}{0.031}{ 500}{1.55}{
1026}{1.2CA}
\punkt{03/10/88}{19:19}{7438.305}{ -0.007}{0.022}{ 500}{1.65}{
679}{1.2CA}
\punkt{07/10/88}{19:30}{7442.313}{ -0.021}{0.027}{ 500}{1.84}{
689}{1.2CA}
\punkt{16/10/88}{20:26}{7451.352}{  0.024}{0.057}{ 500}{2.93}{
453}{1.2CA}
\punkt{16/10/88}{20:37}{7451.359}{  0.006}{0.031}{ 500}{2.64}{
827}{1.2CA}
\punkt{09/06/89}{12:40}{7687.028}{ -0.051}{0.011}{****}{1.63}{
1010}{1.2CA}
\punkt{11/06/89}{00:11}{7688.508}{ -0.046}{0.008}{1000}{1.68}{
1529}{1.2CA}
\punkt{13/06/89}{02:18}{7690.596}{ -0.045}{0.010}{1000}{1.14}{
1408}{1.2CA}
\punkt{24/06/89}{00:26}{7701.519}{ -0.064}{0.006}{ 500}{1.89}{
997}{1.2CA}
\punkt{25/06/89}{01:15}{7702.552}{ -0.043}{0.009}{ 500}{1.24}{
1054}{1.2CA}
\punkt{27/06/89}{01:13}{7704.551}{ -0.040}{0.017}{ 500}{1.83}{
847}{1.2CA}
\punkt{19/07/89}{23:30}{7727.479}{ -0.023}{0.055}{ 500}{1.61}{
2396}{1.2CA}
\punkt{22/07/89}{21:45}{7730.407}{ -0.068}{0.020}{ 500}{1.84}{
767}{1.2CA}
\punkt{24/07/89}{01:18}{7731.555}{ -0.031}{0.040}{ 500}{2.21}{
1039}{1.2CA}
\punkt{24/07/89}{23:26}{7732.477}{ -0.041}{0.023}{1000}{1.28}{
1910}{1.2CA}
\punkt{10/08/89}{21:36}{7749.401}{ -0.083}{0.039}{ 500}{1.89}{
1608}{1.2CA}
\punkt{14/08/89}{02:19}{7752.597}{ -0.081}{0.031}{1000}{2.18}{
1203}{1.2CA}
\punkt{16/08/89}{22:48}{7755.450}{ -0.039}{0.012}{ 500}{1.73}{
2257}{1.2CA}
\punkt{02/09/89}{20:48}{7772.367}{ -0.016}{0.009}{ 500}{1.72}{
912}{1.2CA}
\punkt{03/09/89}{20:59}{7773.374}{ -0.110}{0.006}{ 500}{2.72}{
751}{1.2CA}
\punkt{01/10/89}{07:53}{7800.829}{ -0.007}{0.044}{ 500}{0.00}{
543}{1.2CA}
\punkt{16/10/89}{20:44}{7816.364}{ -0.040}{0.028}{1165}{3.50}{
1710}{1.2CA}
\punkt{31/10/89}{19:09}{7831.298}{ -0.010}{0.011}{ 500}{1.93}{
590}{1.2CA}
\punkt{01/11/89}{18:54}{7832.288}{ -0.017}{0.040}{ 500}{1.67}{
593}{1.2CA}
\punkt{20/12/89}{05:46}{7880.740}{  0.005}{0.041}{ 500}{2.30}{
739}{1.2CA}
\punkt{11/02/90}{01:58}{7933.583}{  0.041}{0.045}{ 500}{2.18}{
1662}{1.2CA}
\punkt{22/05/90}{02:26}{8033.602}{  0.085}{0.009}{ 400}{1.84}{
1505}{1.2CA}
\punkta{24/05/90}{03:02}{8035.627}{  0.058}{0.023}{ 500}{3.63}{
1520}{1.2CA}
\punkt{31/07/90}{00:11}{8103.508}{ -0.010}{0.006}{ 500}{1.47}{
708}{1.2CA}
\punkt{24/09/90}{20:09}{8159.340}{  0.014}{0.028}{ 500}{4.00}{
1017}{1.2CA}
\punkt{25/09/90}{19:31}{8160.314}{  0.025}{0.022}{ 500}{1.64}{
1000}{1.2CA}
\punkt{26/09/90}{19:31}{8161.314}{  0.033}{0.012}{ 500}{1.64}{
1001}{1.2CA}
\punkt{01/10/90}{19:42}{8166.321}{  0.010}{0.031}{ 500}{1.81}{
1288}{1.2CA}
\punkt{16/10/90}{20:38}{8181.360}{  0.019}{0.022}{ 500}{2.65}{
996}{1.2CA}
\punkt{17/10/90}{20:38}{8182.360}{  0.030}{0.010}{ 500}{2.60}{
915}{1.2CA}
\punkt{18/10/90}{20:24}{8183.350}{  0.041}{0.015}{ 500}{2.07}{
852}{1.2CA}
\punkt{19/10/90}{20:24}{8184.350}{  0.043}{0.016}{ 500}{2.10}{
820}{1.2CA}
\punkt{19/10/90}{20:27}{8184.352}{  0.033}{0.015}{ 500}{2.00}{
891}{1.2CA}
\punkt{20/10/90}{20:27}{8185.352}{  0.037}{0.016}{ 500}{1.98}{
842}{1.2CA}
\punkt{10/02/91}{06:02}{8297.752}{  0.014}{0.014}{ 300}{1.40}{
568}{1.2CA}
\punkt{20/03/91}{04:44}{8335.697}{  0.029}{0.010}{ 500}{0.95}{
527}{1.2CA}
\punkt{23/05/91}{00:04}{8399.503}{  0.034}{0.020}{ 500}{1.88}{
721}{1.2CA}
\punkt{24/05/91}{01:53}{8400.579}{  0.018}{0.032}{ 300}{1.19}{
483}{1.2CA}
\punkt{25/05/91}{00:47}{8401.533}{  0.019}{0.025}{ 500}{1.96}{
1086}{1.2CA}
\punkt{27/05/91}{01:03}{8403.544}{  0.043}{0.029}{ 500}{1.87}{
1833}{1.2CA}
\punkt{29/05/91}{02:06}{8405.588}{  0.050}{0.046}{ 200}{1.82}{
1251}{1.2CA}
\punkt{23/07/91}{21:16}{8461.387}{  0.042}{0.024}{ 500}{1.20}{
1385}{1.2CA}
\punkt{25/07/91}{23:40}{8463.486}{  0.035}{0.025}{ 500}{1.26}{
1808}{1.2CA}
\punkt{28/07/91}{22:48}{8466.450}{  0.044}{0.029}{ 500}{1.96}{
1948}{1.2CA}
\punkta{30/07/91}{22:54}{8468.455}{  0.028}{0.024}{ 100}{2.97}{
433}{1.2CA}
\punkt{03/08/91}{23:43}{8472.489}{  0.054}{0.020}{ 500}{1.70}{
620}{1.2CA}
\punkt{11/08/91}{22:43}{8480.447}{  0.065}{0.028}{ 500}{1.67}{
577}{1.2CA}
\punkt{18/09/91}{21:07}{8518.380}{  0.060}{0.016}{ 500}{1.50}{
1058}{1.2CA}
\punkt{20/09/91}{20:33}{8520.357}{  0.086}{0.029}{ 500}{1.57}{
1476}{1.2CA}
\punkt{17/10/91}{19:19}{8547.305}{  0.065}{0.030}{ 500}{2.24}{
1195}{1.2CA}
\punkt{28/10/91}{20:06}{8558.338}{  0.074}{0.025}{ 500}{1.18}{
709}{1.2CA}
\punkt{17/02/92}{05:17}{8669.720}{  0.117}{0.019}{ 500}{1.34}{
1218}{1.2CA}
\punkt{17/02/92}{05:17}{8669.720}{  0.117}{0.019}{ 500}{1.34}{
1218}{1.2CA}

\end{picture}}

\end{picture}

\vspace*{-0.02cm}

\begin{picture}(18 ,2.5 )(0,0)
\put(0,0){\setlength{\unitlength}{0.01059cm}%
\begin{picture}(1700, 236.111)(7350,0)
\put(7350,0){\framebox(1700, 236.111)[tl]{\begin{picture}(0,0)(0,0)
        \put(1700,0){\makebox(0,0)[tr]{\bf{1715+535}\T{0.4}
                                 \hspace*{0.5cm}}}
    \end{picture}}}

\thicklines
\put(7350,0){\setlength{\unitlength}{5cm}\begin{picture}(0,0)(0,-0.25)
   \put(0,0){\setlength{\unitlength}{1cm}\begin{picture}(0,0)(0,0)
        \put(0,0){\line(1,0){0.3}}
        \end{picture}}
   \end{picture}}

\put(9050,0){\setlength{\unitlength}{5cm}\begin{picture}(0,0)(0,-0.25)
   \put(0,0){\setlength{\unitlength}{1cm}\begin{picture}(0,0)(0,0)
        \put(0,0){\line(-1,0){0.3}}
        \end{picture}}
   \end{picture}}

\thinlines
\put(7350,0){\setlength{\unitlength}{5cm}\begin{picture}(0,0)(0,-0.25)
   \multiput(0,0)(0,0.1){3}{\setlength{\unitlength}{1cm}%
\begin{picture}(0,0)(0,0)
        \put(0,0){\line(1,0){0.12}}
        \end{picture}}
   \end{picture}}

\put(7350,0){\setlength{\unitlength}{5cm}\begin{picture}(0,0)(0,-0.25)
   \multiput(0,0)(0,-0.1){3}{\setlength{\unitlength}{1cm}%
\begin{picture}(0,0)(0,0)
        \put(0,0){\line(1,0){0.12}}
        \end{picture}}
   \end{picture}}

\put(9050,0){\setlength{\unitlength}{5cm}\begin{picture}(0,0)(0,-0.25)
   \multiput(0,0)(0,0.1){3}{\setlength{\unitlength}{1cm}%
\begin{picture}(0,0)(0,0)
        \put(0,0){\line(-1,0){0.12}}
        \end{picture}}
   \end{picture}}

\put(9050,0){\setlength{\unitlength}{5cm}\begin{picture}(0,0)(0,-0.25)
   \multiput(0,0)(0,-0.1){3}{\setlength{\unitlength}{1cm}%
\begin{picture}(0,0)(0,0)
        \put(0,0){\line(-1,0){0.12}}
        \end{picture}}
   \end{picture}}

   \put(7527.5, 236.111){\setlength{\unitlength}{1cm}\begin{picture}(0,0)(0,0)
        \put(0,0){\line(0,-1){0.2}}
   \end{picture}}
   \put(7892.5, 236.111){\setlength{\unitlength}{1cm}\begin{picture}(0,0)(0,0)
        \put(0,0){\line(0,-1){0.2}}
   \end{picture}}
   \put(8257.5, 236.111){\setlength{\unitlength}{1cm}\begin{picture}(0,0)(0,0)
        \put(0,0){\line(0,-1){0.2}}
   \end{picture}}
   \put(8622.5, 236.111){\setlength{\unitlength}{1cm}\begin{picture}(0,0)(0,0)
        \put(0,0){\line(0,-1){0.2}}
   \end{picture}}
   \put(8987.5, 236.111){\setlength{\unitlength}{1cm}\begin{picture}(0,0)(0,0)
        \put(0,0){\line(0,-1){0.2}}
   \end{picture}}
    \multiput(7350,0)(50,0){33}%
        {\setlength{\unitlength}{1cm}\begin{picture}(0,0)(0,0)
        \put(0,0){\line(0,1){0.12}}
    \end{picture}}
    \put(7500,0){\setlength{\unitlength}{1cm}\begin{picture}(0,0)(0,0)
        \put(0,0){\line(0,1){0.2}}
    \end{picture}}
    \put(7750,0){\setlength{\unitlength}{1cm}\begin{picture}(0,0)(0,0)
        \put(0,0){\line(0,1){0.2}}
    \end{picture}}
    \put(8000,0){\setlength{\unitlength}{1cm}\begin{picture}(0,0)(0,0)
        \put(0,0){\line(0,1){0.2}}
    \end{picture}}
    \put(8250,0){\setlength{\unitlength}{1cm}\begin{picture}(0,0)(0,0)
        \put(0,0){\line(0,1){0.2}}
    \end{picture}}
    \put(8500,0){\setlength{\unitlength}{1cm}\begin{picture}(0,0)(0,0)
        \put(0,0){\line(0,1){0.2}}
    \end{picture}}
    \put(8750,0){\setlength{\unitlength}{1cm}\begin{picture}(0,0)(0,0)
        \put(0,0){\line(0,1){0.2}}
    \end{picture}}
    \put(9000,0){\setlength{\unitlength}{1cm}\begin{picture}(0,0)(0,0)
        \put(0,0){\line(0,1){0.2}}
    \end{picture}}

\punkt{03/09/88}{21:51}{7408.411}{ -0.041}{0.080}{ 600}{1.45}{
695}{1.2CA}
\punkt{03/09/88}{22:33}{7408.440}{ -0.069}{0.085}{2000}{1.50}{
1460}{1.2CA}
\punkt{05/09/88}{22:31}{7410.439}{ -0.023}{0.075}{ 500}{1.38}{
686}{1.2CA}
\punkt{09/10/88}{17:57}{7444.248}{ -0.010}{0.047}{ 500}{1.60}{
1023}{1.2CA}
\punkt{10/06/89}{00:49}{7687.534}{  0.044}{0.054}{1000}{1.39}{
1328}{1.2CA}
\punkt{11/06/89}{01:28}{7688.561}{  0.040}{0.064}{1000}{1.62}{
1544}{1.2CA}
\punkt{25/06/89}{01:29}{7702.562}{  0.073}{0.047}{ 500}{1.18}{
1101}{1.2CA}
\punkt{26/06/89}{23:43}{7704.489}{  0.063}{0.049}{1000}{1.28}{
1329}{1.2CA}
\punkt{27/06/89}{01:38}{7704.568}{  0.032}{0.075}{ 500}{1.84}{
831}{1.2CA}
\punkt{22/07/89}{23:10}{7730.466}{  0.033}{0.009}{ 500}{1.70}{
1108}{1.2CA}
\punkt{24/07/89}{01:34}{7731.565}{  0.043}{0.008}{ 500}{1.65}{
1337}{1.2CA}
\punkt{02/09/89}{21:05}{7772.379}{  0.058}{0.035}{ 500}{1.81}{
748}{1.2CA}
\punkt{03/09/89}{21:19}{7773.389}{  0.058}{0.011}{ 500}{1.32}{
735}{1.2CA}
\punkta{}{}{8033.624}{-0.089}{0.05}{}{}{}{}
\punkt{31/07/90}{23:54}{8104.496}{  0.020}{0.035}{ 500}{1.48}{
958}{1.2CA}
\punkt{25/09/90}{19:46}{8160.324}{ -0.036}{0.032}{ 500}{1.65}{
966}{1.2CA}
\punkt{26/09/90}{19:46}{8161.324}{ -0.037}{0.080}{ 500}{1.64}{
976}{1.2CA}
\punkt{16/02/91}{05:41}{8303.737}{ -0.012}{0.034}{ 300}{1.72}{
408}{1.2CA}
\punkt{24/05/91}{02:30}{8400.605}{ -0.013}{0.033}{ 300}{1.26}{
431}{1.2CA}
\punkt{23/07/91}{21:56}{8461.414}{ -0.033}{0.029}{ 500}{1.22}{
1356}{1.2CA}
\punkt{06/08/91}{23:23}{8475.475}{ -0.056}{0.072}{ 500}{1.46}{
484}{1.2CA}
\punkt{06/08/91}{23:23}{8475.475}{ -0.056}{0.072}{ 500}{1.46}{
484}{1.2CA}
\end{picture}}

\end{picture}

\vspace*{-0.02cm}

\begin{picture}(18 ,2.5 )(0,0)
\put(0,0){\setlength{\unitlength}{0.01059cm}%
\begin{picture}(1700, 236.111)(7350,0)
\put(7350,0){\framebox(1700, 236.111)[tl]{\begin{picture}(0,0)(0,0)
        \put(1700,0){\makebox(0,0)[tr]{\bf{1857+566}\T{0.4}
                                 \hspace*{0.5cm}}}
    \end{picture}}}

\thicklines
\put(7350,0){\setlength{\unitlength}{5cm}\begin{picture}(0,0)(0,-0.25)
   \put(0,0){\setlength{\unitlength}{1cm}\begin{picture}(0,0)(0,0)
        \put(0,0){\line(1,0){0.3}}
        \end{picture}}
   \end{picture}}

\put(9050,0){\setlength{\unitlength}{5cm}\begin{picture}(0,0)(0,-0.25)
   \put(0,0){\setlength{\unitlength}{1cm}\begin{picture}(0,0)(0,0)
        \put(0,0){\line(-1,0){0.3}}
        \end{picture}}
   \end{picture}}

\thinlines
\put(7350,0){\setlength{\unitlength}{5cm}\begin{picture}(0,0)(0,-0.25)
   \multiput(0,0)(0,0.1){3}{\setlength{\unitlength}{1cm}%
\begin{picture}(0,0)(0,0)
        \put(0,0){\line(1,0){0.12}}
        \end{picture}}
   \end{picture}}

\put(7350,0){\setlength{\unitlength}{5cm}\begin{picture}(0,0)(0,-0.25)
   \multiput(0,0)(0,-0.1){3}{\setlength{\unitlength}{1cm}%
\begin{picture}(0,0)(0,0)
        \put(0,0){\line(1,0){0.12}}
        \end{picture}}
   \end{picture}}

\put(9050,0){\setlength{\unitlength}{5cm}\begin{picture}(0,0)(0,-0.25)
   \multiput(0,0)(0,0.1){3}{\setlength{\unitlength}{1cm}%
\begin{picture}(0,0)(0,0)
        \put(0,0){\line(-1,0){0.12}}
        \end{picture}}
   \end{picture}}

\put(9050,0){\setlength{\unitlength}{5cm}\begin{picture}(0,0)(0,-0.25)
   \multiput(0,0)(0,-0.1){3}{\setlength{\unitlength}{1cm}%
\begin{picture}(0,0)(0,0)
        \put(0,0){\line(-1,0){0.12}}
        \end{picture}}
   \end{picture}}

   \put(7527.5, 236.111){\setlength{\unitlength}{1cm}\begin{picture}(0,0)(0,0)
        \put(0,0){\line(0,-1){0.2}}
   \end{picture}}
   \put(7892.5, 236.111){\setlength{\unitlength}{1cm}\begin{picture}(0,0)(0,0)
        \put(0,0){\line(0,-1){0.2}}
   \end{picture}}
   \put(8257.5, 236.111){\setlength{\unitlength}{1cm}\begin{picture}(0,0)(0,0)
        \put(0,0){\line(0,-1){0.2}}
   \end{picture}}
   \put(8622.5, 236.111){\setlength{\unitlength}{1cm}\begin{picture}(0,0)(0,0)
        \put(0,0){\line(0,-1){0.2}}
   \end{picture}}
   \put(8987.5, 236.111){\setlength{\unitlength}{1cm}\begin{picture}(0,0)(0,0)
        \put(0,0){\line(0,-1){0.2}}
   \end{picture}}
    \multiput(7350,0)(50,0){33}%
        {\setlength{\unitlength}{1cm}\begin{picture}(0,0)(0,0)
        \put(0,0){\line(0,1){0.12}}
    \end{picture}}
    \put(7500,0){\setlength{\unitlength}{1cm}\begin{picture}(0,0)(0,0)
        \put(0,0){\line(0,1){0.2}}
    \end{picture}}
    \put(7750,0){\setlength{\unitlength}{1cm}\begin{picture}(0,0)(0,0)
        \put(0,0){\line(0,1){0.2}}
    \end{picture}}
    \put(8000,0){\setlength{\unitlength}{1cm}\begin{picture}(0,0)(0,0)
        \put(0,0){\line(0,1){0.2}}
    \end{picture}}
    \put(8250,0){\setlength{\unitlength}{1cm}\begin{picture}(0,0)(0,0)
        \put(0,0){\line(0,1){0.2}}
    \end{picture}}
    \put(8500,0){\setlength{\unitlength}{1cm}\begin{picture}(0,0)(0,0)
        \put(0,0){\line(0,1){0.2}}
    \end{picture}}
    \put(8750,0){\setlength{\unitlength}{1cm}\begin{picture}(0,0)(0,0)
        \put(0,0){\line(0,1){0.2}}
    \end{picture}}
    \put(9000,0){\setlength{\unitlength}{1cm}\begin{picture}(0,0)(0,0)
        \put(0,0){\line(0,1){0.2}}
    \end{picture}}

\punkt{}{}{7701.582}{-0.109}{0.03}{}{}{}{}
\punkt{20/07/89}{00:19}{7727.514}{ -0.132}{0.017}{ 500}{1.39}{
1992}{1.2CA}
\punkt{22/07/89}{23:26}{7730.477}{  0.002}{0.017}{ 500}{1.93}{
1105}{1.2CA}
\punkt{24/07/89}{02:17}{7731.595}{ -0.125}{0.012}{ 500}{1.76}{
1108}{1.2CA}
\punkt{12/08/89}{10:18}{7750.930}{ -0.167}{0.019}{****}{1.57}{
964}{1.2CA}
\punkt{14/08/89}{23:54}{7753.496}{ -0.113}{0.012}{ 500}{1.69}{
1189}{1.2CA}
\punkt{01/09/89}{00:24}{7770.517}{ -0.095}{0.015}{ 500}{2.84}{
943}{1.2CA}
\punkt{02/11/89}{21:23}{7833.391}{  0.003}{0.037}{ 500}{1.70}{
587}{1.2CA}
\punkt{22/05/90}{01:47}{8033.575}{  0.022}{0.017}{ 250}{1.39}{
1117}{1.2CA}
\punkt{01/08/90}{00:34}{8104.524}{  0.051}{0.013}{ 500}{1.61}{
888}{1.2CA}
\punkt{24/09/90}{21:10}{8159.382}{  0.089}{0.018}{ 500}{2.18}{
1078}{1.2CA}
\punkt{25/09/90}{21:10}{8160.382}{  0.083}{0.017}{ 500}{2.17}{
1081}{1.2CA}
\punkt{}{}{8162.320}{0.091}{0.02}{}{}{}{}
\punkt{30/09/90}{00:29}{8164.520}{  0.129}{0.035}{ 500}{2.81}{
1671}{1.2CA}
\punkt{}{}{8400.614}{0.017}{0.03}{}{}{}{}
\punkta{26/05/91}{02:58}{8402.624}{ -0.007}{0.038}{ 254}{2.05}{
452}{1.2CA}
\punkt{24/07/91}{00:06}{8461.505}{  0.032}{0.022}{ 100}{1.09}{
741}{1.2CA}
\punkt{}{}{8464.532}{-0.036}{0.03}{}{}{}{}
\punkt{29/07/91}{00:42}{8466.529}{ -0.046}{0.036}{ 300}{2.27}{
1073}{1.2CA}
\punkt{03/08/91}{23:44}{8472.489}{ -0.031}{0.010}{ 500}{1.46}{
530}{1.2CA}
\punkt{12/08/91}{23:29}{8481.479}{ -0.029}{0.012}{ 500}{1.52}{
534}{1.2CA}
\punkt{22/09/91}{20:41}{8522.362}{ -0.019}{0.027}{ 500}{1.53}{
1507}{1.2CA}
\punkt{17/10/91}{20:20}{8547.347}{  0.000}{0.032}{ 500}{2.54}{
1074}{1.2CA}
\punkt{17/10/91}{20:20}{8547.347}{  0.000}{0.032}{ 500}{2.54}{
1074}{1.2CA}

\end{picture}}

\end{picture}

\begin{picture}(18 ,3.5 )(0,0)
\put(0,0){\setlength{\unitlength}{0.01059cm}%
\begin{picture}(1700, 330.555)(7350,0)
\put(7350,0){\framebox(1700, 330.555)[tl]{\begin{picture}(0,0)(0,0)
        \put(1700,0){\makebox(0,0)[tr]{\bf{2215-$\!$-037}\T{0.4}
                                 \hspace*{0.5cm}}}
        \put(1700,-
330.555){\setlength{\unitlength}{1cm}\begin{picture}(0,0)(0,0)
            \put(0,-1){\makebox(0,0)[br]{\bf J.D.\,2,440,000\,+}}
        \end{picture}}
    \end{picture}}}

\thicklines
\put(7350,0){\setlength{\unitlength}{5cm}\begin{picture}(0,0)(0,-0.35)
   \put(0,0){\setlength{\unitlength}{1cm}\begin{picture}(0,0)(0,0)
        \put(0,0){\line(1,0){0.3}}
        \end{picture}}
   \end{picture}}

\put(9050,0){\setlength{\unitlength}{5cm}\begin{picture}(0,0)(0,-0.35)
   \put(0,0){\setlength{\unitlength}{1cm}\begin{picture}(0,0)(0,0)
        \put(0,0){\line(-1,0){0.3}}
        \end{picture}}
   \end{picture}}

\thinlines
\put(7350,0){\setlength{\unitlength}{5cm}\begin{picture}(0,0)(0,-0.35)
   \multiput(0,0)(0,0.1){4}{\setlength{\unitlength}{1cm}%
\begin{picture}(0,0)(0,0)
        \put(0,0){\line(1,0){0.12}}
        \end{picture}}
   \end{picture}}

\put(7350,0){\setlength{\unitlength}{5cm}\begin{picture}(0,0)(0,-0.35)
   \multiput(0,0)(0,-0.1){4}{\setlength{\unitlength}{1cm}%
\begin{picture}(0,0)(0,0)
        \put(0,0){\line(1,0){0.12}}
        \end{picture}}
   \end{picture}}

\put(9050,0){\setlength{\unitlength}{5cm}\begin{picture}(0,0)(0,-0.35)
   \multiput(0,0)(0,0.1){4}{\setlength{\unitlength}{1cm}%
\begin{picture}(0,0)(0,0)
        \put(0,0){\line(-1,0){0.12}}
        \end{picture}}
   \end{picture}}

\put(9050,0){\setlength{\unitlength}{5cm}\begin{picture}(0,0)(0,-0.35)
   \multiput(0,0)(0,-0.1){4}{\setlength{\unitlength}{1cm}%
\begin{picture}(0,0)(0,0)
        \put(0,0){\line(-1,0){0.12}}
        \end{picture}}
   \end{picture}}

   \put(7527.5, 330.555){\setlength{\unitlength}{1cm}\begin{picture}(0,0)(0,0)
        \put(0,0){\line(0,-1){0.2}}
   \end{picture}}
   \put(7892.5, 330.555){\setlength{\unitlength}{1cm}\begin{picture}(0,0)(0,0)
        \put(0,0){\line(0,-1){0.2}}
   \end{picture}}
   \put(8257.5, 330.555){\setlength{\unitlength}{1cm}\begin{picture}(0,0)(0,0)
        \put(0,0){\line(0,-1){0.2}}
   \end{picture}}
   \put(8622.5, 330.555){\setlength{\unitlength}{1cm}\begin{picture}(0,0)(0,0)
        \put(0,0){\line(0,-1){0.2}}
   \end{picture}}
   \put(8987.5, 330.555){\setlength{\unitlength}{1cm}\begin{picture}(0,0)(0,0)
        \put(0,0){\line(0,-1){0.2}}
   \end{picture}}
    \multiput(7350,0)(50,0){33}%
        {\setlength{\unitlength}{1cm}\begin{picture}(0,0)(0,0)
        \put(0,0){\line(0,1){0.12}}
    \end{picture}}
    \put(7500,0){\setlength{\unitlength}{1cm}\begin{picture}(0,0)(0,0)
        \put(0,0){\line(0,1){0.2}}
        \put(0,-0.2){\makebox(0,0)[t]{\bf 7500}}
    \end{picture}}
    \put(7750,0){\setlength{\unitlength}{1cm}\begin{picture}(0,0)(0,0)
        \put(0,0){\line(0,1){0.2}}
        \put(0,-0.2){\makebox(0,0)[t]{\bf 7750}}
    \end{picture}}
    \put(8000,0){\setlength{\unitlength}{1cm}\begin{picture}(0,0)(0,0)
        \put(0,0){\line(0,1){0.2}}
        \put(0,-0.2){\makebox(0,0)[t]{\bf 8000}}
    \end{picture}}
    \put(8250,0){\setlength{\unitlength}{1cm}\begin{picture}(0,0)(0,0)
        \put(0,0){\line(0,1){0.2}}
        \put(0,-0.2){\makebox(0,0)[t]{\bf 8250}}
    \end{picture}}
    \put(8500,0){\setlength{\unitlength}{1cm}\begin{picture}(0,0)(0,0)
        \put(0,0){\line(0,1){0.2}}
        \put(0,-0.2){\makebox(0,0)[t]{\bf 8500}}
    \end{picture}}
    \put(8750,0){\setlength{\unitlength}{1cm}\begin{picture}(0,0)(0,0)
        \put(0,0){\line(0,1){0.2}}
        \put(0,-0.2){\makebox(0,0)[t]{\bf 8750}}
    \end{picture}}
    \put(9000,0){\setlength{\unitlength}{1cm}\begin{picture}(0,0)(0,0)
        \put(0,0){\line(0,1){0.2}}
        \put(0,-0.2){\makebox(0,0)[t]{\bf 9000}}
    \end{picture}}

\put(7000,0){\begin{picture}(0,0)(7000,-47.214)
\punkta{31/08/88}{22:37}{7405.443}{  0.074}{0.040}{ 100}{2.37}{
1008}{1.2CA}
\punkt{05/09/88}{23:38}{7410.485}{  0.028}{0.024}{1000}{1.92}{
1005}{1.2CA}
\punkt{04/10/88}{20:52}{7439.370}{ -0.006}{0.033}{ 500}{1.64}{
565}{1.2CA}
\punkt{07/10/88}{20:29}{7442.354}{ -0.002}{0.046}{ 500}{1.58}{
601}{1.2CA}
\punkt{10/10/88}{19:37}{7445.318}{  0.101}{0.035}{ 500}{1.99}{
608}{1.2CA}
\punkt{16/10/88}{21:50}{7451.410}{  0.020}{0.051}{ 500}{2.59}{
632}{1.2CA}
\punkt{11/06/89}{02:33}{7688.607}{ -0.134}{0.022}{ 500}{1.89}{
758}{1.2CA}
\punkt{26/06/89}{01:56}{7703.581}{ -0.120}{0.032}{ 500}{1.58}{
1368}{1.2CA}
\punkt{27/06/89}{02:09}{7704.590}{ -0.096}{0.060}{ 500}{2.00}{
840}{1.2CA}
\punkt{10/08/89}{23:49}{7749.492}{ -0.070}{0.024}{1000}{2.03}{
1008}{1.2CA}
\punkt{01/09/89}{23:55}{7771.497}{  0.038}{0.031}{ 500}{2.62}{
871}{1.2CA}
\punkt{02/09/89}{23:54}{7772.496}{ -0.044}{0.016}{ 538}{2.06}{
745}{1.2CA}
\punkta{}{}{7816.429}{0.006}{0.05}{}{}{}{}
\punkt{31/10/89}{20:22}{7831.349}{ -0.122}{0.016}{ 500}{1.83}{
591}{1.2CA}
\punkt{01/11/89}{19:42}{7832.321}{ -0.093}{0.022}{ 500}{1.94}{
588}{1.2CA}
\punkt{02/11/89}{19:24}{7833.309}{ -0.134}{0.076}{ 500}{2.06}{
594}{1.2CA}
\punkt{15/12/89}{19:08}{7876.298}{ -0.109}{0.026}{ 500}{2.43}{
672}{1.2CA}
\punkt{31/07/90}{00:46}{8103.532}{ -0.189}{0.089}{ 500}{1.79}{
631}{1.2CA}
\punkt{24/09/90}{21:51}{8159.411}{  0.011}{0.043}{ 500}{2.49}{
819}{1.2CA}
\punkt{25/09/90}{21:51}{8160.411}{  0.010}{0.018}{ 500}{2.48}{
823}{1.2CA}
\punkt{16/10/90}{22:07}{8181.422}{  0.093}{0.048}{ 500}{3.16}{
837}{1.2CA}
\punkt{16/10/90}{22:07}{8181.422}{  0.093}{0.002}{ 500}{3.16}{
837}{1.2CA}
\punkt{17/10/90}{22:07}{8182.422}{  0.085}{0.021}{ 500}{3.15}{
850}{1.2CA}
\punkt{18/10/90}{19:10}{8183.299}{  0.116}{0.039}{ 500}{2.60}{
837}{1.2CA}
\punkt{19/10/90}{19:10}{8184.299}{  0.118}{0.018}{ 500}{2.62}{
837}{1.2CA}
\punkt{20/10/90}{22:31}{8185.439}{  0.019}{0.036}{ 100}{2.49}{
551}{1.2CA}
\punkt{21/10/90}{22:00}{8186.417}{  0.081}{0.038}{ 500}{4.17}{
1014}{1.2CA}
\punkt{22/10/90}{22:00}{8187.417}{  0.068}{0.039}{ 500}{4.17}{
1015}{1.2CA}
\punkta{23/05/91}{03:51}{8399.661}{  0.176}{0.092}{ 100}{2.22}{
463}{1.2CA}
\punkt{25/05/91}{03:33}{8401.649}{  0.132}{0.037}{ 300}{1.82}{
522}{1.2CA}
\punkt{23/07/91}{23:33}{8461.481}{  0.095}{0.039}{ 500}{1.79}{
2130}{1.2CA}
\punkt{25/07/91}{00:54}{8462.538}{  0.022}{0.029}{ 500}{1.42}{
2499}{1.2CA}
\punkt{26/07/91}{01:49}{8463.576}{  0.074}{0.066}{ 200}{1.76}{
1880}{1.2CA}
\punkt{01/08/91}{23:42}{8470.488}{  0.013}{0.022}{ 300}{1.30}{
696}{1.2CA}
\punkt{01/08/91}{23:48}{8470.492}{  0.034}{0.014}{ 300}{1.29}{
694}{1.2CA}
\punkt{02/08/91}{23:16}{8471.470}{  0.053}{0.014}{ 500}{1.49}{
779}{1.2CA}
\punkt{04/08/91}{01:58}{8472.582}{ -0.046}{0.011}{ 500}{1.15}{
645}{1.2CA}
\punkt{05/08/91}{02:23}{8473.600}{ -0.037}{0.010}{ 500}{1.18}{
655}{1.2CA}
\punkt{07/08/91}{02:14}{8475.593}{ -0.038}{0.032}{ 500}{1.28}{
490}{1.2CA}
\punkt{12/08/91}{01:35}{8480.567}{ -0.013}{0.012}{ 500}{1.63}{
601}{1.2CA}
\punkt{18/09/91}{23:14}{8518.469}{ -0.164}{0.051}{ 500}{1.83}{
1806}{1.2CA}
\punkta{26/10/91}{21:51}{8556.411}{ -0.123}{0.058}{ 500}{2.68}{
1140}{1.2CA}
\punkt{27/10/91}{22:31}{8557.438}{ -0.189}{0.053}{ 500}{1.32}{
965}{1.2CA}
\punkt{27/10/91}{22:31}{8557.438}{ -0.189}{0.053}{ 500}{1.32}{
965}{1.2CA}

    \put(7350,0){\setlength{\unitlength}{1cm}\begin{picture}(0,0)(0,0)
        \put(0,-1.3){\makebox(0,0)[tl]{\footnotesize{\bf Fig.~1.} (continued)}}
    \end{picture}}

\end{picture}}

\end{picture}}

\end{picture}

\end{figure*}

\subsection{Narrow-absorption-line quasars}

{}From a uniform spectroscopic survey of 55 high redshift QSOs, Sargent,
Boksenberg \& Steidel (\cite{SBS88}, hereafter SBS) obtained 40
Mg\,II-absorption systems. The large scale distribution of the redshifts is
random and therefore consistent with the intervening hypothesis for the
origin of the lines. Very probably, most MgII-absorption systems arise from
intervening galaxies. This assumption is confirmed by the work of Bergeron
(\cite{Ber88a}) who found an image of a galaxy $<\! 10\arcsec$ from the
quasar in 10 out of 14 cases of MgII absorption. All galaxies agree in
redshift with the absorption systems. The large average number of about 0.7
Mg\,II-absorption systems per quasar found by SBS and the result of Bergeron
indicate that in most cases the absorption occurs far away from the centre
of the foreground galaxy, in its extended halo where one would expect a low
density of stars. In our opinion, only MgII systems having large equivalent
widths make it more probable that the quasar light passes regions of higher
star density. We have included in our sample some quasars with
MgII-absorption lines of rest frame equivalent widths $W_{\rm r}>\!1$\AA\
(together for both lines of the doublet); these objects are denoted {\em
A1\/}.

\subsection{Highly luminous quasars}

There is increasing observational evidence for a so-called amplification
bias (AB, sometimes ``magnification bias'' is also used) by gravitational
lensing; i.e.\ an enhancement of foreground galaxies around quasars out of
flux (or luminosity) limited samples. Many authors discuss the AB as a
result of microlensing (for Refs.\ see Borgeest et al.~\cite{BLR91}).
We have included in our sample a number of highly luminous quasars ({\em
HL\/}s, $M_V < -29.0$). Probably, the
{\em HLs} are subject to a strong AB. If microlensing indeed
contributes significantly to the AB in our sample of highly luminous
quasars, strong microlensing variability must be found in the
lightcurves of these quasars.

\begin{table}
\caption[ ]{{\em A priori\/} selection criteria for the sample}
\begin{flushleft}
\begin{tabular}{lp{7.2cm}}
\hline
Class & Definition \T{0.1}\\
\hline
{\em A1}\T{0.1}  & Mg\,II absorption quasars with
            $W_{\rm r} \ga 1\,$\AA , together for both lines of the
            doublet \\
{\em G1}  & (i) Objects having the image of an associated foreground galaxy
             with $z_{\rm gal} \ga 0.05$ closer than 50\,kpc to
             the line of sight, or\\
          & (ii) objects located within 1 optical radii, $\theta_{\rm a}$ (see
            Table~1), of a foreground galaxy \\
{\em G2}  & (i) Objects having the image of an associated foreground galaxy
             with $z_{\rm gal} \ga 0.05$ between 50 and 100\,kpc from
             the line of sight, or\\
          & (ii) objects located between 1 and 2 optical radii from
            a foreground galaxy \\
{\em HL}  & Optically highly luminous quasars, $M_V \leq -28.0$
            (for known variable objects, the maximum flux values
            found in the literature have been taken) \\
{\em Ra}  & Quasars from large-area radio surveys\\
\hline
\end{tabular}
\end{flushleft}
\end{table}

\section{Discussion of the lightcurves}

\subsection{Literature data}

Extensive monitoring programs which include low amplitude variability
quasars have been carried out by several investigators:
\begin{itemize}
\item
At the Rosemary Hill Observatory more than 200, mostly radio-selected
quasars were monitored since 1968, however not all objects
over the total period (Pica et al.~\cite{PPSL80}, hereafter PPSL;
Pica \& Smith \cite{PS83}, hereafter PS83; Smith et al.\ \cite{SNLC93},
hereafter SNLC, and Refs.\ therein).
\item
Lloyd (\cite{Llo84}, hereafter L84) reports on lightcurves of 36 radio sources
from the Herstmonceux Optical Monitoring program for the period
1966-1980 (see also Tritton \& Selmes \cite{TS71}, hereafter TS71;
Selmes et al.\ \cite{STW75}, hereafter STW).
\item
Another program has been carried out at the Asagio Observatory (e.g.\
Barbieri et al.~\cite{BRZ79}, hereafter BRZ) over the period 1967 to 1977.
\item
Monitoring data obtained until 1973 are reviewed and critically discussed
by Grandi \& Tifft (\cite{GT74}, hereafter GT74).
\item
Moore \& Stockman (\cite{MS84}, hereafter MS84) have collected a catalog
of the observational properties of 239 quasars, including variability data.
\item
Lightcurves of many bright quasars have been obtained from the Harvard
historical plate collection (e.g.\ Angione \cite{Ang73}, hereafter A73)
spanning periods of up to 100 years.
\item
Netzer \& Sheffer (\cite{NS83}, hereafter NS83) compared the 1981 and
$\sim$\,1950 POSS magnitudes of 64 optically selected UM quasars. They
found that 39\,\% of the quasars have varied by more than 0.45\,mag.
\end{itemize}
More recent publications on optical variability of quasars generally deal
with violent variable sources.

\subsection{HQM lightcurves}

In this section, we compare the lightcurves
plotted in this paper with those of previous work. The other HQM lightcurves
are discussed in Paper~II.

\noindent
{\bf S5\,0014$+$813}. There are no variability data in the literature. The
only feature in the well sampled HQM lightcurve is a very weak linear
brightening of 0.014$\pm$0.002\,mag\,yr$^{-1}$. To our knowledge, it has
up to now not been possible to determine a long-term trend in an optical
quasar lightcurve with this accuracy.

\noindent
{\bf PKS\,0038$-$019}. There are no variability data in the literature. Our
HQM data are well described by a second order polynomial.

\noindent
{\bf S5\,0153$+$744}. There are no variability data in the literature. The
HQM lightcurve is consistent with a constant flux. A hint on moderate
variability comes only from the POSS photometry. On our best frames, an
enhancement of faint galaxies is obvious in the vicinity of the quasar.

\noindent
{\bf GC\,0248$+$430}. The only optical variability data in the literature
are from our HQM program (see BDHKS). Besides
a steady increase until the beginning of 1990, we reported the possible
detection of short-term variability. A reevaluation of the data has however
shown that there is probably no short-term activity: The photometric points
which deviate from the main trend were all obtained during poor atmospheric
conditions. The more recent data show that the quasar is fading again. BDHKS
have collected literature data on radio measurements
showing that the object is much more active in the radio wavelength range.

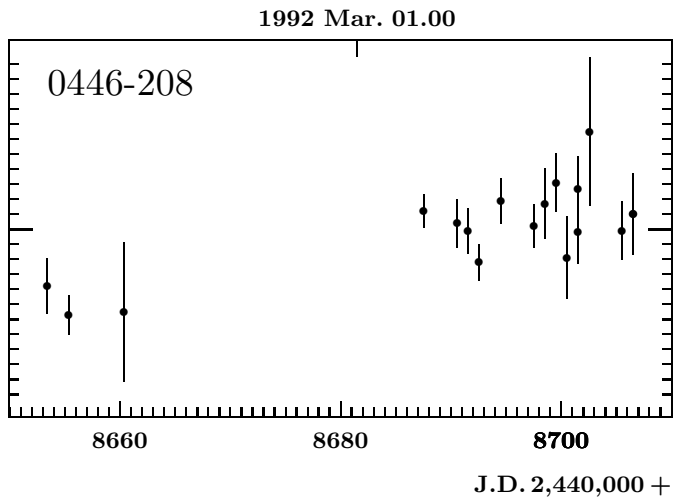
\begin{figure}

\vspace*{0.4cm}

\begin{picture}(8.8 ,5 )(0,0)
\put(0,0){\setlength{\unitlength}{0.1466cm}%
\begin{picture}(60,  34.106)(650,0)
\put(650,0){\framebox(60,  34.106)[tl]{\begin{picture}(0,0)(0,0)
        \put(0,0){\makebox(0,0)[tl]{\hspace*{0.5cm}\large{0446-208}\T{0.4}
                                 }}
        \put(60,-
34.106){\setlength{\unitlength}{1cm}\begin{picture}(0,0)(0,0)
            \put(0,-1){\makebox(0,0)[br]{\bf J.D.\,2,440,000\,+}}
        \end{picture}}
    \end{picture}}}

\thicklines
\put(650,0){\setlength{\unitlength}{20cm}\begin{picture}(0,0)(0,-0.125)
   \put(0,0){\setlength{\unitlength}{1cm}\begin{picture}(0,0)(0,0)
        \put(0,0){\line(1,0){0.3}}
        \end{picture}}
   \end{picture}}

\put(710,0){\setlength{\unitlength}{20cm}\begin{picture}(0,0)(0,-0.125)
   \put(0,0){\setlength{\unitlength}{1cm}\begin{picture}(0,0)(0,0)
        \put(0,0){\line(-1,0){0.3}}
        \end{picture}}
   \end{picture}}

\thinlines
\put(650,0){\setlength{\unitlength}{20cm}\begin{picture}(0,0)(0,-0.125)
   \multiput(0,0)(0,0.01){12}{\setlength{\unitlength}{1cm}%
\begin{picture}(0,0)(0,0)
        \put(0,0){\line(1,0){0.12}}
        \end{picture}}
   \end{picture}}

\put(650,0){\setlength{\unitlength}{20cm}\begin{picture}(0,0)(0,-0.125)
   \multiput(0,0)(0,-0.01){12}{\setlength{\unitlength}{1cm}%
\begin{picture}(0,0)(0,0)
        \put(0,0){\line(1,0){0.12}}
        \end{picture}}
   \end{picture}}

\put(710,0){\setlength{\unitlength}{20cm}\begin{picture}(0,0)(0,-0.125)
   \multiput(0,0)(0,0.01){12}{\setlength{\unitlength}{1cm}%
\begin{picture}(0,0)(0,0)
        \put(0,0){\line(-1,0){0.12}}
        \end{picture}}
   \end{picture}}

\put(710,0){\setlength{\unitlength}{20cm}\begin{picture}(0,0)(0,-0.125)
   \multiput(0,0)(0,-0.01){12}{\setlength{\unitlength}{1cm}%
\begin{picture}(0,0)(0,0)
        \put(0,0){\line(-1,0){0.12}}
        \end{picture}}
   \end{picture}}

   \put(681.5,  34.106){\setlength{\unitlength}{1cm}\begin{picture}(0,0)(0,0)
        \put(0,0){\line(0,-1){0.2}}
        \put(0,0.2){\makebox(0,0)[b]{\bf 1992 Mar.~01.00}}
   \end{picture}}

    \multiput(650,0)(1,0){60}%
        {\setlength{\unitlength}{1cm}\begin{picture}(0,0)(0,0)
        \put(0,0){\line(0,1){0.12}}
    \end{picture}}
    \put(660,0){\setlength{\unitlength}{1cm}\begin{picture}(0,0)(0,0)
        \put(0,0){\line(0,1){0.2}}
        \put(0,-0.2){\makebox(0,0)[t]{\bf 8660}}
    \end{picture}}
    \put(680,0){\setlength{\unitlength}{1cm}\begin{picture}(0,0)(0,0)
        \put(0,0){\line(0,1){0.2}}
        \put(0,-0.2){\makebox(0,0)[t]{\bf 8680}}
    \end{picture}}
   \put(700,0){\setlength{\unitlength}{1cm}\begin{picture}(0,0)(0,0)
        \put(0,0){\line(0,1){0.2}}
        \put(0,-0.2){\makebox(0,0)[t]{\bf 8700}}
    \end{picture}}
   \put(700,0){\setlength{\unitlength}{1cm}\begin{picture}(0,0)(0,0)
        \put(0,0){\line(0,1){0.2}}
        \put(0,-0.2){\makebox(0,0)[t]{\bf 8700}}
    \end{picture}}
    \put(700,0){\setlength{\unitlength}{1cm}\begin{picture}(0,0)(0,0)
        \put(0,0){\line(0,1){0.2}}
        \put(0,-0.2){\makebox(0,0)[t]{\bf 8700}}
    \end{picture}}
    \put(700,0){\setlength{\unitlength}{1cm}\begin{picture}(0,0)(0,0)
       \put(0,0){\line(0,1){0.2}}
        \put(0,-0.2){\makebox(0,0)[t]{\bf 8700}}
    \end{picture}}
    \put(700,0){\setlength{\unitlength}{1cm}\begin{picture}(0,0)(0,0)
       \put(0,0){\line(0,1){0.2}}
        \put(0,-0.2){\makebox(0,0)[t]{\bf 8700}}
    \end{picture}}

\punktb{31/01/92}{21:27}{653.394}{ -0.038}{0.018}{ 500}{2.16}{
612}{1.2CA}
\punktb{02/02/92}{20:01}{655.334}{ -0.057}{0.013}{ 500}{1.87}{
634}{1.2CA}
\punktb{07/02/92}{20:34}{660.357}{ -0.055}{0.046}{ 500}{1.58}{
600}{1.2CA}
\punktb{06/03/92}{00:49}{687.534}{  0.012}{0.011}{ 600}{1.26}{
454}{1.2CA}
\punktb{09/03/92}{01:10}{690.549}{  0.004}{0.016}{ 300}{1.75}{
314}{1.2CA}
\punktb{10/03/92}{01:14}{691.551}{ -0.001}{0.015}{ 300}{1.65}{
363}{1.2CA}
\punktb{11/03/92}{00:48}{692.534}{ -0.022}{0.012}{ 300}{1.49}{
390}{1.2CA}
\punktb{13/03/92}{00:59}{694.542}{  0.019}{0.015}{ 500}{1.41}{
1207}{1.2CA}
\punktb{16/03/92}{00:28}{697.520}{  0.002}{0.014}{ 300}{1.54}{
569}{1.2CA}
\punktb{17/03/92}{00:46}{698.533}{  0.017}{0.023}{ 300}{1.50}{
614}{1.2CA}
\punktb{18/03/92}{00:51}{699.536}{  0.031}{0.019}{ 300}{1.96}{
694}{1.2CA}
\punktb{19/03/92}{00:15}{700.511}{ -0.019}{0.027}{ 500}{1.87}{
1817}{1.2CA}
\punktb{20/03/92}{00:06}{701.505}{  0.027}{0.021}{ 200}{1.49}{
469}{1.2CA}
\punktb{20/03/92}{00:12}{701.509}{ -0.002}{0.021}{ 300}{1.66}{
543}{1.2CA}
\punktb{21/03/92}{02:00}{702.583}{  0.065}{0.049}{ 200}{2.50}{
539}{1.2CA}
\punktb{24/03/92}{00:35}{705.524}{ -0.001}{0.019}{ 300}{1.82}{
305}{1.2CA}
\punktb{25/03/92}{00:34}{706.524}{  0.010}{0.027}{ 300}{1.68}{
307}{1.2CA}
\punktb{25/03/92}{00:34}{706.524}{  0.010}{0.027}{ 300}{1.68}{
307}{1.2CA}

\end{picture}}

\end{picture}

\vspace*{1cm}
\caption{Enlarged presentation of the winter '92 lightcurve
         of the quasar MC\,1 (0446-208)
         in Johnson-$R\/$. Here, dashes on the vertical
         axes indicate 0.01\,mag\,-\,steps. The only significant feature
         is an overall brightening}
\end{figure}

\noindent
{\bf 0446$-$208 (MC\,1)}. There are no variability data in the literature.
The HQM lightcurve is sufficiently sampled to show the structure of a
weak variability which is well described by a third order polynomial.
The winter 1992 lightcurve is plotted with higher resolution in Fig.~2.

\noindent
{\bf PKS\,0454$+$039}. There are no variability data in the literature.
The HQM data show that the object is brightening by
$\sim$\,0.06\,mag\,yr$^{-1}$. Our POSS photometry yields also some
evidence for moderate variability.

\noindent
{\bf Ton\,469 (0955$+$326)}. L84 recorded a smooth lightcurve between 1967
and 1979, showing a broad minimum and a total amplitude
$\Delta B\simeq$0.5\,mag. Xie et al.\ (\cite{XLZL88}) searched for
variations during one single night, Jan.~1, 1987, with a negative result.
In the UV, however, a change in flux by a factor of 2 was recorded during
one day (Bruhweiler et al.\ \cite{BKS86}).
The historical lightcurve measured by A73 has large scatter
with $\sigma_0=0.27$\,mag. The only significant feature in the HQM lightcurve
is a linear brightening by $\sim$\,0.04\,mag\,yr$^{-1}$.

\begin{figure}

\vspace*{0.4cm}

\begin{picture}(8.8 ,7.5 )(0,0)
\put(0,0){\setlength{\unitlength}{0.1466cm}%
\begin{picture}(60,  51.160)(630,0)
\put(630,0){\framebox(60,  51.160)[tl]{\begin{picture}(0,0)(0,0)
        \put(0,0){\makebox(0,0)[tl]{\hspace*{0.5cm}\large{1150+497}\T{0.4}
                                 }}
        \put(0,-20){\makebox(0,0)[tl]{\hspace*{0.5cm}\bf{1209+107}\T{0.4}
                                 }}
        \put(60,-
51.160){\setlength{\unitlength}{1cm}\begin{picture}(0,0)(0,0)
            \put(0,-1){\makebox(0,0)[br]{\bf J.D.\,2,440,000\,+}}
        \end{picture}}
    \end{picture}}}

\thicklines
\put(630,0){\setlength{\unitlength}{5cm}\begin{picture}(0,0)(0,-0.75)
   \put(0,0){\setlength{\unitlength}{1cm}\begin{picture}(0,0)(0,0)
        \put(0,0){\line(1,0){0.3}}
        \end{picture}}
   \end{picture}}

\put(690,0){\setlength{\unitlength}{5cm}\begin{picture}(0,0)(0,-0.75)
   \put(0,0){\setlength{\unitlength}{1cm}\begin{picture}(0,0)(0,0)
        \put(0,0){\line(-1,0){0.3}}
        \end{picture}}
   \end{picture}}

\thinlines
\put(630,0){\setlength{\unitlength}{5cm}\begin{picture}(0,0)(0,-0.75)
   \multiput(0,0)(0,0.1){8}{\setlength{\unitlength}{1cm}%
\begin{picture}(0,0)(0,0)
        \put(0,0){\line(1,0){0.12}}
        \end{picture}}
   \end{picture}}

\put(630,0){\setlength{\unitlength}{5cm}\begin{picture}(0,0)(0,-0.75)
   \multiput(0,0)(0,-0.1){8}{\setlength{\unitlength}{1cm}%
\begin{picture}(0,0)(0,0)
        \put(0,0){\line(1,0){0.12}}
        \end{picture}}
   \end{picture}}

\put(690,0){\setlength{\unitlength}{5cm}\begin{picture}(0,0)(0,-0.75)
   \multiput(0,0)(0,0.1){8}{\setlength{\unitlength}{1cm}%
\begin{picture}(0,0)(0,0)
        \put(0,0){\line(-1,0){0.12}}
        \end{picture}}
   \end{picture}}

\put(690,0){\setlength{\unitlength}{5cm}\begin{picture}(0,0)(0,-0.75)
   \multiput(0,0)(0,-0.1){8}{\setlength{\unitlength}{1cm}%
\begin{picture}(0,0)(0,0)
        \put(0,0){\line(-1,0){0.12}}
        \end{picture}}
   \end{picture}}

   \put(668.5,  51.160){\setlength{\unitlength}{1cm}\begin{picture}(0,0)(0,0)
        \put(0,0){\line(0,-1){0.2}}
        \put(0,0.2){\makebox(0,0)[b]{\bf 1992 Feb.~16.00}}
   \end{picture}}

    \multiput(630,0)(1,0){60}%
        {\setlength{\unitlength}{1cm}\begin{picture}(0,0)(0,0)
        \put(0,0){\line(0,1){0.12}}
    \end{picture}}
    \put(660,0){\setlength{\unitlength}{1cm}\begin{picture}(0,0)(0,0)
        \put(0,0){\line(0,1){0.2}}
        \put(0,-0.2){\makebox(0,0)[t]{\bf 8660}}
    \end{picture}}
    \put(680,0){\setlength{\unitlength}{1cm}\begin{picture}(0,0)(0,0)
        \put(0,0){\line(0,1){0.2}}
        \put(0,-0.2){\makebox(0,0)[t]{\bf 8680}}
    \end{picture}}
   \put(640,0){\setlength{\unitlength}{1cm}\begin{picture}(0,0)(0,0)
        \put(0,0){\line(0,1){0.2}}
        \put(0,-0.2){\makebox(0,0)[t]{\bf 8640}}
    \end{picture}}

\punktd{18/01/92}{21:35}{640.400}{ -0.577}{0.015}{ 500}{1.67}{
6324}{1.2CA}
\punktd{12/02/92}{04:46}{664.699}{ -0.443}{0.009}{ 500}{1.14}{
474}{1.2CA}
\punktd{15/02/92}{02:55}{667.622}{  0.244}{0.008}{ 500}{1.26}{
1180}{1.2CA}
\punktd{15/02/92}{03:34}{667.649}{  0.287}{0.013}{ 500}{1.34}{
881}{1.2CA}
\punktd{15/02/92}{04:56}{667.706}{  0.293}{0.013}{ 500}{1.79}{
507}{1.2CA}
\punktd{15/02/92}{23:50}{668.493}{  0.534}{0.035}{ 500}{1.82}{
1849}{1.2CA}
\punktd{16/02/92}{00:38}{668.527}{  0.469}{0.014}{ 500}{1.03}{
1808}{1.2CA}
\punktd{28/02/92}{02:54}{680.621}{ -0.334}{0.026}{ 500}{3.02}{
524}{1.2CA}
\punktc{02/02/92}{04:48}{654.700}{ -0.046}{0.018}{ 500}{1.91}{
525}{1.2CA}
\punktc{03/02/92}{04:30}{655.688}{ -0.075}{0.046}{ 500}{1.70}{
513}{1.2CA}
\punktc{07/02/92}{03:47}{659.658}{ -0.028}{0.023}{ 500}{2.30}{
485}{1.2CA}
\punktc{09/02/92}{04:17}{661.679}{ -0.058}{0.011}{ 500}{1.04}{
474}{1.2CA}
\punktc{14/02/92}{01:59}{666.583}{  0.005}{0.034}{ 500}{2.39}{
803}{1.2CA}
\punktc{15/02/92}{03:17}{667.637}{ -0.073}{0.013}{ 500}{1.41}{
947}{1.2CA}
\punktc{16/02/92}{03:13}{668.634}{ -0.058}{0.024}{ 500}{1.46}{
1646}{1.2CA}

\end{picture}}

\end{picture}

\vspace*{1cm}
\caption{$R\/$-lightcurve
         of LB\,2126 (1150+497) in winter '92. For comparison, data for
         Q\,1209+107 (error bars only) are also shown; the latter object
         was observed between the exposures of LB\,2126 on
         Feb.~15 (40\,min difference), it shows no indication of variability}
\end{figure}
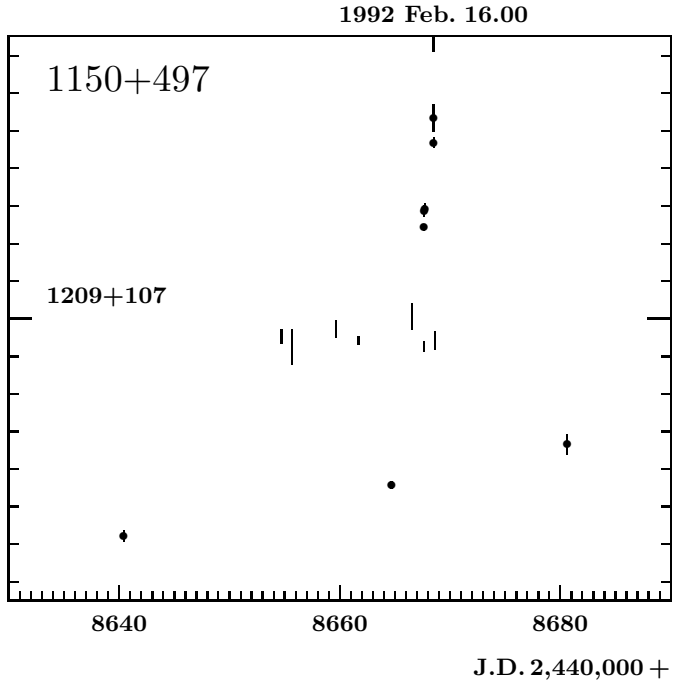

\noindent
{\bf LB\,2136 (1150+497)}. BRZ found, between 1967 and 1976, a total range of
variability $\Delta B\simeq1.4$\,mag. A 0.25\,mag fading was recorded by TS71
between 1967 and 1968. Radio variability was detected by Moore et
al.\ (\cite{MBDN81}). Besides 0836+710 (see v.~Linde et al.\ 1993)
LB\,2126 is the only object for which violent variability was detected
during the HQM program {\em and\/} which was not known to be an OVV before. A
rapid flare was recorded on Feb.~15/16, 1992 (Fig.~3). Interestingly, a flare
in 0836+710 occurred nearly simultaneously (Feb.~16/17, 1992). One may
therefore suspect that a systematic error had ``produced'' the flares. We
looked very carefully for possible sources of such an error, without any
result. In Fig.~3, we have also plotted the lightcurve for 1209+107. One
measurement for this object was made between two exposures of
LB~2126 on Feb.~15; there is absolutely no indication for an error above
our estimate. With the knowledge that flares are occurring in the
lightcurve of LB~2126, one may assume in the data of BRZ another flare,
in early summer 1972. The radio spectral index of LB\,2126 in the catalogue
of VV is $\alpha=0.57$, close to the value which separates steep spectra
from flat ones. In addition, the object is a core-jet VLBI source. One
may therefore assume that LB\,2126 belongs to the blazar class. This
hypothesis can only be tested by further observations.

\noindent
{\bf Mkn\,205 (1219$+$755)}. Zamorani et al.\ (\cite{ZGMT84}) detected
20\% X-ray variability on a timescale of about one day. Our optical data
show significant variations, too, however on a longer timescale. The
lightcurve can be fitted by a fourth order polynomial or a sine function of
a period not much longer than the total timespan of observations. POSS
photometry also indicates variability.

\begin{table*}
\caption[ ]{Results of HQM and POSS photometry. In column 2, those objects
are marked which were previously known to be variable; in column 3, we give
the {\em a priori} selection criteria through which the corresponding
object entered our sample (cf. Table~3). For definitions of the quantities
listed in the following columns, see text}
\scriptsize
\begin{flushleft}
\begin{tabular}{l|cl|rrcrcc|cccccc|c|l}
\hline
\hline
&&&&&&&&&&&&& \\
Object     & $\!\!\!\!\!\!$Var$\!\!\!\!\!\!$
           & Class
           & \multicolumn{1}{c}{$n'$}
           & \multicolumn{1}{c}{$n$}
           & \multicolumn{1}{c}{$\Delta t$}
           & \multicolumn{1}{c}{$\hat{n}$}
           & \multicolumn{1}{c}{$\bar{\sigma\T{0}}$}
           & \multicolumn{1}{c|}{$\tilde{\sigma\T{0}}$}
           & \multicolumn{1}{c}{$\sigma_0$}
           & \multicolumn{1}{c}{$\sigma_1$}
           & \multicolumn{1}{c}{$\sigma_2$}
           & \multicolumn{1}{c}{$\sigma_3$}
           & \multicolumn{1}{c}{$\sigma_4$}
           & \multicolumn{1}{c|}{$\sigma_5$}
           & \multicolumn{1}{c|}{$R'_{\rm obs}$\U{0.1}}
           & \multicolumn{1}{c}{$\Delta R_{40}$} \\
           &
           &
           &
           &
           & \multicolumn{1}{c}{[yrs]}
           &
           & \multicolumn{2}{c|}{[mag]}
           & \multicolumn{6}{c|}{[mag]}
           & \multicolumn{1}{c|}{[mag\,yr$^{-1}$]}
           & \multicolumn{1}{c}{[mag]}\U{0.1}      \\
\hline
\hline
0003+158   \T{0.1}
           & $\times$           & {\em Ra}
           & 17 & 16 & 4.04  & 4.6 & .025 & .026
           & .077 & .066 & .064 & .014 & --- & ---
           & $-$0.102$\pm$.006
           & $+$0.72$\pm$.12 \\
0007$-$000
           & $\times$           & {\em G2\/}
           & 5 & 4 & 4.05  & 3.5 & .013 & .014
           & .146 & .007 & --- & --- & --- & ---
           & $-$0.088$\pm$.002
           & $-$0.01$\pm$.30$^{\star ,\, \triangleleft}$ \\
0013$-$004
           & $\times$ & {\em A1}
           & 10 & 10 & 3.10  & 4.4 & .028 & .031
           & .054 & .019 & .019 & .016 & --- & ---
           & $-$0.075$\pm$.012
           & $+$0.09$\pm$.23$^{\triangleleft}$ \\
0014$+$813
           &            & {\em A1}
           & 63 & 54 & 4.06  & 10.8 & .022 & .023
           & .020 & .015 & .015 & .015 & .014 & .014
           & +0.014$\pm$.002
           & $-$0.08$\pm$.21 \\
0038$-$019
           &            & {\em G2}
           & 30 & 23 & 4.05  & 7.3 & .025 & .027
           & .045 & .040 & .017 & --- & --- & ---
           & $+$0.063$\pm$.008
           & $+$0.00$\pm$.31 \U{0.1}\\
\hline
0058$+$019\T{0.1}
           & $\times$           & {\em A1}
           & 15 & 14 & 4.04  & 5.6 & .022 & .024
           & .053 & .048 & .033 & .033 & .017 & ---
           & $+$0.123$\pm$.034
           & $+$0.04$\pm$.41$^{\star}$ \\
0104$+$318
           & $\times$           & {\em G1}
           & 12 & 8 & 3.25  & 5.3 & .063 & .072
           & .091 & .087 & --- & --- & --- & ---
           & $-$0.272$\pm$.050
           & $-$0.57$\pm$.23$^{\triangleleft}$ \\
0151$+$045
           & $\times$           & {\em G1}
           & 39 & 25 & 4.05  & 6.8 & .020 & .022
           & .138 & .096 & .094 & .045 & .031 & ---
           & $+$0.227$\pm$.021
           & $+$0.35$\pm$.36 \\
0153$+$744
           &            & {\em HL}
           & 24 & 24 & 4.05  & 8.0 & .052 & .056
           & .035 & .035 & .033 & .033 & .033 & .029
           & $-$0.001$\pm$.006
           & $-$0.65$\pm$.20 \\
0248$+$430
           &    & {\em G1\/}
           & 160 & 79 & 3.97  & 13.4 & .027 & .029
           & .046 & .043 & .027 & .025 & .023 & .023
           & $+$0.084$\pm$.006
           & $-$0.10$\pm$.18 \U{0.1}\\
\hline
0446$-$208\T{0.1}
           &            & {\em G1}
           & 77 & 56 & 3.96  & 9.2 & .026 & .028
           & .042 & .040 & .030 & .026 & .026 & .023
           & $+$0.134$\pm$.017
           & $+$0.23$\pm$.20 \\
0454$+$039
           &            & {\em A1}
           & 36 & 31 & 2.98  & 7.6 & .027 & .029
           & .060 & .027 & .026 & .026 & .026 & .022
           & $+$0.057$\pm$.005
           & $-$0.23$\pm$.10 \\
0731$+$653
           &            & {\em A1}
           & 26 & 16 & 2.29  & 5.0 & .024 & .026
           & .088 & .060 & .035 & .034 & --- & ---
           & $+$0.185$\pm$.021
           & $-$0.37$\pm$.18 \\
0745$+$557
           &            & {\em G2}
           & 15 & 13 & 2.27  & 4.3 & .034 & .040
           & .105 & .045 & .038 & --- & --- & ---
           & $-$0.469$\pm$.044
           & $-$0.12$\pm$.26 \\
0805$+$046
           & $\times$           & {\em A1}
           & 7 & 6 & 2.39  & 3.8 & .042 & .046
           & .103 & .057 & --- & --- & --- & ---
           & $+$0.181$\pm$.044
           & $-$0.09$\pm$.23 \U{0.1}\\
\hline
0809$+$483\T{0.1}
           & $\times$           & {\em G1}
           & 23 & 20 & 2.29  & 5.2 & .046 & .054
           & .079 & .036 & .019 & .018 & .017 & ---
           & $+$0.147$\pm$.013
           & $-$0.32$\pm$.44$^{\star}$ \\
0903$+$175
           &               & {\em G1}
           & 28 & 15 & 2.39  & 5.4 & .040 & .045
           & .034 & .033 & .028 & .023 & .023 & ---
           & $-$0.006$\pm$.011
           & $+$0.08$\pm$.20 \\
0955$+$326
           & $\times$           & {\em G2}
           & 45 & 42 & 2.29  & 6.8 & .018 & .020
           & .028 & .013 & .013 & .013 & .013 & .013
           & $+$0.042$\pm$.003
           & $-$0.07$\pm$.40$^{\star}$ \\
1011$+$250
           & $\times$           & {\em HL}
           & 19 & 18 & 2.35  & 5.6 & .023 & .026
           & .022 & .020 & .020 & .019 & .014 & ---
           & $+$0.011$\pm$.005
           & $-$0.10$\pm$.51$^{\star ,\, \triangleleft}$\\
1109+357
           &                & {\em G1}
           & 14 & 6 & 2.16  & 4.1 & .067 & .079
           & .049 & .031 & --- & --- & --- & ---
           & $-$0.046$\pm$.018
           & $-$0.10$\pm$.32$^{\triangleleft}$\U{0.1}\\
\hline
1150$+$497\T{0.1}
           & $\times$           & {\em G2}
           & 19 & 19 & 2.71  & 6.0 & .022 & .028
           & \multicolumn{6}{l|}{OVV}
           & $+$91$\pm$11
           & $+$0.17$\pm$.23\\
1209$+$107
           &                   & {\em G1}
           & 18 & 10 & 2.02  & 4.3 & .028 & .032
           & .081 & .022 & --- & --- & --- & ---
           & $-$0.093$\pm$.014
           & $+$0.01$\pm$.22\\
1219$+$755
           &                   & {\em G1}
           & 46 & 33 & 3.45  & 9.4 & .033 & .036
           & .071 & .071 & .071 & .065 & .035 & .023
           & $-$0.579$\pm$.050
           & $-$0.76$\pm$.30\\
1222$+$228
           &                   & {\em HL}
           & 12 & 10 & 1.92  & 4.6 & .019 & .020
           & .019 & .017 & .009 & --- & --- & ---
           & $-$0.052$\pm$.008
           & $+$0.28$\pm$.24\\
1332$+$552
           & $\times$           & {\em G2}
           & 9 & 6 & 1.77  & 4.3 & .046 & .058
           & .023 & .020 & .018 & --- & --- & ---
           & $-$0.018$\pm$.017
           & $-$0.74$\pm$.15\U{0.1}\\
\hline
1421$+$330\T{0.1}
           &                    & {\em HL}
           & 11 & 11 & 2.68  & 4.8 & .009 & .009
           & .023 & .021 & .011 & --- & --- & ---
           & $-$0.034$\pm$.009
           & $+$0.44$\pm$.46$^{\star}$\\
1435$+$638
           &                    & {\em HL}
           & 14 & 13 & 3.45  & 3.9 & .016 & .019
           & .053 & .018 & .013 & --- & --- & ---
           & $+$0.038$\pm$.004
           & $+$1.57$\pm$.20$^{\triangleleft}$\\
1520$+$413
           &                    & {\em HL}
           & 13 & 11 & 2.20  & 4.5 & .033 & .034
           & .054 & .049 & .044 & --- & --- & ---
           & $-$0.028$\pm$.019
           & $-$0.17$\pm$.45\\
1522$+$101
           &                    & {\em G1\/}
           & 29 & 27 & 2.73  & 7.1 & .021 & .022
           & .047 & .046 & .018 & .018 & .017 & .017
           & $+$0.130$\pm$.026
           & $+$0.17$\pm$.39\\
1604$+$290
           &                    & {\em HL}
           & 27 & 16 & 2.68  & 5.8 & .041 & .048
           & .102 & .071 & .042 & .042 & --- & ---
           & $-$0.223$\pm$.063
           & $-$0.29$\pm$.29\U{0.1}\\
\hline
1630$+$377\T{0.1}
           &                    & {\em HL}
           & 24 & 23 & 2.38  & 6.6 & .021 & .023
           & .030 & .020 & .019 & .018 & .018 & ---
           & $+$0.028$\pm$.005
           & $+$0.03$\pm$.22\\
1633$+$267
           &                    & {\em HL}
           & 13 & 11 & 2.16  & 4.9 & .039 & .046
           & .065 & .035 & .022 & --- & --- & ---
           & $-$0.070$\pm$.015
           & $-$0.39$\pm$.15\\
1634$+$706
           &                       & {\em A1}
           & 17 & 17 & 2.16  & 6.1 & .013 & .014
           & .022 & .019 & .019 & .019 & .018 & ---
           & $+$0.014$\pm$.006
           & $+$0.35$\pm$.20\\
1640$+$396
           &                    & {\em Ra}
           & 37 & 22 & 2.61  & 5.4 & .034 & .041
           & .126 & .126 & .122 & .071 & --- & ---
           & $+$0.626$\pm$.118
           & $+$0.02$\pm$.23\\
1700$+$642
           &                     & {\em G1}
           & 79 & 42 & 3.45  & 14.5 & .022 & .024
           & .049 & .030 & .027 & .026 & .023 & .022
           & $+$0.100$\pm$.013
           & $+$0.09$\pm$.17\U{0.1}\\
\hline
1701$+$610\T{0.1}
           &                    & {\em G2}
           & 22 & 12 & 2.38  & 4.4 & .043 & .046
           & .084 & .066 & .064 & --- & --- & ---
           & $-$0.054$\pm$.021
           & $-$0.91$\pm$.37$^{\star}$\\
1704$+$608
           & $\times$           & {\em Ra}
           & 51 & 45 & 2.18  & 6.2 & .018 & .019
           & .071 & .026 & .025 & .012 & --- & ---
           & $+$0.237$\pm$.012
           & \multicolumn{1}{c}{---} \\
1715$+$535
           &                    & {\em HL}
           & 38 & 11 & 2.92  & 9.1 & .030 & .034
           & .038 & .029 & .023 & .015 & .015 & ---
           & $+$0.103$\pm$.013
           & $-$0.13$\pm$.24\\
1718$+$481
           &                    & {\em HL}
           & 21 & 21 & 2.94  & 4.6 & .018 & .019
           & .012 & .011 & .009 & --- & --- & ---
           & $-$0.004$\pm$.003
           & $+$0.01$\pm$.25\\
1821$+$643
           &                     &
           & 15 & 14 & 2.09  & 5.4 & .014 & .015
           & .032 & .026 & .010 & .005 & --- & ---
           & $-$0.169$\pm$.021
           & $-$0.70$\pm$.28\U{0.1}\\
\hline
1857$+$566\T{0.1}
           &                    & {\em A1}
           & 36 & 31 & 2.32  & 7.2 & .025 & .028
           & .081 & .074 & .038 & .037 & .032 & .031
           & $+$0.170$\pm$.016
           & $+$0.70$\pm$.19\\
2126$-$158
           &                 & {\em A1}
           & 14 & 11 & 2.92  & 4.6 & .022 & .024
           & .021 & .016 & .016 & --- & --- & ---
           & $+$0.012$\pm$.005
           & $-$0.20$\pm$.23\\
2134$+$004
           & $\times$           & {\em HL}
           & 21 & 18 & 2.93  & 7.0 & .019 & .020
           & .032 & .027 & .027 & .027 & .024 & .021
           & $-$0.015$\pm$.006
           & $-$0.08$\pm$.34\\
2215$-$037
           &                    & {\em G1}
           & 67 & 43 & 3.15  & 8.8 & .033 & .037
           & .089 & .088 & .088 & .064 & .057 & .057
           & $-$0.775$\pm$.113
           & $+$0.21$\pm$.25\\
2251$+$244
           &                      & {\em G1}
           & 44 & 21 & 1.97  & 6.3 & .032 & .042
           & .050 & .026 & .017 & --- & --- & ---
           & $-$0.100$\pm$.010
           & $-$0.22$\pm$.37\U{0.1}\\
\hline
2308$+$098\T{0.1}
           &                    & {\em G2}
           & 11 & 10 & 2.84  & 4.0 & .014 & .016
           & .056 & .019 & .016 & --- & --- & ---
           & $+$0.041$\pm$.005
           & $+$0.30$\pm$.24\\
2354$+$144
           & $\times$           & {\em G1}
           & 17 & 13 & 2.28  & 5.2 & .028 & .033
           & .105 & .099 & .028 & .027 & --- & ---
           & $-$0.248$\pm$.016
           & $+$1.08$\pm$.32$^{\triangleleft}$\U{0.1}\\
\hline
\multicolumn{17}{p{15.5cm}}{$^{\star}$\  There are only less than five
                            stars in the CCD frames which are useful
                            for POSS photometry\T{0.2}} \\
\multicolumn{17}{p{15.5cm}}{$^{\triangleleft}$\  All useful stars are
                            brighter than the quasar\U{0.2}} \\
\hline
\hline
\end{tabular}
\end{flushleft}
\end{table*}

\noindent
{\bf PG\,1522$+$101}. There are no variability data in the literature. The
HQM data clearly indicate variations, well fitted by a second order
polynomial. Very close to the quasar lies a probably interacting pair of
galaxies which we detected in May 1989 on a deep frame taken with the 2.2m
telescope; after subtraction of the pointspread function on an exposure
obtained at ESO, Magain et al.\ (\cite{MRSS90}) found a third galaxy just
on the top of the quasar image.

\noindent
{\bf HS\,1700$+$642}. There are no variability data in the literature for
this very luminous object. Due to the good sampling some details in the
weak variability are seen in the HQM lightcurve.

\noindent
{\bf PG\,1715$+$535}. There are no variability data in the literature. Our
lightcurve is consistent with a constant flux, although the 1989/90 data
seem to indicate a slight brightening. The photometric errors are
relatively large due to an unresolved object 6\arcsec\ from the quasar.
Due to a ``quick-look'' spectrum obtained at the MPIA 3.5m telescope,
kindly provided by H.~Hagen, this object is identified as a star .

\noindent
{\bf 4C\,56.28 (1857+566)}. There are no variability data in the
literature. The main trend in the HQM lightcurve is well fitted by a second
order polynomial. A few isolated points show significant deviations from
the best fit; it is not clear whether this is a real effect. POSS
photometry indicates variations on a very long time scale, too.

\noindent
{\bf 2215$-$037}. Pica et al.\ (\cite{PWSL87}) recorded a long-term
steady increase of $\Delta B \simeq 1$\,mag over 17 years and also
found some evidence for short-term variability. The HQM data also indicate
short-term variations on a timescale comparable
with the time lags between the observing campaigns.

\section{Statistical results}

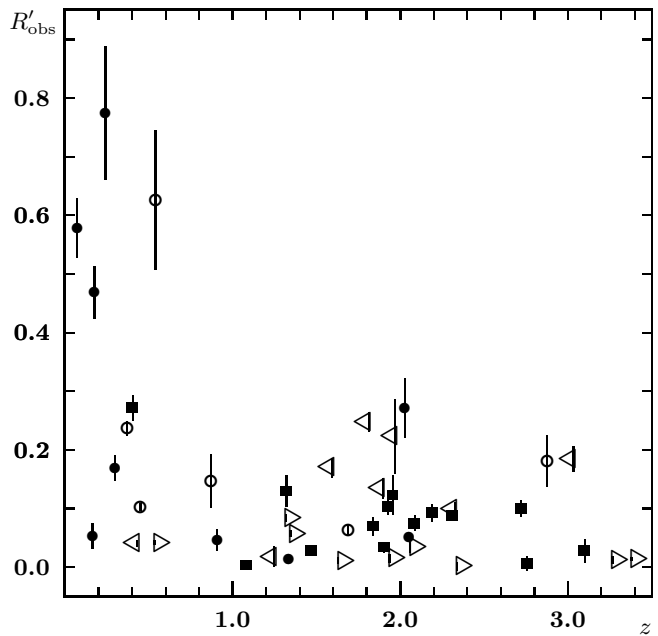
\begin{figure}

\hspace*{0.9cm}
\begin{picture}(7.8 ,7.8 )(0,0)
\put(0,0){\setlength{\unitlength}{2.228cm}%
\begin{picture}(3.5,   3.500)(0,0)
\put(0,0){\framebox(3.5,   3.500)[tl]{\begin{picture}(0,0)(0,0)
        \put(0,0){\makebox(0,0)[tr]{$R'_{\rm obs}$\hspace*{0.1cm}}}
        \put(3.5,-
3.500){\setlength{\unitlength}{1cm}\begin{picture}(0,0)(0,0)
            \put(0,-0.5){\makebox(0,0)[br]{$z$}}
        \end{picture}}
    \end{picture}}}

\put(0,0){\setlength{\unitlength}{7.8cm}\begin{picture}(0,0)(0,-0.05)
   \multiput(0,0)(0,0.1){10}{\setlength{\unitlength}{1cm}%
\begin{picture}(0,0)(0,0)
        \put(0,0){\line(1,0){0.12}}
        \end{picture}}
   \end{picture}}

\put(3.5,0){\setlength{\unitlength}{7.8cm}\begin{picture}(0,0)(0,-0.05)
   \multiput(0,0)(0,0.1){10}{\setlength{\unitlength}{1cm}%
\begin{picture}(0,0)(0,0)
        \put(0,0){\line(-1,0){0.12}}
        \end{picture}}
   \end{picture}}

\put(0,0){\setlength{\unitlength}{7.8cm}\begin{picture}(0,0)(0,-0.05)
   \put(0,0.8){\setlength{\unitlength}{1cm}\begin{picture}(0,0)(0,0)
        \put(0,0){\line(1,0){0.12}}
        \put(-0.2,0){\makebox(0,0)[r]{\bf 0.8}}
        \end{picture}}
   \put(0,0.6){\setlength{\unitlength}{1cm}\begin{picture}(0,0)(0,0)
        \put(0,0){\line(1,0){0.12}}
        \put(-0.2,0){\makebox(0,0)[r]{\bf 0.6}}
        \end{picture}}
   \put(0,0.4){\setlength{\unitlength}{1cm}\begin{picture}(0,0)(0,0)
        \put(0,0){\line(1,0){0.12}}
        \put(-0.2,0){\makebox(0,0)[r]{\bf 0.4}}
        \end{picture}}
   \put(0,0.2){\setlength{\unitlength}{1cm}\begin{picture}(0,0)(0,0)
        \put(0,0){\line(1,0){0.12}}
        \put(-0.2,0){\makebox(0,0)[r]{\bf 0.2}}
        \end{picture}}
   \put(0,0.0){\setlength{\unitlength}{1cm}\begin{picture}(0,0)(0,0)
        \put(0,0){\line(1,0){0.12}}
        \put(-0.2,0){\makebox(0,0)[r]{\bf 0.0}}
        \end{picture}}
   \put(0,0.2){\setlength{\unitlength}{1cm}\begin{picture}(0,0)(0,0)
        \put(0,0){\line(1,0){0.12}}
        \put(-0.2,0){\makebox(0,0)[r]{\bf 0.2}}
        \end{picture}}
   \put(0,0.4){\setlength{\unitlength}{1cm}\begin{picture}(0,0)(0,0)
        \put(0,0){\line(1,0){0.12}}
        \put(-0.2,0){\makebox(0,0)[r]{\bf 0.4}}
        \end{picture}}
   \put(0,0.6){\setlength{\unitlength}{1cm}\begin{picture}(0,0)(0,0)
        \put(0,0){\line(1,0){0.12}}
        \put(-0.2,0){\makebox(0,0)[r]{\bf 0.6}}
        \end{picture}}
   \put(0,0.8){\setlength{\unitlength}{1cm}\begin{picture}(0,0)(0,0)
        \put(0,0){\line(1,0){0.12}}
        \end{picture}}
   \end{picture}}

    \multiput(0,3.5)(0.2,0){18}%
        {\setlength{\unitlength}{1cm}\begin{picture}(0,0)(0,0)
        \put(0,0){\line(0,-1){0.12}}
    \end{picture}}
    \multiput(0,0)(0.2,0){18}%
        {\setlength{\unitlength}{1cm}\begin{picture}(0,0)(0,0)
        \put(0,0){\line(0,1){0.12}}
    \end{picture}}
    \put(1.0,0){\setlength{\unitlength}{1cm}\begin{picture}(0,0)(0,0)
        \put(0,0){\line(0,1){0.2}}
        \put(0,-0.2){\makebox(0,0)[t]{\bf 1.0}}
    \end{picture}}
    \put(2.0,0){\setlength{\unitlength}{1cm}\begin{picture}(0,0)(0,0)
        \put(0,0){\line(0,1){0.2}}
        \put(0,-0.2){\makebox(0,0)[t]{\bf 2.0}}
    \end{picture}}
   \put(3.0,0){\setlength{\unitlength}{1cm}\begin{picture}(0,0)(0,0)
        \put(0,0){\line(0,1){0.2}}
        \put(0,-0.2){\makebox(0,0)[t]{\bf 3.0}}
    \end{picture}}
   \put(3.0,0){\setlength{\unitlength}{1cm}\begin{picture}(0,0)(0,0)
        \put(0,0){\line(0,1){0.2}}
    \end{picture}}
    \put(3.0,0){\setlength{\unitlength}{1cm}\begin{picture}(0,0)(0,0)
        \put(0,0){\line(0,1){0.2}}
    \end{picture}}
    \put(3.0,0){\setlength{\unitlength}{1cm}\begin{picture}(0,0)(0,0)
       \put(0,0){\line(0,1){0.2}}
    \end{picture}}
    \put(3.0,0){\setlength{\unitlength}{1cm}\begin{picture}(0,0)(0,0)
       \put(0,0){\line(0,1){0.2}}
    \end{picture}}
\quadrat{2.310}{0.088}{0.002}
\quadrat{2.086}{0.075}{0.012}
\quadrat{1.955}{0.123}{0.034}
\quadrat{0.404}{0.272}{0.021}
\quadrat{2.756}{0.006}{0.011}
\quadrat{2.191}{0.093}{0.014}
\quadrat{1.904}{0.034}{0.009}
\quadrat{3.100}{0.028}{0.019}
\quadrat{1.321}{0.130}{0.026}
\quadrat{1.471}{0.028}{0.005}
\quadrat{1.840}{0.070}{0.015}
\quadrat{2.720}{0.100}{0.013}
\quadrat{1.929}{0.103}{0.013}
\quadrat{1.084}{0.004}{0.003}
\rfahne{3.380}{0.014}{0.002}
\rfahne{2.338}{0.001}{0.006}
\rfahne{1.316}{0.084}{0.006}
\rfahne{1.345}{0.057}{0.005}
\rfahne{0.533}{0.042}{0.003}
\rfahne{1.631}{0.011}{0.005}
\rfahne{2.060}{0.034}{0.009}
\rfahne{3.266}{0.012}{0.005}
\rfahne{1.936}{0.015}{0.006}
\lfahne{1.896}{0.134}{0.017}
\lfahne{3.035}{0.185}{0.021}
\lfahne{1.249}{0.018}{0.017}
\lfahne{1.970}{0.223}{0.063}
\lfahne{1.595}{0.170}{0.016}
\lfahne{2.328}{0.100}{0.010}
\lfahne{0.432}{0.041}{0.005}
\lfahne{1.813}{0.248}{0.016}
\kreisa{0.450}{0.102}{0.006}
\kreisa{1.690}{0.063}{0.008}
\kreisa{2.877}{0.181}{0.044}
\kreisa{0.871}{0.147}{0.044}
\kreisa{0.540}{0.626}{0.118}
\kreisa{0.371}{0.237}{0.012}
\kreis{2.027}{0.272}{0.050}
\kreis{0.174}{0.469}{0.044}
\kreis{0.909}{0.046}{0.018}
\kreis{0.072}{0.579}{0.050}
\kreis{2.051}{0.052}{0.008}
\kreis{1.334}{0.014}{0.006}
\kreis{0.164}{0.054}{0.021}
\kreis{0.297}{0.169}{0.021}
\kreis{0.241}{0.775}{0.113}

\end{picture}}

\end{picture}

\vspace*{0.5cm}
\caption{Maximum gradients $R'_{\rm obs}$ in the
observer's restframe plotted versus
$z$. {\em Filled squares:} radio quiet, X-ray quiet; $\bullet$:
radio quiet, X-ray loud; $\circ$: radio loud, X-ray loud;
$\triangleleft$: steep-spectrum radio quasars; $\triangleright$:
flat-spectrum radio quasars}
\end{figure}

As noted above, the sub-sample of quasars discussed in this paper results
from excluding those objects which were known to be of OVV type.
In Table~4, we list some reasonable parameters which describe the
lightcurves quantitatively. We give the total number $n'$ of data points
and the total timespan $\Delta t$ of observation; $n$ is the number of
points used for further analyses, data with the largest errors have been
dropped. The number $\hat{n}$ describes the sampling:
\begin{equation}
\hat{n}=\frac{(t_n-t_1)^2}{\sum_{i=1}^{n-1}(t_{i+1}-t_i)^2}+1
\end{equation}
For equally distributed measurements we have $\hat{n}=n$;
whereas $\hat{n}=2$ means that measurements were only carried out at
$t_1$ or $t_n$. The photometric errors in a lightcurve are described by two
parameters
\begin{equation}
\bar{\sigma}=\frac{1}{n}\sum_{i=1}^{n}\sigma_i\;\;\;\mbox{and}\;\;\;
\tilde{\sigma}^2=\frac{1}{n}\sum_{i=1}^{n}\sigma_i^2\;,
\end{equation}
where $\sigma_i$ is the error of a single measurement.
The standard deviation of the $n$ data points is denoted $\sigma_0$. We have
calculated polynomial fits of order $j<\hat{n}-1$ for each lightcurve. The
standard deviation of the points from such fits is then denoted $\sigma_j$.
Obviously $\sigma_j$ decreases with increasing $j$. A lightcurve is well
described by a polynomial fit of order $j$ when $\sigma_j\simeq\bar{\sigma}$.
$R'_{\rm obs}$ is the {\em maximum\/} observed gradient in each lightcurve;
the value is a conservative one, well determined by several data.
$\Delta R_{40}$ is the result of our POSS photometry.

The maximum lightcurve gradients $R'_{\rm obs}$ are plotted versus
redshift $z$ in Fig.~4. It is obvious that there is an
anti-correlation. When transforming the variations into the quasar restframes
\begin{equation}
R'_{\rm qso}=(1+z)R'_{\rm obs}\;,
\end{equation}
and plotting $R'_{\rm qso}$ versus $z$, cf.\ Fig.~5, there seems to
be no correlation any more. Thus we have no indication for a redshift
dependence of the intrinsic variability behaviour. It is also interesting
that there is no dependence on the absolute luminosity, $M_V$ (not plotted
here).

\begin{figure}

\hspace*{0.9cm}
\begin{picture}(7.8 ,9.3 )(0,0)
\put(0,0){\setlength{\unitlength}{2.228cm}%
\begin{picture}(3.5,   4.173)(0,0)
\put(0,0){\framebox(3.5,   4.173)[tl]{\begin{picture}(0,0)(0,0)
        \put(0,0){\makebox(0,0)[tr]{$R'_{\rm qso}$\hspace*{0.1cm}}}
        \put(3.5,-4.173){\setlength{\unitlength}{1cm}\begin{picture}(0,0)(0,0)
            \put(0,-0.5){\makebox(0,0)[br]{$z$}}
        \end{picture}}
    \end{picture}}}

\put(0,0){\setlength{\unitlength}{7.8cm}\begin{picture}(0,0)(0,-0.05)
   \multiput(0,0)(0,0.1){12}{\setlength{\unitlength}{1cm}%
\begin{picture}(0,0)(0,0)
        \put(0,0){\line(1,0){0.12}}
        \end{picture}}
   \end{picture}}

\put(3.5,0){\setlength{\unitlength}{7.8cm}\begin{picture}(0,0)(0,-0.05)
   \multiput(0,0)(0,0.1){12}{\setlength{\unitlength}{1cm}%
\begin{picture}(0,0)(0,0)
        \put(0,0){\line(-1,0){0.12}}
        \end{picture}}
   \end{picture}}

\put(0,0){\setlength{\unitlength}{7.8cm}\begin{picture}(0,0)(0,-0.05)
   \put(0,0.8){\setlength{\unitlength}{1cm}\begin{picture}(0,0)(0,0)
        \put(0,0){\line(1,0){0.12}}
        \put(-0.2,0){\makebox(0,0)[r]{\bf 0.8}}
        \end{picture}}
   \put(0,0.6){\setlength{\unitlength}{1cm}\begin{picture}(0,0)(0,0)
        \put(0,0){\line(1,0){0.12}}
        \put(-0.2,0){\makebox(0,0)[r]{\bf 0.6}}
        \end{picture}}
   \put(0,0.4){\setlength{\unitlength}{1cm}\begin{picture}(0,0)(0,0)
        \put(0,0){\line(1,0){0.12}}
        \put(-0.2,0){\makebox(0,0)[r]{\bf 0.4}}
        \end{picture}}
   \put(0,1.0){\setlength{\unitlength}{1cm}\begin{picture}(0,0)(0,0)
        \put(0,0){\line(1,0){0.12}}
        \put(-0.2,0){\makebox(0,0)[r]{\bf 1.0}}
        \end{picture}}
   \put(0,0.0){\setlength{\unitlength}{1cm}\begin{picture}(0,0)(0,0)
        \put(0,0){\line(1,0){0.12}}
        \put(-0.2,0){\makebox(0,0)[r]{\bf 0.0}}
        \end{picture}}
   \put(0,0.2){\setlength{\unitlength}{1cm}\begin{picture}(0,0)(0,0)
        \put(0,0){\line(1,0){0.12}}
        \put(-0.2,0){\makebox(0,0)[r]{\bf 0.2}}
        \end{picture}}
   \put(0,0.4){\setlength{\unitlength}{1cm}\begin{picture}(0,0)(0,0)
        \put(0,0){\line(1,0){0.12}}
        \put(-0.2,0){\makebox(0,0)[r]{\bf 0.4}}
        \end{picture}}
   \put(0,0.6){\setlength{\unitlength}{1cm}\begin{picture}(0,0)(0,0)
        \put(0,0){\line(1,0){0.12}}
        \put(-0.2,0){\makebox(0,0)[r]{\bf 0.6}}
        \end{picture}}
   \put(0,0.8){\setlength{\unitlength}{1cm}\begin{picture}(0,0)(0,0)
        \put(0,0){\line(1,0){0.12}}
        \end{picture}}
   \end{picture}}

    \multiput(0,4.173)(0.2,0){18}%
        {\setlength{\unitlength}{1cm}\begin{picture}(0,0)(0,0)
        \put(0,0){\line(0,-1){0.12}}
    \end{picture}}
    \multiput(0,0)(0.2,0){18}%
        {\setlength{\unitlength}{1cm}\begin{picture}(0,0)(0,0)
        \put(0,0){\line(0,1){0.12}}
    \end{picture}}
    \put(1.0,0){\setlength{\unitlength}{1cm}\begin{picture}(0,0)(0,0)
        \put(0,0){\line(0,1){0.2}}
        \put(0,-0.2){\makebox(0,0)[t]{\bf 1.0}}
    \end{picture}}
    \put(2.0,0){\setlength{\unitlength}{1cm}\begin{picture}(0,0)(0,0)
        \put(0,0){\line(0,1){0.2}}
        \put(0,-0.2){\makebox(0,0)[t]{\bf 2.0}}
    \end{picture}}
   \put(3.0,0){\setlength{\unitlength}{1cm}\begin{picture}(0,0)(0,0)
        \put(0,0){\line(0,1){0.2}}
        \put(0,-0.2){\makebox(0,0)[t]{\bf 3.0}}
    \end{picture}}
   \put(3.0,0){\setlength{\unitlength}{1cm}\begin{picture}(0,0)(0,0)
        \put(0,0){\line(0,1){0.2}}
        \put(0,-0.2){\makebox(0,0)[t]{\bf 3.0}}
    \end{picture}}
    \put(3.0,0){\setlength{\unitlength}{1cm}\begin{picture}(0,0)(0,0)
        \put(0,0){\line(0,1){0.2}}
        \put(0,-0.2){\makebox(0,0)[t]{\bf 3.0}}
    \end{picture}}
    \put(3.0,0){\setlength{\unitlength}{1cm}\begin{picture}(0,0)(0,0)
       \put(0,0){\line(0,1){0.2}}
        \put(0,-0.2){\makebox(0,0)[t]{\bf 3.0}}
    \end{picture}}
    \put(3.0,0){\setlength{\unitlength}{1cm}\begin{picture}(0,0)(0,0)
       \put(0,0){\line(0,1){0.2}}
        \put(0,-0.2){\makebox(0,0)[t]{\bf 3.0}}
    \end{picture}}

\quadrat{2.310}{0.291}{0.007}
\quadrat{2.086}{0.231}{0.037}
\quadrat{1.955}{0.363}{0.100}
\quadrat{0.404}{0.382}{0.029}
\quadrat{2.756}{0.023}{0.041}
\quadrat{2.191}{0.297}{0.045}
\quadrat{1.904}{0.099}{0.026}
\quadrat{3.100}{0.115}{0.078}
\quadrat{1.321}{0.302}{0.060}
\quadrat{1.471}{0.069}{0.012}
\quadrat{1.840}{0.199}{0.043}
\quadrat{2.720}{0.372}{0.048}
\quadrat{1.929}{0.302}{0.038}
\quadrat{1.084}{0.008}{0.006}
\rfahne{3.380}{0.061}{0.009}
\rfahne{2.338}{0.003}{0.020}
\rfahne{1.316}{0.195}{0.014}
\rfahne{1.345}{0.134}{0.012}
\rfahne{0.533}{0.064}{0.005}
\rfahne{1.631}{0.029}{0.013}
\rfahne{2.060}{0.104}{0.028}
\rfahne{3.266}{0.051}{0.021}
\rfahne{1.936}{0.044}{0.018}
\lfahne{1.896}{0.388}{0.049}
\lfahne{3.035}{0.746}{0.085}
\lfahne{1.249}{0.040}{0.038}
\lfahne{1.970}{0.662}{0.187}
\lfahne{1.595}{0.441}{0.042}
\lfahne{2.328}{0.333}{0.033}
\lfahne{0.432}{0.059}{0.007}
\lfahne{1.813}{0.698}{0.045}
\kreisa{0.450}{0.148}{0.009}
\kreisa{1.690}{0.169}{0.022}
\kreisa{2.877}{0.702}{0.171}
\kreisa{0.871}{0.275}{0.082}
\kreisa{0.540}{0.964}{0.182}
\kreisa{0.371}{0.325}{0.016}
\kreis{2.027}{0.823}{0.151}
\kreis{0.174}{0.551}{0.052}
\kreis{0.909}{0.088}{0.034}
\kreis{0.072}{0.621}{0.054}
\kreis{2.051}{0.159}{0.024}
\kreis{1.334}{0.033}{0.014}
\kreis{0.164}{0.063}{0.024}
\kreis{0.297}{0.219}{0.027}
\kreis{0.241}{0.962}{0.140}

\end{picture}}

\end{picture}

\vspace*{0.5cm}
\caption{Maximum gradients $R'_{\rm qso}$ in
the quasars' restframes plotted versus $z$. Same symbols as in the
previous figure}
\end{figure}
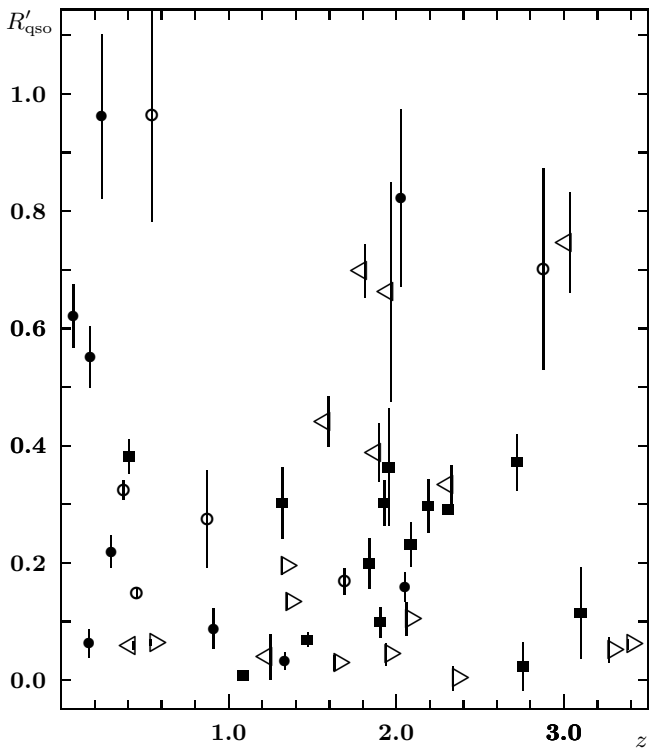

\subsection{Correlation with radio properties}

The intrinsic lightcurve gradients seem, however, to be correlated with the
radio and X-ray properties of the quasars. None of the radio {\em and\/}
X-ray quiet objects, marked by filled squares in Fig.~5, has an intrinsic
lightcurve gradient above 0.4\,mag\,yr$^{-1}$. Even less variable are the
flat-spectrum radio quasars, marked in Fig.~5 by triangles pointing to the
right; none has an intrinsic lightcurve gradient above 0.2\,mag\,yr$^{-1}$.
In Table~5, we list the average intrinsic variation indices for three
different sub-samples.

\begin{table}
\caption[ ]{Average intrinsic variation indices $<\!\! R'_{\rm qso} \!\!>$
           for different quasar sub-samples}
\begin{flushleft}
\begin{tabular}{l|rc}
\hline
Sub-sample & $N$& $<\!\! R'_{\rm qso} \!\!>$ \\
\hline
Radio quiet, X-ray quiet & 14 & 0.218$\pm$0.037\\
Radio-Flat  &  9 & 0.076$\pm$0.021\\
Others      &  23 & 0.412$\pm$0.066\\
\hline
\end{tabular}
\end{flushleft}
\end{table}

Bregman (\cite{Bre90}) argues that all quasars can be divided into two
broad classes: radio quiet objects whose continuum emission is dominated by
thermal emission and blazars showing predominately non-thermal emission.
Note that many objects may represent mixed types. Angel and Stockman
(\cite{AS80}) defined blazars as objects showing flat radio spectra,
violent radio and optical variability, high and strongly variable optical
polarization and steep non-thermal optical continua. All objects of our
total HQM sample which show violent optical variability are radio loud.
Most of them have a flat radio spectrum and fulfill also other blazar
criteria if one has looked for them; the class of OVVs will be discussed in
more detail in a subsequent paper. None of the radio-quiet objects in our
sample show violent optical variability. Following Bregman, the optical
emission of these objects is thermal radiation from an accretion disk. We
conclude from our data that {\em the thermal radiation from an accretion
disk is only weakly variable.\/} X-ray loud and/or steep-spectrum radio
quasars are a bit more strongly variable. We therefore assume an additional
radiation mechanism in the optical. Possibly, these objects represent
certain mixed types as noted above.

Our lightcurves clearly show that not all flat-spectrum radio quasars are
blazars. The flat-spectrum sources discussed in this paper are even less
variable in the optical than radio quiet quasars. One should keep in mind
that our observations only cover a relatively short time-span; however, L84
already noted that there is ``a population of well-observed compact
flat-spectrum radio sources which undergo only modest variations over
periods of $\sim$\,80\,yr''. For some objects it is known that they are
more strongly
variable in the radio than in the optical (e.g.\ GC\,0248+430). If
the variable radio emission originates in synchrotron radiation from a jet,
the jet seems to be quiet in the optical for these objects. The optical
spectrum should therefore be similar to that of radio quiet objects.

We can only speculate about the origin of the very low degree of optical
variability for this class of flat-spectrum radio sources. If one believes
that the radio properties indicate a jet roughly orientated towards the
observer, the accretion disk will be viewed face on. This orientation
effect may be responsible for the extremely low optical variability.

\subsection{Microlensing}

Chang \& Refsdal (\cite{CR79}) were the first to show theoretically that
stars in a foreground galaxy may induce significant non-intrinsic
variability in the lightcurve of a quasar due to gravitational
microlensing. Gott (\cite{Got81}) proposed that it should be possible to
detect low-mass stars in galaxy haloes by this effect. Chang \& Refsdal
(\cite{CR84}), Paczynski (\cite{Pac86}) and KRS developed a set
of parameters to describe the microlensing properties for a given star
field. The first realistic light-curve simulations were carried out by
Paczynski (\cite{Pac86}), KRS and Schneider \& Weiss (\cite{SW87}). In
these papers, the authors assumed for simplicity that all stars in front of
the quasar are of equal mass. Kayser et al. (\cite{KWRS89}) computed light
curves for star fields containing objects of different masses.

Microlensing is now well-understood theoretically, and there is already
some observational evidence for it. The most exciting data are those
of Corrigan et al. (\cite{CIHW90,CIAF91}) who detected differences between
the lightcurves of the four components of the ``Einstein cross'' 2237+030
which must be a result of microlensing; however, the lightcurves are not
sufficiently resolved in time and the measurement errors are relatively
large. Vanderriest et al.\ (\cite{VSHC89}) tried to determine the time
delay $\Delta T$ between the components A and B of the ``Double Quasar''
0957+561 and found a probable value of $\Delta T = 415 \pm 20 \,$days.
After shifting the B-lightcurve by $\Delta T$ there remained some
differences which Vanderriest et al.\ interpreted in terms of microlensing.
(There is also a difference in the $A/B$ flux ratios between the radio, the
emission lines and the optical continuum, which might be a result of
microlensing.) Angonin et al. (\cite{ARSV90}) found differences in the
equivalent widths and/or the profiles of the broad emission as well as the
broad absorption lines between the four images of the ``Clover Leaf''
H\,1413+117 and discussed this in terms of microlensing.

\begin{figure}

\hspace*{0.9cm}
\begin{picture}(7.8 ,9.3 )(0,0)
\put(0,0){\setlength{\unitlength}{2.228cm}%
\begin{picture}(3.5,   4.173)(0,0)
\put(0,0){\framebox(3.5,   4.173)[tl]{\begin{picture}(0,0)(0,0)
        \put(3.15,-0.2){\makebox(0,0)[tl]{\large{\em }}}
        \put(0,0){\makebox(0,0)[tr]{$q$\hspace*{0.1cm}}}
        \put(3.5,-4.173){\setlength{\unitlength}{1cm}\begin{picture}(0,0)(0,0)
            \put(0,-0.5){\makebox(0,0)[br]{$z$}}
        \end{picture}}
    \end{picture}}}

\put(0,0){\setlength{\unitlength}{7.8cm}\begin{picture}(0,0)(0,-0.05)
   \put(0,0.824){\setlength{\unitlength}{1cm}\begin{picture}(0,0)(0,0)
        \put(0,0){\line(1,0){0.12}}
        \put(-0.2,0){\makebox(0,0)[r]{\bf 2.0}}
        \end{picture}}
   \put(0,0.618){\setlength{\unitlength}{1cm}\begin{picture}(0,0)(0,0)
        \put(0,0){\line(1,0){0.12}}
        \put(-0.2,0){\makebox(0,0)[r]{\bf 1.5}}
        \end{picture}}
   \put(0,0.412){\setlength{\unitlength}{1cm}\begin{picture}(0,0)(0,0)
        \put(0,0){\line(1,0){0.12}}
        \put(-0.2,0){\makebox(0,0)[r]{\bf 1.0}}
        \end{picture}}
   \put(0,0.0){\setlength{\unitlength}{1cm}\begin{picture}(0,0)(0,0)
        \put(0,0){\line(1,0){0.12}}
        \put(-0.2,0){\makebox(0,0)[r]{\bf 0.0}}
        \end{picture}}
   \put(0,0.206){\setlength{\unitlength}{1cm}\begin{picture}(0,0)(0,0)
        \put(0,0){\line(1,0){0.12}}
        \put(-0.2,0){\makebox(0,0)[r]{\bf 0.5}}
        \end{picture}}
   \end{picture}}

\put(3.5,0){\setlength{\unitlength}{7.8cm}\begin{picture}(0,0)(0,-0.05)
   \put(0,0.824){\setlength{\unitlength}{1cm}\begin{picture}(0,0)(0,0)
        \put(0,0){\line(-1,0){0.12}}
        \end{picture}}
   \put(0,0.618){\setlength{\unitlength}{1cm}\begin{picture}(0,0)(0,0)
        \put(0,0){\line(-1,0){0.12}}
        \end{picture}}
   \put(0,0.412){\setlength{\unitlength}{1cm}\begin{picture}(0,0)(0,0)
        \put(0,0){\line(-1,0){0.12}}
        \end{picture}}
   \put(0,0.0){\setlength{\unitlength}{1cm}\begin{picture}(0,0)(0,0)
        \put(0,0){\line(-1,0){0.12}}
        \end{picture}}
   \put(0,0.206){\setlength{\unitlength}{1cm}\begin{picture}(0,0)(0,0)
        \put(0,0){\line(-1,0){0.12}}
        \end{picture}}
   \end{picture}}

    \multiput(0,4.173)(0.2,0){18}%
        {\setlength{\unitlength}{1cm}\begin{picture}(0,0)(0,0)
        \put(0,0){\line(0,-1){0.12}}
    \end{picture}}
    \multiput(0,0)(0.2,0){18}%
        {\setlength{\unitlength}{1cm}\begin{picture}(0,0)(0,0)
        \put(0,0){\line(0,1){0.12}}
    \end{picture}}
    \put(1.0,0){\setlength{\unitlength}{1cm}\begin{picture}(0,0)(0,0)
        \put(0,0){\line(0,1){0.2}}
        \put(0,-0.2){\makebox(0,0)[t]{\bf 1.0}}
    \end{picture}}
    \put(2.0,0){\setlength{\unitlength}{1cm}\begin{picture}(0,0)(0,0)
        \put(0,0){\line(0,1){0.2}}
        \put(0,-0.2){\makebox(0,0)[t]{\bf 2.0}}
    \end{picture}}
   \put(3.0,0){\setlength{\unitlength}{1cm}\begin{picture}(0,0)(0,0)
        \put(0,0){\line(0,1){0.2}}
        \put(0,-0.2){\makebox(0,0)[t]{\bf 3.0}}
    \end{picture}}
   \put(3.0,0){\setlength{\unitlength}{1cm}\begin{picture}(0,0)(0,0)
        \put(0,0){\line(0,1){0.2}}
        \put(0,-0.2){\makebox(0,0)[t]{\bf 3.0}}
    \end{picture}}
    \put(3.0,0){\setlength{\unitlength}{1cm}\begin{picture}(0,0)(0,0)
        \put(0,0){\line(0,1){0.2}}
        \put(0,-0.2){\makebox(0,0)[t]{\bf 3.0}}
    \end{picture}}
    \put(3.0,0){\setlength{\unitlength}{1cm}\begin{picture}(0,0)(0,0)
       \put(0,0){\line(0,1){0.2}}
        \put(0,-0.2){\makebox(0,0)[t]{\bf 3.0}}
    \end{picture}}
    \put(3.0,0){\setlength{\unitlength}{1cm}\begin{picture}(0,0)(0,0)
       \put(0,0){\line(0,1){0.2}}
        \put(0,-0.2){\makebox(0,0)[t]{\bf 3.0}}
    \end{picture}}

\kreis{2.310}{0.550}{0.013}
\quadrat{2.086}{0.437}{0.070}
\quadrat{1.955}{0.686}{0.189}
\kreis{0.404}{0.722}{0.055}
\kreis{2.756}{0.043}{0.077}
\kreis{2.191}{0.561}{0.085}
\quadrata{1.904}{0.187}{0.049}
\quadrata{3.100}{0.217}{0.147}
\kreis{1.321}{0.571}{0.113}
\quadrata{1.471}{0.130}{0.023}
\quadrata{1.840}{0.376}{0.081}
\kreis{2.720}{0.703}{0.091}
\quadrata{1.929}{0.571}{0.072}
\quadrata{1.084}{0.015}{0.011}

\quadrat{3.380}{0.331}{0.049}
\quadrata{2.338}{0.016}{0.018}
\kreis{1.316}{1.057}{0.076}
\quadrata{1.345}{0.726}{0.065}
\kreisa{0.533}{0.347}{0.027}
\quadrata{1.631}{0.157}{0.070}
\quadrata{2.060}{0.564}{0.152}
\quadrat{3.266}{0.276}{0.114}
\quadrata{1.936}{0.239}{0.096}

\kreis{1.896}{0.388}{0.049}
\quadrat{3.035}{0.746}{0.085}
\kreisa{1.249}{0.040}{0.038}
\quadrata{1.970}{0.662}{0.187}
\quadrat{1.595}{0.441}{0.042}
\kreis{2.328}{0.333}{0.033}
\kreisa{0.432}{0.059}{0.007}
\kreis{1.813}{0.698}{0.045}
\quadrata{0.450}{0.148}{0.009}
\kreisa{1.690}{0.169}{0.022}
\quadrat{2.877}{0.702}{0.171}
\kreis{0.871}{0.275}{0.082}
\quadrata{0.540}{0.964}{0.182}
\quadrata{0.371}{0.325}{0.016}
\kreis{2.027}{0.823}{0.151}
\kreisa{0.174}{0.551}{0.052}
\kreis{0.909}{0.088}{0.034}
\kreis{0.072}{0.621}{0.054}
\quadrata{2.051}{0.159}{0.024}
\quadrat{1.334}{0.033}{0.014}
\kreisa{0.164}{0.063}{0.024}
\quadrata{0.297}{0.219}{0.027}
\kreis{0.241}{0.962}{0.140}

\end{picture}}

\end{picture}

\vspace*{0.5cm}
\caption{Maximum variation relative to the average variation of the
         corresponding quasar sub-sample, $q$, versus redshift $z$.
         {\em Filled circles:} {\em G1\/}-objects; {\em open circles:}
         {\em G2\/}-objects; {\em open squares:}
         {\em A1\/}-objects; {\em Filled squares:} other objects}
\end{figure}
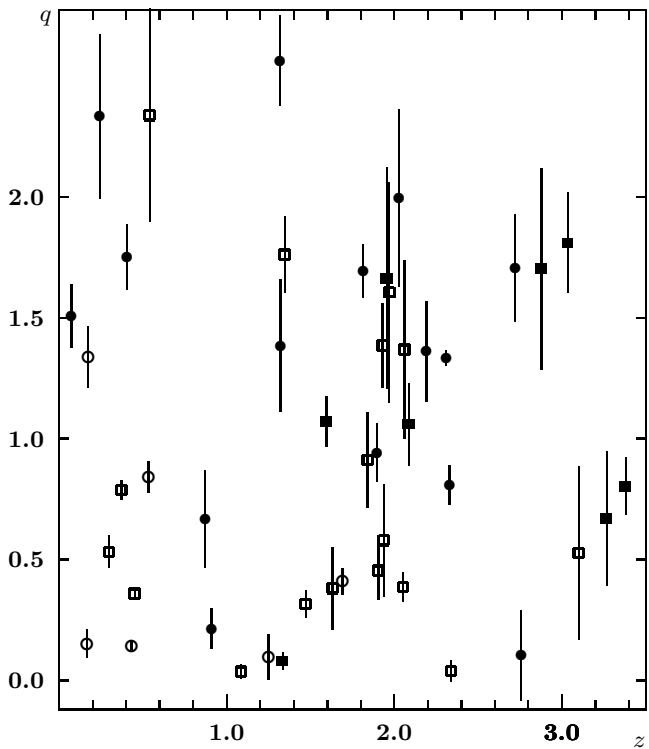

The HQM quasar sample has been selected to include good candidates for
microlensing variability. We did, however, not
succeed in discovering a typical {\em high amplification event\/} (cf.\
KRS) in any of our lightcurves. All flares recorded up to now occurred in
radio loud objects and were very probably of intrinsic origin (cf.\ the
detailed discussion of the 3C\,345 lightcurve in Schramm et al.\
\cite{SBCW93}). Thus, if microlensing is present in our sample it leads
only to low amplitude variations. Such variations occur if the source is
larger than the Einstein radii of the foreground stars; see, e.g.,
Refsdal \& Stabell (\cite{RS91}). In this case the microlensing variations
do not show a typical shape so that they can only be found statistically.
In Fig.~6, we have plotted the variability index
\begin{equation}
q=R'_{\rm qso}/<\!\! R'_{\rm qso}\!\!>
\end{equation}
for each quasar. Our sample includes 14 {\em G1\/}-objects for which we
assume the highest microlensing probability; 9 of them are more strongly
variable than the average in the corresponding sub-class. We regard this
result as a statistical hint of microlensing. Note that there is no
indication of enhanced variability for the {\em A1\/}-objects. Only further
observations and a better theoretical understanding of the intrinsic
variability mechanisms may lead to a better knowledge on the importance
of microlensing.

\bigskip

\acknowledgements{It would have been impossible to carry out our
          program without the excellent support by the MPIA and the Calar
          Alto staff members. We like to thank J.~v.~Linde, M.-D.~Linnert,
          D.~Mehlert, L.~Nieser, M.~Schaaf, T.~Schramm
          and H.J.~Witt for their help during lots of long and
          sometimes painful cold nights at Calar Alto and P.~King and
          S.~Refsdal for carefully reading the manuscript. This work has
          been supported by the Deutsche Forschungsgemeinschaft
          under Bo$\,$904/1, Re$\,$439/5 and Schr\,292/6.}

\end{document}